\documentclass[aps,preprint,floatfix,nofootinbib,showpacs]{revtex4-1}
\pdfoutput=1
\usepackage{graphicx,color}
\usepackage{hyperref}

\def\lsim{\:\raisebox{-0.5ex}{$\stackrel{\textstyle<}{\sim}$}\:}

\begin{document}

{\small
\begin{flushright}
IUEP-HEP-18-03
\end{flushright} }

\title{New Emerging Results in
Higgs Precision Analysis Updates 2018 \\
after Establishment of Third--Generation Yukawa Couplings}

\renewcommand{\thefootnote}{\arabic{footnote}}

\author{
Kingman Cheung$^{1,2,3,4}$, Jae Sik Lee$^{5,6,1}$, and
Po-Yan Tseng$^{7,1}$}
\affiliation{
$^1$ Physics Division, National Center for Theoretical Sciences,
Hsinchu, Taiwan \\
$^2$ Department of Physics, National Tsing Hua University,
Hsinchu 300, Taiwan \\
$^3$ Division of Quantum Phases and Devices, School of Physics, 
Konkuk University, Seoul 143-701, Republic of Korea \\
$^4$ Department of Physics, National Central University, Chungli, Taiwan\\
$^5$ Department of Physics 
Chonnam National University, \\
300 Yongbong-dong, Buk-gu, Gwangju, 500-757, Republic of Korea \\
$^6$ Institute for Universe and Elementary Particles, Chonnam National University, \\
300 Yongbong-dong, Buk-gu, Gwangju, 500-757, Republic of Korea\\
$^7$Kavli IPMU (WPI), UTIAS, The University of Tokyo, 
Kashiwa, Chiba 277-8583, Japan
}
\date{October 5, 2018}

\begin{abstract}
We perform global fits of the Higgs boson couplings to 
all the 7 TeV, 8 TeV, and 13 TeV data available up to the Summer 2018.
New measurements at 13 TeV extend to include the Higgs signal strengths
exclusively measured in associated Higgs production with top-quark pair and
the third-generation Yukawa couplings now have been established.
Some important consequences emerge from the global fits. (i) The overall
average signal strength of the Higgs boson stands at $2\sigma$ above the
SM value ($\mu = 1.10 \pm 0.05 $). (ii) For the first time
the bottom-quark Yukawa
coupling shows a preference of the positive sign to the negative one.
(iii) The negative top-quark Yukawa coupling is completely ruled out 
unless there exist additional particles running in the 
$H$-$\gamma$-$\gamma$ loop
with contributions equal to two times
the SM top-quark contribution within about 10 \%.
(iv) The branching ratio for nonstandard decays of the
Higgs boson is now below 8.4\% at the 95\% confidence level.

\end{abstract}

\maketitle

\section{Introduction}

Ever since the discovery of a standard model (SM) like Higgs boson in 2012
\cite{atlas,cms},
the main focus of the LHC experiments has been 
put on fully establishing its identity.
Though the initial data sets till the summer 2013
indicated that it might be different from the SM
Higgs boson \cite{higgcision}, the data sets collected till the summer 2014 
showed that the data is best described by the SM Higgs boson 
\cite{update2014}.
Ever since then more production channels and decay channels of the Higgs boson
are established. 
On the production side, in addition to gluon fusion (ggF),
vector-boson fusion (VBF), the associated production
with a $V=W/Z$ boson (VH), and the associated production with a 
top-quark pair (ttH) have been extensively investigated 
\cite{ICHEP_tth_CMS_VVtautau,Aaboud:2018urx}.
On the decay side, $H \to b \bar b$ \cite{Sirunyan:2018kst,Aaboud:2017xsd}
and $H\to\tau\tau$ \cite{Sirunyan:2017khh,ICHEP_htautau_ATLAS}
were also very recently established in single measurements
\footnote{
We note that, in combined measurements of CMS and ATLAS,
$H\to\tau\tau$ was already established in Run I
\cite{Khachatryan:2016vau}.}.
It is the right timing to perform the global fits to all Higgs-boson 
signal strengths in various scenarios of new physics, generically labeled by
{\bf CPC$n$} and {\bf CPV$n$} in this work
with $n$ standing for the number of fitting parameters.

In this work, we analyze the direct Higgs data collected at the
Tevatron and the LHC adopting the formalism suggested in
Ref.~\cite{higgcision} to study the impact of the established
third--generation Yukawa couplings on the nature of 125--GeV Higgs.
More precisely, we use 3 signal strengths measured at 
the Tevatron~\cite{tevatron_aa_ww,tevatron_bb}
and, for the Higgs-boson data at 7 and 8 TeV, we use
20 signal strengths and the correlation matrix obtained
in the combined ATLAS and CMS analysis~\cite{Khachatryan:2016vau}.
On the other hand, for the 13 TeV data, we use
41 signal strengths in total.
Since any information on correlations between ATLAS and CMS data and
those among various channels is not currently available,
it is assumed that each data at 13 TeV is Gaussian distributed 
and correlations among them are ignored accordingly. 
And, when we combine them, we simply take a $\chi^2$ method.
For the details of the 13 TeV data, we refer to Appendix B.

Some very interesting results emerge from the new global fits, which 
were not realized previously.
\begin{enumerate}

\item
 The combined average signal strength of the Higgs boson now stands
at a $2$-$\sigma$ deviation from the SM value, namely $\mu_{\rm exp} 
= 1.10 \pm 0.05$.

\item 
For the first time the bottom-Yukawa coupling shows statistical difference
between the positive and negative signs. Thanks to the discriminating power
of the Higgs-gluon vertex $S^g$ the positive sign of the bottom-Yukawa
is more preferred than the negative one.

\item 
Previously in 2014 the fits still allowed the negative sign of the
top-Yukawa coupling at the 95\% confidence level (CL).  
Now with more precisely measured signal 
strengths together with the establishment of the associated production with
the top-quark pair, the negative island of the top-Yukawa is now 
entirely
ruled out, except in the scenarios with non-zero
$\Delta S^\gamma$.
Even with $\Delta S^\gamma\neq 0$, it 
has to be adjusted within 10 \% of two times 
the SM top-quark contribution.
This tuning is going to be more and more 
severe as more data accumulate.

\item
The nonstandard (or invisible decay) branching ratio of the Higgs boson
is now reduced to less than 8.4\% at the 95\% CL
which improves substantially from
the previous value of 19\%. This is obtained by varying only 
$\Delta \Gamma_{\rm tot}$.

\end{enumerate}

The organization of the paper is as follows. 
In the next section, we describe briefly our formalism to make
this work more self-contained.
In Sec. III, we show the data for the Higgs signal strengths.
In Sec. IV, we show the results for all the fits.
We conclude in Sec. V.
In Appendix A, 
we describe the correspondence between the coupling modifiers
of our work with those of the LHCHXSWG \cite{YR3,YR4} and of
a recent  ATLAS paper \cite{atlas-k}.
In Appendix B, we list all the 13 TeV Higgs boson data that we
use in our global fitting.

\section{Formalism} 
In order to make the current presentation more self-contained, we 
include here brief description of the formalism that we use in 
calculating the signal strengths and chi-squares.
We follow the conventions and notations of 
{\tt CPsuperH}~\cite{Lee:2003nta,Lee:2007gn,Lee:2012wa}
for the Higgs couplings to the SM particles assuming the Higgs boson is 
a generally CP-mixed state without carrying any definite CP--parity.

\begin{itemize}
%
\item\underline{Higgs couplings to fermions}:
\begin{equation}
\label{eq1}
{\cal L}_{H\bar{f}f}\ =\ - \sum_{f=u,d,l}\,\frac{g m_f}{2 M_W}\,
\sum_{i=1}^3\, H\, \bar{f}\,\Big( g^S_{H\bar{f}f}\, +\,
ig^P_{H\bar{f}f}\gamma_5 \Big)\, f\ .
\end{equation}
For the SM couplings, $g^S_{H\bar{f}f}=1$ and $g^P_{H\bar{f}f}=0$.
\item\underline{Higgs couplings to the massive vector bosons}:
\begin{equation}
\label{eq2}
{\cal L}_{HVV}  =  g\,M_W \, \left(
g_{_{HWW}} W^+_\mu W^{- \mu}\ + \
g_{_{HZZ}} \frac{1}{2c_W^2}\,Z_\mu Z^\mu\right) \, H\,.
\end{equation}
For the SM couplings, 
we have $g_{_{HWW}}=g_{_{HZZ}}\equiv g_{_{HVV}}=1$, respecting 
the custodial symmetry.

\item\underline{Higgs couplings to two photons}:
The amplitude for the decay process
$H \rightarrow \gamma\gamma$ can be written as
\begin{equation} \label{hipp}
{\cal M}_{\gamma\gamma H}=-\frac{\alpha M_{H}^2}{4\pi\,v}
\bigg\{S^\gamma(M_{H})\,
\left(\epsilon^*_{1\perp}\cdot\epsilon^*_{2\perp}\right)
 -P^\gamma(M_{H})\frac{2}{M_{H}^2}
\langle\epsilon^*_1\epsilon^*_2 k_1k_2\rangle
\bigg\}\,,
\end{equation}
where $k_{1,2}$ are the momenta of the two photons and
$\epsilon_{1,2}$ the wave vectors of the corresponding photons,
$\epsilon^\mu_{1\perp} = \epsilon^\mu_1 - 2k^\mu_1 (k_2 \cdot
\epsilon_1) / M^2_{H}$, $\epsilon^\mu_{2\perp} = \epsilon^\mu_2 -
2k^\mu_2 (k_1 \cdot \epsilon_2) / M^2_{H}$ and $\langle \epsilon_1
\epsilon_2 k_1 k_2 \rangle \equiv \epsilon_{\mu\nu\rho\sigma}\,
\epsilon_1^\mu \epsilon_2^\nu k_1^\rho k_2^\sigma$. 
The decay rate of $H\to \gamma\gamma$ is 
proportional to $|S^\gamma|^2 + |P^\gamma|^2$.
Including some additional loop contributions from new particles,
the scalar and
pseudoscalar form factors, retaining only the dominant loop
contributions from the third--generation fermions and $W^\pm$,
are given by
\footnote{
For the loop functions of $F_{sf,pf,1}(\tau)$, 
we refer to, for example, Ref.~\cite{Lee:2003nta}.}
\begin{eqnarray}
S^\gamma(M_{H})&=&2\sum_{f=b,t,\tau} N_C\,
Q_f^2\, g^{S}_{H\bar{f}f}\,F_{sf}(\tau_{f}) 
- g_{_{HWW}}F_1(\tau_{W}) 
+ \Delta S^\gamma \,, \nonumber \\
P^\gamma(M_{H})&=&2\sum_{f=b,t,\tau}
N_C\,Q_f^2\,g^{P}_{H\bar{f}f}\,F_{pf}(\tau_{f})
+ \Delta P^\gamma \,, 
\end{eqnarray}
where $\tau_{x}=M_{H}^2/4m_x^2$, $N_C=3$ for quarks and $N_C=1$ for
taus, respectively.
The additional contributions $\Delta S^\gamma$ and $\Delta P^\gamma$
are assumed to be real in our work,
as there are unlikely any new
charged particles lighter than $M_H/2$.

Taking $M_H=125.09$ GeV, we find that
\begin{eqnarray}
\label{eq:haa}
S^\gamma &\simeq& 
-8.34\,g_{HWW} + 1.76\, g^S_{H\bar{t}t}
+(-0.015+0.017\,i)\,g^S_{H\bar{b}b} \nonumber \\ &&
+(-0.024+0.022\,i)\,g^S_{H\bar{\tau}\tau}
+(-0.007+0.005 \,i)\,g^S_{H\bar{c}c}+\Delta S^\gamma
\nonumber \\[2mm]
P^\gamma &\simeq& 2.78\, g^P_{H\bar{t}t} 
+(-0.018+0.018\,i)\,g^P_{H\bar{b}b} \nonumber \\ &&
+(-0.025+0.022\,i)\,g^P_{H\bar{\tau}\tau}
+(-0.007+0.005\,i)\,g^P_{H\bar{c}c}+\Delta P^\gamma  \label{5}
\end{eqnarray}
giving
$S^\gamma_{\rm SM}=-6.62+0.044\,i$ and $P^\gamma_{\rm SM}=0$.

%
\item\underline{Higgs couplings to two gluons}: 
Similar to $H\to\gamma\gamma$, 
the amplitude for the decay process
$H \rightarrow gg$ can be written as
\begin{equation} \label{higg}
{\cal M}_{gg H}=-\frac{\alpha_s\,M_{H}^2\,\delta^{ab}}{4\pi\,v}
\bigg\{S^g(M_{H})
\left(\epsilon^*_{1\perp}\cdot\epsilon^*_{2\perp}\right)
 -P^g(M_{H})\frac{2}{M_{H}^2}
\langle\epsilon^*_1\epsilon^*_2 k_1k_2\rangle
\bigg\}\,,
\end{equation}
where $a$ and $b$ ($a,b=1$ to 8) are indices of the eight $SU(3)$
generators in the adjoint representation.
The decay rate of $H\to gg $ is 
proportional to $|S^g|^2 + |P^g|^2$
\footnote{Note that the production rate of $gg\to H$ at the Higgs peak
is also proportional to $|S^g|^2 + |P^g|^2$ in our formalism.}.
Again, including some additional loop contributions from new particles,
the scalar and pseudoscalar form factors are given by
\begin{eqnarray}
S^g(M_{H})&=&\sum_{f=b,t}
g^{S}_{H\bar{f}f}\,F_{sf}(\tau_{f}) +  
\Delta S^g\,,
\nonumber \\
P^g(M_{H})&=&\sum_{f=b,t}
g^{P}_{H\bar{f}f}\,F_{pf}(\tau_{f}) +
\Delta P^g
\,.
\end{eqnarray}
The additional contributions $\Delta S^g$ and $\Delta P^g$
are assumed to be real again. 

Taking $M_H=125.09$ GeV, we find that
\begin{eqnarray}
\label{eq:hgg}
S^g &\simeq& 
0.688 \, g^S_{H\bar{t}t}
+(-0.037+0.050\,i)\,g^S_{H\bar{b}b} + \Delta S^g\,, \nonumber \\ 
P^g &\simeq& 
1.047\, g^P_{H\bar{t}t} 
+(-0.042+0.051\,i)\,g^P_{H\bar{b}b} + \Delta P^g\,, \label{8}
\end{eqnarray}
giving $S^g_{\rm SM}=0.651+0.050\,i$ and $P^g_{\rm SM}=0$.


\item\underline{Higgs couplings to $Z$ and $\gamma$}: 
The amplitude for the decay process $H \to
Z(k_1,\epsilon_1)\
\gamma(k_2,\epsilon_2)$ can be written as
\begin{equation}
{\cal M}_{Z\gamma H} = -\,\frac{\alpha}{2\pi v}\left\{
S^{Z\gamma}(M_{H})\,
\left[ k_1\cdot k_2\,\epsilon_1^*\cdot\epsilon_2^*
-k_1\cdot\epsilon_2^*\,k_2\cdot\epsilon_1^* \right] \ - \
P^{Z\gamma}(M_{H})\,
\langle \epsilon_1^*\epsilon_2^* k_1 k_2\rangle
\right\}
\end{equation}
where $k_{1,2}$ are the momenta of the $Z$ boson and the photon (we note that
$2k_1\cdot k_2 = M_{H}^2-M_Z^2$),
$\epsilon_{1,2}$ are their polarization vectors.
The scalar and pseudoscalar form factors can be found in Ref.~\cite{higgcision}.
\end{itemize}
Finally, we define
the ratios of the effective 
Higgs couplings to $gg$, $\gamma\gamma$, and
$Z\gamma$ relative to the SM ones as follows:
\begin{equation}
C_g\equiv\sqrt{\frac{\left|S^g\right|^2+\left|P^g\right|^2}
{\left|S^g_{\rm SM}\right|^2}}\,; \ \
C_\gamma\equiv\sqrt{\frac{\left|S^\gamma\right|^2+\left|P^\gamma\right|^2}
{\left|S^\gamma_{\rm SM}\right|^2}}\,; \ \
C_{Z\gamma}\equiv\sqrt{\frac{\left|S^{Z\gamma}\right|^2+\left|P^{Z\gamma}\right|^2}
{\left|S^{Z\gamma}_{\rm SM}\right|^2}}\,. 
\end{equation}
Note that the ratios of decay rates relative to the SM are 
given by $|C_g|^2$, $|C_\gamma|^2$, and $|C_{Z\gamma}|^2$, respectively.

The theoretical signal strength may be written  as the product
\begin{equation}
\widehat\mu({\cal P},{\cal D}) \simeq
\widehat\mu({\cal P})\ \widehat\mu({\cal D}) 
\end{equation}
where ${\cal P}={\rm ggF}, {\rm VBF}, {\rm VH}, {\rm ttH}$ denote the production mechanisms
and ${\cal D}=\gamma\gamma , ZZ, WW, b\bar{b}, \tau\bar\tau$
the decay channels.
More explicitly, we are taking
\begin{eqnarray}
\widehat\mu({\rm ggF}) &=&
\frac{\left|S^g(M_H)\right|^2+\left|P^g(M_H)\right|^2}
{\left|S^g_{\rm SM}(M_H)\right|^2}\,, \nonumber \\[2mm]
\widehat\mu({\rm VBF}) &=& g_{_{HWW,HZZ}}^2\,, \nonumber \\[2mm]
\widehat\mu({\rm VH}) &=& g_{_{HWW,HZZ}}^2\,, \nonumber \\[2mm]
\widehat\mu({\rm ttH}) &=& \left(g^S_{H\bar{t}t}\right)^2 
+\left(g^P_{H\bar{t}t}\right)^2\,; 
\end{eqnarray}
and
\begin{equation}
\widehat\mu({\cal D}) = \frac{B(H\to {\cal D})}{B(H_{\rm SM}\to {\cal D})}
\end{equation}
with
\begin{equation}
\label{eq:dgam}
B(H\to {\cal D})=\frac{\Gamma(H\to{\cal D})}
{\Gamma_{\rm tot}(H)+\Delta\Gamma_{\rm tot}}
\end{equation}
Note that we introduce an arbitrary non-SM contribution $\Delta\Gamma_{\rm tot}$
to the total decay width. Incidentally,
$\Gamma_{\rm tot}(H)$ becomes the SM total decay width
when 
$g^S_{H\bar{f}f}=1$,
$g^P_{H\bar{f}f}=0$,
$g_{_{HWW,HZZ}}=1$,
$\Delta S^{\gamma,g,Z\gamma}=
\Delta P^{\gamma,g,Z\gamma}=0$.

The experimentally observed signal strengths 
should be compared to the theoretical ones summed
over all production mechanisms:
\begin{equation}
\label{eq:cpq}
\mu({\cal Q},{\cal D}) =
\sum_{{\cal P}={\rm ggF, VBF, VH, ttH}}\ C_{{\cal Q} {\cal P}}\
\widehat\mu({\cal P},{\cal D}) 
\end{equation}
where ${\cal Q}$ denote the experimentally defined
channel involved with the decay ${\cal D}$ and
the decomposition coefficients $C_{{\cal Q} {\cal P}}$ may depend
on the relative Higgs production cross sections for a given 
Higgs-boson mass, experimental cuts, etc.

The $\chi^2$ associated with an uncorrelated observable is
\begin{equation}
\chi^2({\cal Q},{\cal D}) =
\frac{\left[\mu({\cal Q},{\cal D})-\mu^{\rm EXP}({\cal Q},{\cal D})\right]^2}
{\left[\sigma^{\rm EXP}({\cal Q},{\cal D})\right]^2}\,,
\end{equation}
where $\sigma^{\rm EXP}({\cal Q},{\cal D})$ denotes the experimental error.
For $n$ correlated observables, we use 
\begin{equation}
\chi^2_n = \sum^n_{i,j=1}
(\mu_i-\mu^{\rm EXP}_i)\,\left(V^{-1}\right)_{ij}\,(\mu_j-\mu^{\rm EXP}_j)\,,
\end{equation}
where $V$ is a $n \times n$ covariance matrix  whose $(i,j)$ component is given by
$$V_{ij} = \rho_{ij}\, \sigma^{\rm EXP}_i\, \sigma^{\rm EXP}_j$$
with $\rho$ denoting the relevant $n \times n$ correlation matrix. 
Note $\rho_{ij}=\rho_{ji}$, $\rho_{ii}=1$,
and if $\rho_{ij}=\delta_{ij}$, $\chi^2_n $ reduces to
$$
\chi^2_n=\sum^n_{i=1}
\frac{(\mu_i-\mu^{\rm EXP}_i)^2}{(\sigma^{\rm EXP}_i)^2}\,,
$$
i.e., the sum of $\chi^2$ of each uncorrelated observable.

%
\begin{table}[t!]
\caption{\small \label{tab:tev}
{\bf (Tevatron: 1.96 TeV)}
The signal strengths data from Tevatron (10.0 fb$^{-1}$ at 1.96 TeV).
\smallskip }
\begin{ruledtabular}
\begin{tabular}{cccccccr}
Channel & Signal strength $\mu$ & $M_H$(GeV) & \multicolumn{4}{c}{Production mode}  &
$\chi^2_{\rm SM}$(each)\\
        & c.v $\pm$ error       &            & ggF & VBF & VH & ttH & \\
\hline
\multicolumn{8}{c}{Tevatron (Nov. 2012)} \\
\hline
Combined $H\to \gamma\gamma$\cite{tevatron_aa_ww} & $6.14^{+3.25}_{-3.19}$   & 125 & 78\%
& 5\% & 17\% & - & 2.60 \\
Combined $H\to WW^{(\ast)}$\cite{tevatron_aa_ww}  & $0.85^{+0.88}_{-0.81}$        & 125 &
78\% & 5\% & 17\% & - & 0.03 \\
VH tag $H\to bb$\cite{tevatron_bb}        & $1.59^{+0.69}_{-0.72}$        & 125 & -    &
-   & 100\% & - & 0.67 \\
\hline
&&&&&&& $\chi^2_{\rm SM}$(subtot): 3.30
\end{tabular}
\end{ruledtabular}
\end{table}
%
%
\begin{table}[t!]
\caption{\small \label{tab:78all}
{\bf (LHC: 7$+$8 TeV)}
Combined ATLAS and CMS data on signal strengths from Table 8 of
Ref.~\cite{Khachatryan:2016vau}. \smallskip
}
\begin{ruledtabular}
\begin{tabular}{c|ccccc}
 & \multicolumn{5}{c}{Decay mode} \\
\hline
Production mode & $H\to \gamma\gamma$ & $H\to Z Z^{(\ast)} $ & $H\to W W^{(\ast)} $
& $H\to bb$ & $H\to \tau^+\tau^-$  \\
\hline
\hline
ggF & $1.10^{+0.23}_{-0.22}$  & $1.13^{+0.34}_{-0.31}$ & $0.84^{+0.17}_{-0.17}$
    & -                       & $1.0^{+0.6}_{-0.6}$       \\
VBF & $1.3^{+0.5}_{-0.5}$     & $0.1^{+1.1}_{-0.6}$    & $1.2^{+0.4}_{-0.4}$
    & -                       & $1.3^{+0.4}_{-0.4}$       \\
WH  & $0.5^{+1.3}_{-1.2}$     & -                      & $1.6^{+1.2}_{-1.0}$
    & $1.0^{+0.5}_{-0.5}$     & $-1.4^{+1.4}_{-1.4}$      \\
ZH  & $0.5^{+3.0}_{-2.5}$     & -                      & $5.9^{+2.6}_{-2.2}$
    & $0.4^{+0.4}_{-0.4}$     & $2.2^{+2.2}_{-1.8}$       \\
ttH  & $2.2^{+1.6}_{-1.3}$    & -                      & $5.0^{+1.8}_{-1.7}$
     & $1.1^{+1.0}_{-1.0}$    & $-1.9^{+3.7}_{-3.3}$      \\
\hline
&&&&& $\chi^2_{\rm SM}$(subtot): 19.93
\end{tabular}
\end{ruledtabular}
\end{table}
%
%
\begin{table}[h!]
\caption{\small \label{all13}
{\bf (LHC: 13 TeV)}
Combined ATLAS and CMS (13 TeV) data on signal strengths.
The $\mu^{\rm dec}_{\rm combined}$ ($\mu^{\rm prod}_{\rm combined}$)
represents the combined signal strength 
for a specific decay (production) channel by summing all the 
production (decay) modes,
and $\chi^2_{\rm min}$ are the corresponding minimal chi-square values.
In the VH/WH row,
the production mode for $H\to\gamma\gamma$ and $H\to ZZ^{(*)}$ 
is VH while it is 
WH for $H\to WW^{(*)}$ and $H\to \tau^+\tau^-$;
for the remaining 
decay mode $H\to b \bar b$, we combine the two signal strengths from
WH and VH, see Table~\ref{bb}.
\smallskip
}
\begin{ruledtabular}
\begin{tabular}{c|ccccc|cc}
 & \multicolumn{5}{c}{Decay mode} \\
\hline
Production mode & $H\to \gamma\gamma$ & $H\to Z Z^{(\ast)} $ & $H\to W W^{(\ast)} $ 
& $H\to bb$ & $H\to \tau^+\tau^-$ 
& $\mu^{\rm prod}_{\rm combined}$ & $\chi^2_{\rm SM}(\chi^2_{\rm min})$  \\
\hline
\hline
ggF & $1.02^{+0.12}_{-0.11}$  & $1.09^{+0.11}_{-0.11}$ &  $1.29^{+0.16}_{-0.16}$ 
    & $2.51^{+2.43}_{-2.01}$  & $1.06^{+0.40}_{-0.37}$ &  $1.11^{+0.07}_{-0.07}$ 
    & 5.42(3.15)    \\
VBF & $1.23^{+0.32}_{-0.31}$  & $1.51^{+0.59}_{-0.59}$ &  $0.54^{+0.32}_{-0.31}$ 
    & -                       & $1.15^{+0.36}_{-0.34}$ &  $1.02^{+0.18}_{-0.18}$
    & 7.53(7.51)    \\
VH/WH & $1.42^{+0.51}_{-0.51}$  & $0.71^{+0.65}_{-0.65}$ & $3.27^{+1.88}_{-1.70}$ 
        & $1.07^{+0.23}_{-0.22}$  & $3.39^{+1.68}_{-1.54}$ & $1.15^{+0.20}_{-0.19}$
        & 7.05(6.44) \\
ZH  & -  & - &  $1.00^{+1.57}_{-1.00}$ & $1.20^{+0.33}_{-0.31}$ 
    & $1.23^{+1.62}_{-1.35}$   & $1.19^{+0.32}_{-0.30}$
    & 0.45(0.02)    \\
ttH & $1.36^{+0.38}_{-0.37}$  & $0.00^{+0.53}_{-0.00}$ &  - & $0.91^{+0.45}_{-0.43}$  
    & -                       & $0.93^{+0.24}_{-0.24}$ 
    & 5.96(5.86)    \\
ttH (excl.) & $1.39^{+0.48}_{-0.42}$ &  - &$1.59^{+0.44}_{-0.43}$ & $0.77^{+0.36}_{-0.35}$             
            & $0.87^{+0.73}_{-0.73}$ &  $1.16^{+0.22}_{-0.22}$
            & 4.17(3.62)   \\
\hline
$\mu^{\rm dec}_{\rm combined}$ & $1.10^{+0.10}_{-0.10}$  & $1.05^{+0.11}_{-0.11}$  
                     & $1.20^{+0.14}_{-0.13}$  & $1.05^{+0.19}_{-0.19}$ 
                     & $1.15^{+0.24}_{-0.23}$  & $1.10^{+0.06}_{-0.06}$
                      \\
$\chi^2_{\rm SM}$($\chi^2_{\rm min}$) & 6.83(5.72)  & 9.13(8.88)  
             & 9.48(7.32) & 1.56(1.51) & 3.58(3.20)  & & 30.58(27.56) \\
\end{tabular}
\end{ruledtabular}
\end{table}
\section{Results on Higgs Signal Strength Data}
In our work, we use the direct Higgs data collected at the Tevatron and the LHC.
We use 3 signal strengths measured at the Tevatron, see Table~\ref{tab:tev}.
The Higgs-boson data at 7 and 8 (7+8) TeV used in this analysis
are the signal strengths obtained
from a combined ATLAS and CMS analysis~\cite{Khachatryan:2016vau},
see Table~\ref{tab:78all}.
We also take into account the correlation matrix given in Fig.~27 of
Ref.~\cite{Khachatryan:2016vau}.
On the other hand, the 13 TeV data are
still given separately by ATLAS and CMS and
in different production and decay channels
\footnote{For the details of the 13 TeV data sets 
used in this work, see Appendix B.}.
In total, the number of signal strengths considered is
$3\,(1.96\,{\rm TeV})
+20\,(7+8\,{\rm TeV})
+41\,(13\,{\rm TeV}) = 64$.

Precaution is noted before we show the combined results of ATLAS
and CMS. In each of the data, there are statistical, systematic,
and theoretical uncertainties. Especially, the latter one, e.g.,
uncertainties coming from factorization scale, renormalization scale,
higher order corrections, is correlated between ATLAS and CMS. 
Since no such information is available at the time of writing,
we only combine them with a simple $\chi^2$ method and assuming
each data is Gaussian distributed. Our finding of 2 sigma excess
in the overall signal strength is to be taken cautiously.

At 7+8 TeV, we use the combined average signal strengths given in
Ref.~\cite{Khachatryan:2016vau} in which the experimental correlations 
are considered.
At 1.96 TeV and 13 TeV, we combine signal strengths of various channels
using a simple $\chi^2$ method and assuming each is
Gaussian distributed. 
The combined signal stregnth at 1.96 TeV is $1.44\pm 0.55$ and, at 13 TeV,
the combined ATLAS and CMS signal strengths
for each production and decay channel are presented in Table~\ref{all13}.
Before we go to the global fits,
we would like to point out a few peculiar features in the data sets,
and the average signal strengths.
\begin{enumerate}
\item 
The combined overall signal strength at 7+8 TeV
is $\mu_{7+8 {\rm TeV} } = 1.09\,^{+0.11}_{-0.10}$ 
\cite{Khachatryan:2016vau}, 
which is larger than the SM value by slightly less than $1 \sigma$.

\item At 13 TeV, from Table~\ref{all13}, it is clear that all decay channels 
show 
slight excess over the SM value of 1.0, especially the $H \to \gamma\gamma$
and $H \to W W^*$ channels. 

\item Again from Table~\ref{all13}, almost all production modes, except
for ttH, show excess above the SM, especially the gluon fusion (ggF).

\item 
The 13 TeV data shows similar deviations in both ATLAS and CMS results:
$\mu_{13 {\rm TeV}}^{\rm ATLAS} = 1.09 \pm 0.08$ and 
$\mu_{13 {\rm TeV}}^{\rm CMS} = 1.1 \, ^{+0.09}_{-0.08}$.
By combining these two results we obtain 
\[
 \mu_{13 {\rm TeV}} = 1.10 \pm 0.06
\]
which is about $1.67\sigma$ above the SM.

\item
Finally, we combine 
all the signal strengths for the Tevatron at 1.96 TeV, and 
for $7+8$ and 13 TeV ATLAS and CMS, and thus obtain
\[
  \mu_{\rm All} = 1.10 \pm 0.05
\]
which indicates a $2 \sigma$ deviation from the SM value.

\end{enumerate}
We summarize the results in Table~\ref{tab:mu}.

\begin{table}[t!]
\caption{\small \label{tab:mu}
Combined average signal strengths for the Tevatron
at 1.96 TeV, and for ATLAS and CMS at $7+8$ TeV and 13 TeV.
\smallskip
}
\begin{ruledtabular}
\begin{tabular} {c||c|c||c}
Energy & ATLAS & CMS & Combined \\ \hline
1.96 TeV [Table~\ref{tab:tev}] &
  &   & $1.44\pm 0.55$ \\ \hline
7+8 TeV \cite{Khachatryan:2016vau} &
$1.20^{+0.15}_{-0.14}$  &  $0.97^{+0.14}_{-0.13}$ & $1.09^{+0.11}_{-0.10}$ \\ \hline
13 TeV [Table~\ref{all13}] &
$1.09\pm 0.08$  &  $1.11^{+0.09}_{-0.08}$ & $1.10\pm 0.06$ \\ \hline
& & & $1.10\pm 0.05$
\end{tabular}
\end{ruledtabular}
\end{table}

Here and in the following section, we will present two statistical
measures: (i) goodness of fit quantifying the agreement within
observables in a given fit, and (ii) $p$-value of a given fit hypothesis
against the SM null hypothesis.  Goodness of fit is expressed in terms of
an integral,  which is given by
\[
\mbox{Goodness of fit } = 
\int^\infty_{\chi^2 } f [x, n] \, d x
\]
where the probability density function is given by
\[
f[x,n] = \frac{x^{n/2 - 1} e^{-x/2} }{ 2^{n/2} \Gamma(n/2) }\,,
\]
$n$ is the degree of freedom, and $\Gamma(n/2)$ is the gamma function.
The rule of thumb is that when the value of $\chi^2$ per degree of freedom
is less than around 1,  it is a good fit.

On the other hand, the $p$-value of the given fit hypothesis (test hypothesis)
with $m$ fitting parameters against the SM null hypothesis is given by
\[
\mbox{ $p$-value } = 
\int_{\Delta \chi^2}^{\infty} f[x, m] \, d x  \;,
\]
where $\Delta \chi^2$ in the lower limit of the integral
is equal to chi-square difference
between the best-fit point of the fit hpothesis and the SM one:
$\Delta \chi^2 = \chi^2_{\rm SM} - \chi^2_{\rm min}$.
This $p$-value represents the probability that the test hypothesis
is a fluctuation of the SM null hypothesis.  A large $p$-value means
that the test hypothesis is very similar to the SM null hypothesis.
For example, in Table \ref{CPC} the {\bf CPC1} case (with 1 fitted parameter) has a
$\Delta \chi^2 = 53.81 - 51.44 = 2.37$ corresponding to a $p$-value of 0.124.
For {\bf CPC2} case (with 2 fitted parameters) has
$\Delta \chi^2 = 53.81 - 51.87 = 1.94$ corresponding to a $p$-value of 0.379.
From these two fits we can easily see that the SM null hypothesis is more
similar to the {\bf CPC2} best fit-point.  According to the $p$-values in
Table \ref{CPC}, the SM is more consistent with the fits with more parameters.

\section{Results on Global Fits}

We perform global fits in which 
one or more parameters are varied.
They are categorized into CP-conserving ({\bf CPC})
and CP-violating ({\bf CPV}) fits, because the current 
data still allows the observed Higgs boson
to be a mixture of CP-even and CP-odd states.
Assuming generation independence for the normalized Yukawa couplings of
$g^{S,P}_{H\bar f f}$,
we use the following notation for the
parameters in the fits:
\begin{eqnarray}
&&
C_u^S=g^S_{H\bar uu}\,, \ \
C_d^S=g^S_{H\bar dd}\,, \ \
C_\ell^S=g^S_{H\bar ll}\,; \ \
C_w=g_{_{HWW}}\,, \ \
C_z=g_{_{HZZ}}\,; \nonumber \\
&&
C_u^P=g^P_{H\bar uu}\,, \ \
C_d^P=g^P_{H\bar dd}\,, \ \
C_\ell^P=g^P_{H\bar ll}\,.
\end{eqnarray}
In most of the fits, we keep the custodial symmetry 
between the $W$ and $Z$ bosons
by taking $C_v \equiv C_w = C_z$.
However, in the last CP-conserving scenario ({\bf CPCX4}), 
we adopt $C_w \neq C_z$,
which is motivated by the data.

\subsection{CP Conserving fits}
In CP-conserving fits, we are varying
$C_u^S$, $C_d^S$, $C_\ell^S$,$C_{v;w,z}$, $\Delta S^g$, $\Delta S^\gamma$,
and $\Delta \Gamma_{\rm tot}$ while taking
$C_u^P=C_d^P=C_\ell^P=\Delta P^{\gamma}=\Delta P^g=0$.
All the CP-conserving fits considered in this work are listed here:

\begin{itemize}
\item {\bf CPC1}: vary $\Delta \Gamma_{\rm tot}$ while keeping
$C^S_u=C^S_d=C^S_\ell=C_v=1$ and $\Delta S^\gamma=\Delta S^g=0$.

\item {\bf CPC2}: vary $\Delta S^\gamma$ and $\Delta S^g$ while keeping
$C^S_u=C^S_d=C^S_\ell=C_v=1$ and $\Delta \Gamma_{\rm tot}=0$.

\item {\bf CPC3}: vary $\Delta S^\gamma,~\Delta S^g$ and $\Delta \Gamma_{\rm tot}$ 
while keeping $C^S_u=C^S_d=C^S_\ell=C_v=1$.

\item {\bf CPC4}: vary $C^S_u,~C^S_d,~C^S_\ell,~C_v$ 
while keeping $\Delta S^\gamma=\Delta S^g=\Delta \Gamma_{\rm tot}=0$.

\item {\bf CPC6}: vary $C^S_u,~C^S_d,~C^S_\ell,~C_v,~\Delta S^\gamma,~\Delta S^g$ 
while keeping $\Delta \Gamma_{\rm tot}=0$.

\item {\bf CPCN2}: vary $C^S_u,~C_v$ 
while keeping $C^S_d=C^S_\ell=1$,
 and $\Delta S^\gamma=\Delta S^g=\Delta\Gamma_{\rm tot}=0$.

\item {\bf CPCN3}: vary $C^S_u,~C_v,~\Delta S^\gamma$ 
while keeping $C^S_d=C^S_\ell=1$ and $\Delta S^g=\Delta\Gamma_{\rm tot}=0$.

\item {\bf CPCN4}: vary $C^S_u,~C_v,~\Delta S^\gamma,~\Delta S^g$ 
while keeping $C^S_d=C^S_\ell=1$ and $\Delta\Gamma_{\rm tot}=0$.

\item {\bf CPCX2}: vary $C_v,~\Delta\Gamma_{\rm tot}$ 
while keeping $C^S_u=C^S_d=C^S_\ell=1$, and $\Delta S^\gamma=\Delta S^g=0$.

\item {\bf CPCX3}: vary $C^S_u,~C_v,~\Delta S^g$ 
while keeping $C^S_d=C^S_\ell=1$ and $\Delta S^\gamma=\Delta\Gamma_{\rm tot}=0$.

\item {\bf CPCX4}: vary $C^S_u,~C_w,~C_z,~\Delta S^g$ 
while keeping $C^S_d=C^S_\ell=1$ and $\Delta S^\gamma=\Delta\Gamma_{\rm tot}=0$.
\end{itemize}
Note that {\bf CPC1} to {\bf CPC6} were those originally in our first Higgcision
paper \cite{higgcision} while {\bf CPCN2} to {\bf CPCN4} were those 
studied in our 2014 update paper \cite{update2014}. 
The {\bf CPCX2} to {\bf CPCX4} are new in this
work. 
The reason why we study more scenarios here is because we want to
fully understand the effects of having $\Delta S^g$ 
alone, in order to discriminate the contribution from the bottom-Yukawa
coupling to Higgs production.  
In doing so we find that the effect of the bottom-Yukawa 
coupling becomes sizable in the Higgs-gluon-gluon vertex: numerically flipping
the sign of bottom-Yukawa coupling can cause more than 10\% change in
$|S^g|$ while it is less than 0.5\% in 
$|S^\gamma|$.
%

\begin{table}[t!]
\caption{\small \label{CPC}
{\bf (CPC)}
The best-fitted values in various CP conserving fits and the 
corresponding chi-square per degree of freedom and
goodness of fit.
The $p$-value for each fit hypothesis against the SM null
hypothesis is also shown.
For the SM, we obtain $\chi^2=53.81$, $\chi^2/dof=53.81/64$, and
so the goodness of fit $=0.814$. 
}
\begin{ruledtabular}
\begin{tabular}{c|cccccc}
Cases & {\bf CPC1} & {\bf CPC2}   & {\bf CPC3} & {\bf CPC4} & {\bf CPC6} \\
\hline
      & Vary $\Delta\Gamma_{\rm tot}$ & Vary $\Delta S^\gamma$ 
      & Vary $\Delta S^\gamma$ & Vary $C^S_u,~C^S_d,$ 
      & Vary $C^S_u,~C^S_d,~C^S_\ell,~C_v$   \\
Parameters &   & $\Delta S^g$  & $\Delta S^g,~\Delta\Gamma_{\rm tot}$           
           & $C^S_\ell,~C_v$  
           & $\Delta S^{\gamma},~\Delta S^g$   \\
\hline
\multicolumn{6}{c}
{After ICHEP 2018}\\
\hline
$C^S_u$           & 1 & 1 & 1 & $1.001^{+0.056}_{-0.055}$ & $1.033^{+0.079}_{-0.082}$   \\
$C^S_d$           & 1 & 1 & 1 & $0.962^{+0.101}_{-0.101}$ & $0.945^{+0.109}_{-0.105}$ \\
$C^S_\ell$        & 1 & 1 & 1 & $1.024^{+0.093}_{-0.093}$ & $1.018^{+0.095}_{-0.094}$  \\
$C_v$           & 1 & 1 & 1 & $1.019^{+0.044}_{-0.045}$ & $1.012^{+0.047}_{-0.048}$  \\
$\Delta S^\gamma$ & 0                      & $-0.226^{+0.32}_{-0.32}$   
& $-0.150^{+0.32}_{-0.33}$  & 0 & $-0.128^{+0.368}_{-0.369}$  \\
$\Delta S^g$      & 0                      & $0.016^{+0.025}_{-0.025}$  
& $-0.003^{+0.034}_{-0.031}$  & 0 & $-0.032^{+0.061}_{-0.057}$ \\
$\Delta \Gamma_{\rm tot}$ (MeV) & $-0.285^{+0.18}_{-0.17}$  & 0          
& $-0.247^{+0.31}_{-0.27}$  & 0 & 0 \\
\hline
$\chi^2/dof$ & 51.44/63 & 51.87/62 & 51.23/61 & 50.79/60 & 50.46/58  \\
goodness of fit    & 0.851    & 0.817    & 0.809    & 0.796    & 0.749   \\
$p$-value    & 0.124 & 0.379    & 0.461   & 0.554    & 0.764   \\
\end{tabular}
\end{ruledtabular}
\end{table}

\subsubsection{{\bf CPC1} to {\bf CPC6}}
\begin{figure}[t!]
\centering
\includegraphics[height=3.0in,angle=0]{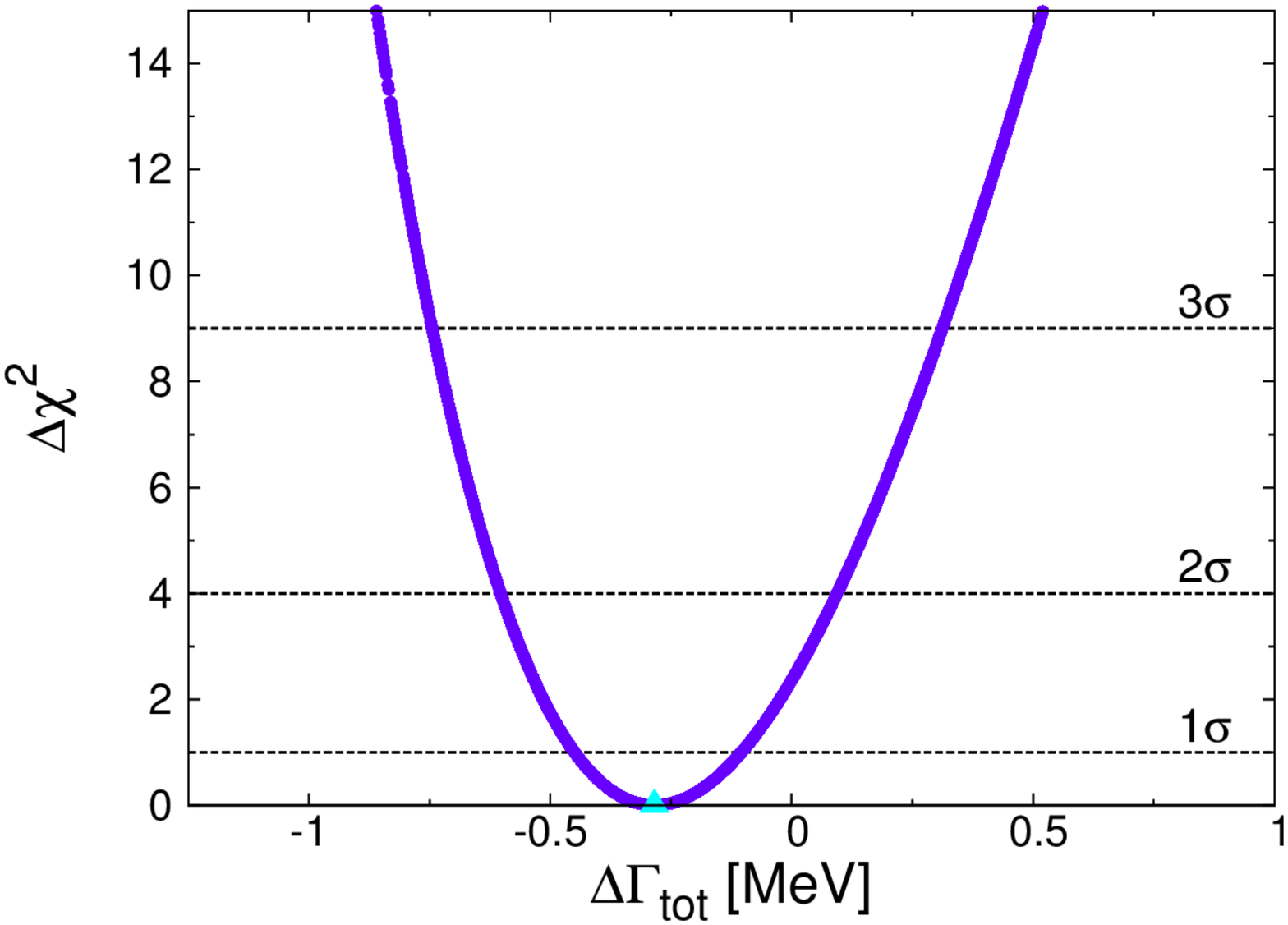}
\caption{\small \label{CPC1}
{\bf CPC1}:
$\Delta \chi^2$ from the minimum versus $\Delta \Gamma_{\rm tot}$
with only $\Delta \Gamma_{\rm tot}$ varying in the fit.
The best-fit point is denoted by the triangle.
}
\end{figure}

\begin{figure}[t!]
\centering
\includegraphics[height=2.2in,angle=0]{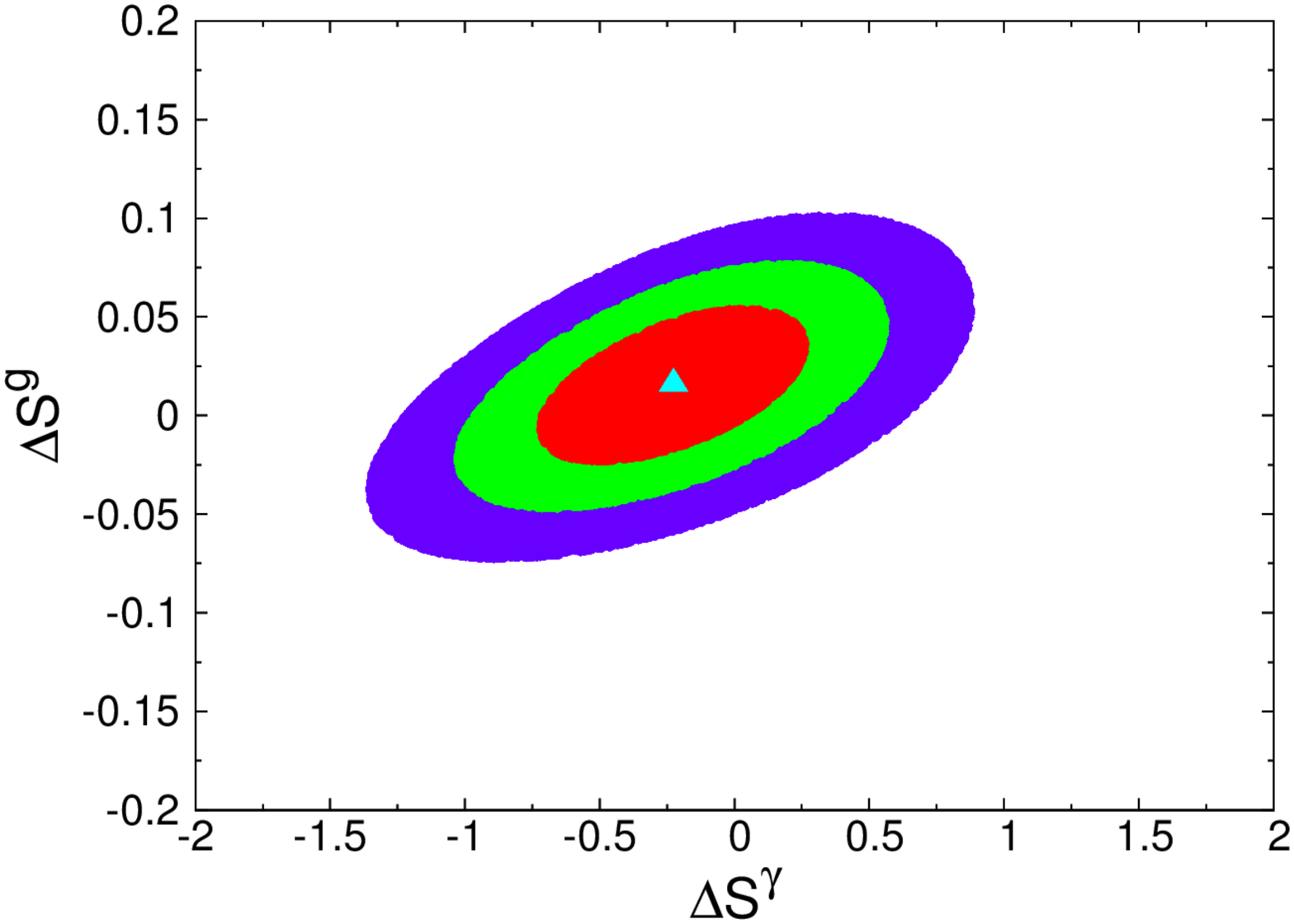}
\includegraphics[height=2.2in,angle=0]{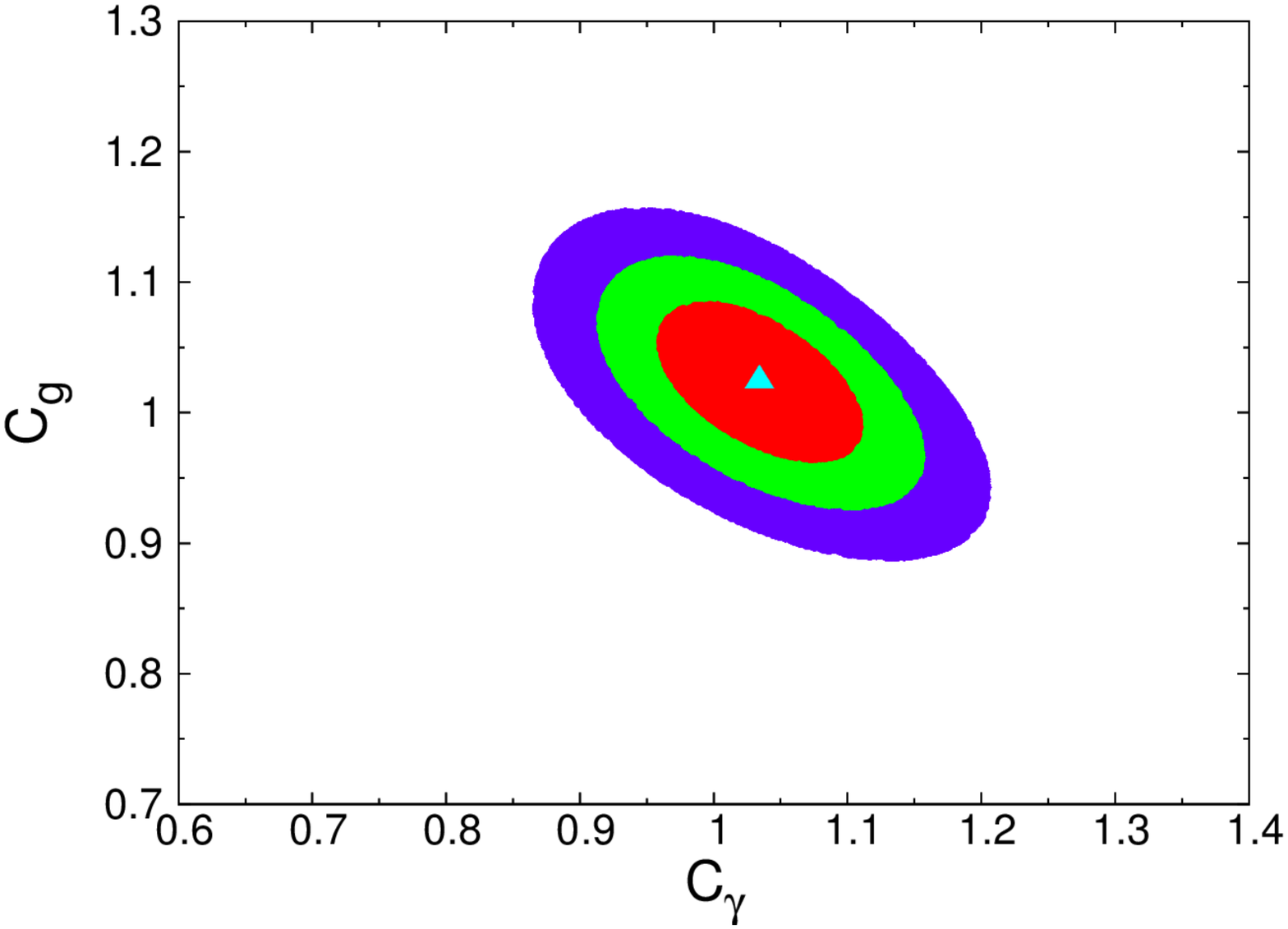}
\caption{\small \label{CPC2}
{\bf CPC2}:
The confidence-level regions of the fit by varying $\Delta S^\gamma$ and
$\Delta S^g$ only in (a) $(\Delta S^\gamma,\, \Delta S^g)$ plane and
(b) in the corresponding $(C_\gamma,\, C_g)$ plane.
The contour regions shown are for $\Delta \chi^2 \le 2.3$ (red), $5.99$
(green), and $11.83$ (blue) above the minimum, which correspond to
confidence levels of 68.3\%, 95\%, and 99.7\%, respectively.
The best-fit point is denoted by the triangle.
}
\end{figure}

\begin{figure}[t!]
\centering
\includegraphics[height=2.2in,angle=0]{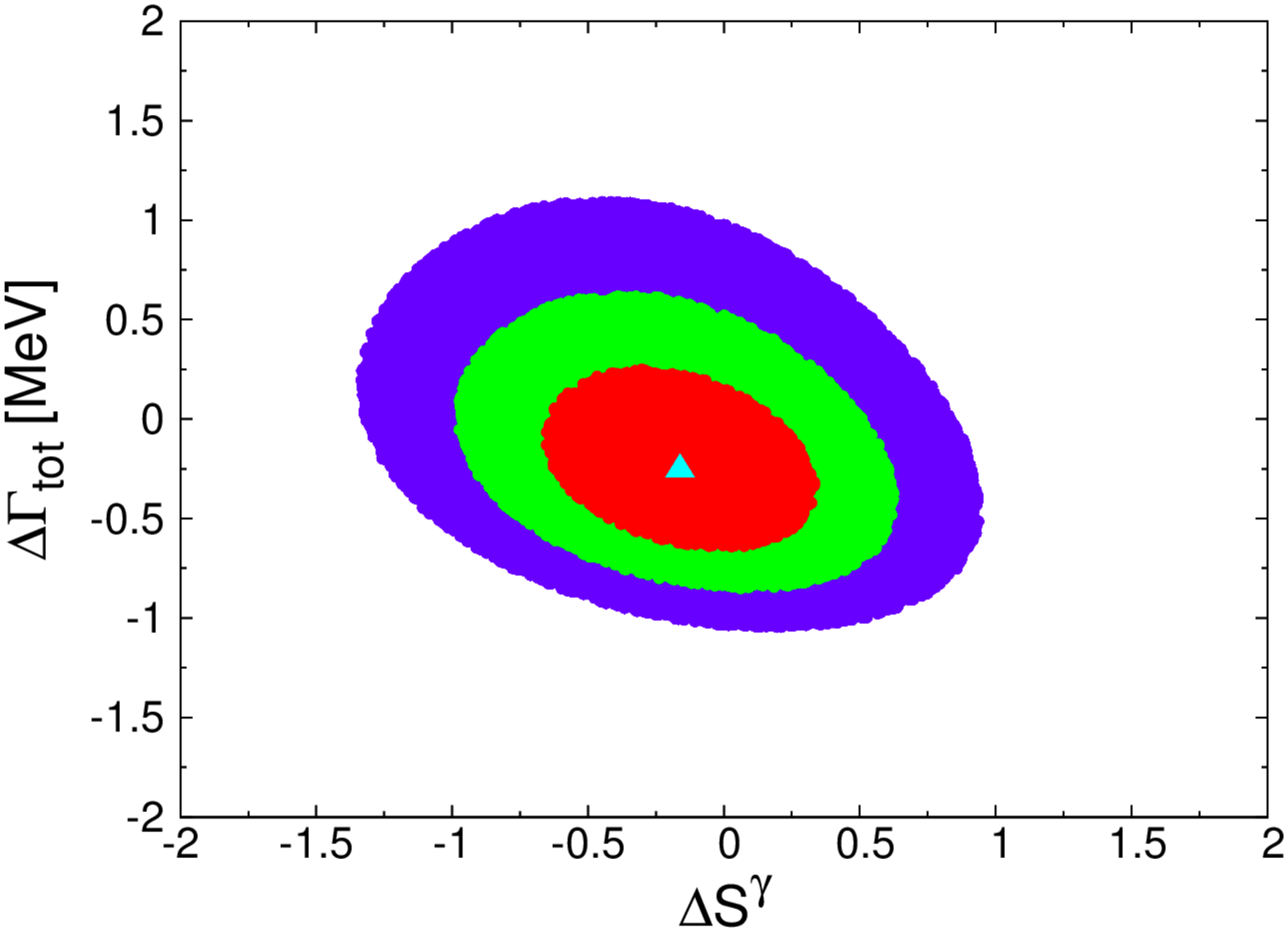}
\includegraphics[height=2.2in,angle=0]{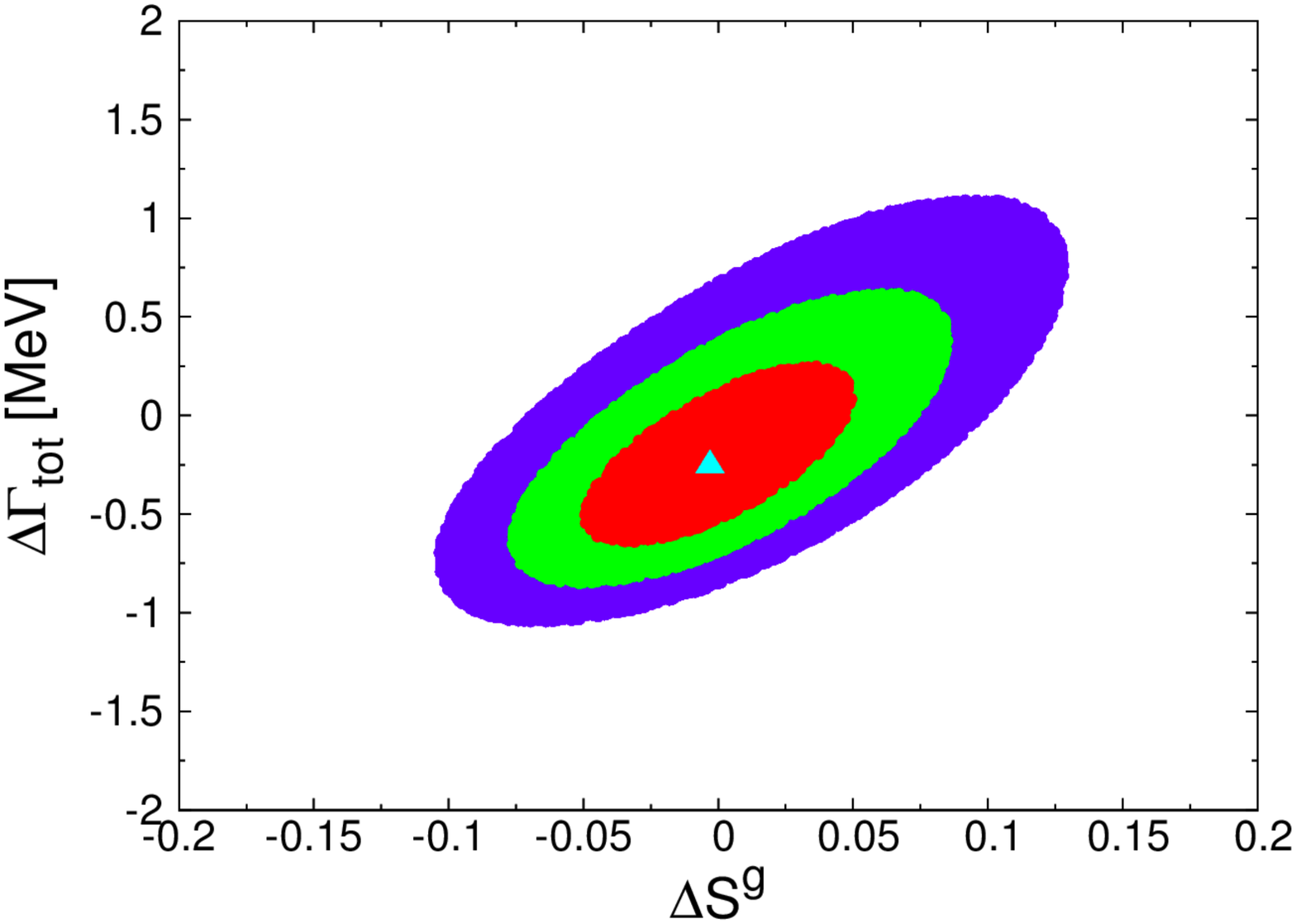}
\includegraphics[height=2.2in,angle=0]{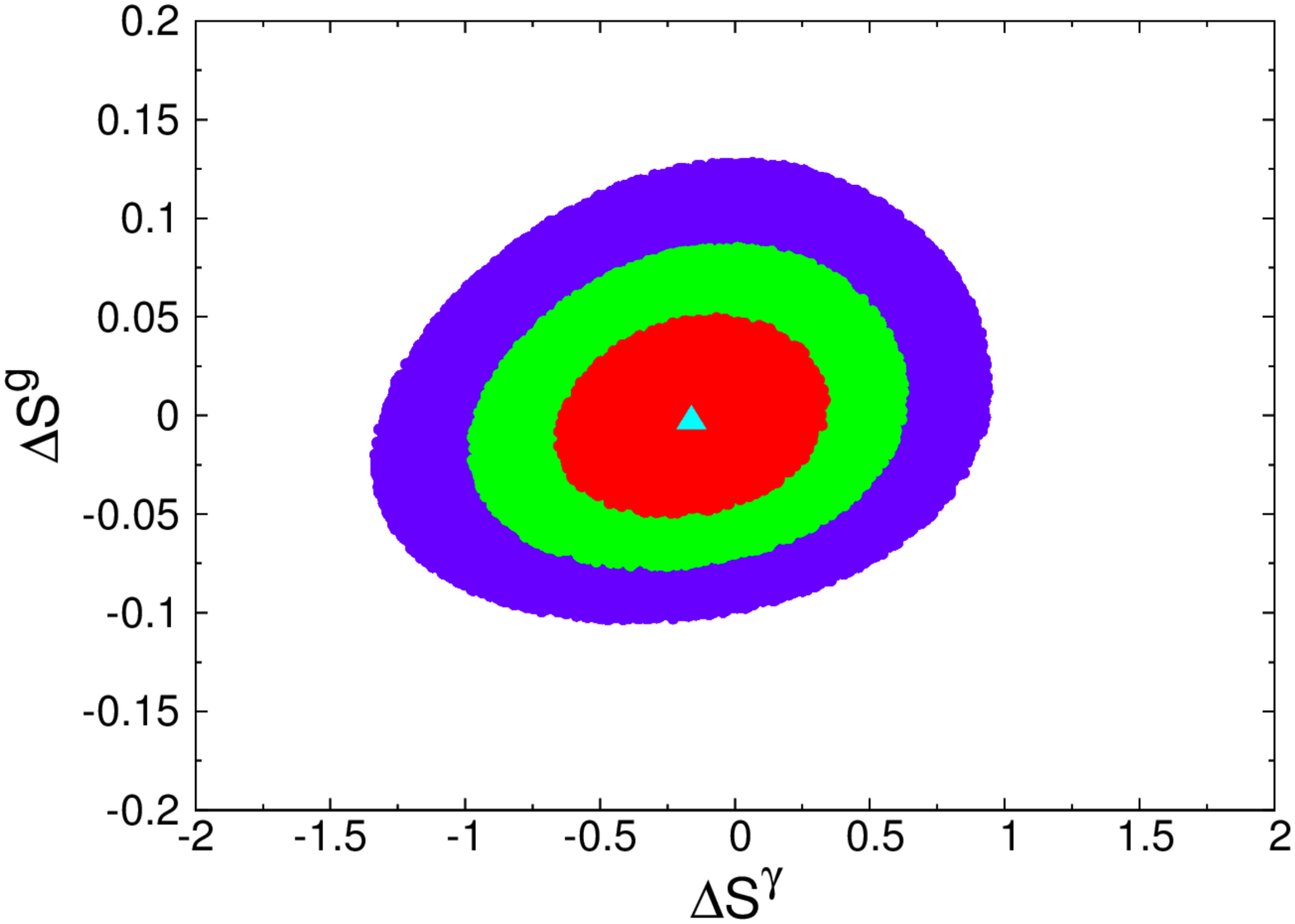}
\includegraphics[height=2.2in,angle=0]{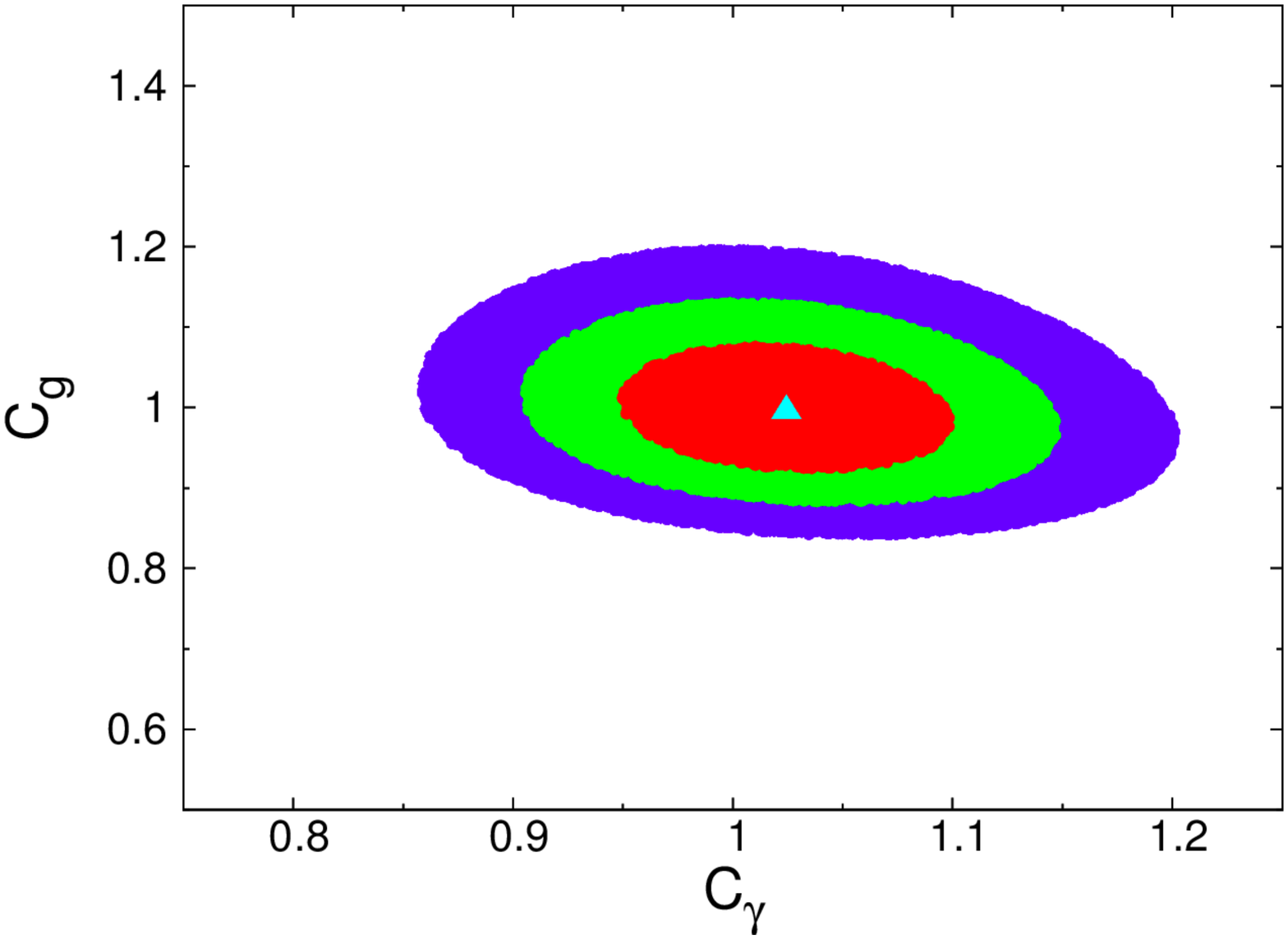}
\caption{\small \label{CPC3}
{\bf CPC3}:
The confidence-level regions of the fit by varying $\Delta S^\gamma$,
$\Delta S^g$, and $\Delta \Gamma_{\rm tot}$. 
The color code is the same as in Fig.~\ref{CPC2}.
}
\end{figure}

\begin{figure}[t!]
\centering
\includegraphics[height=1.5in,angle=0]{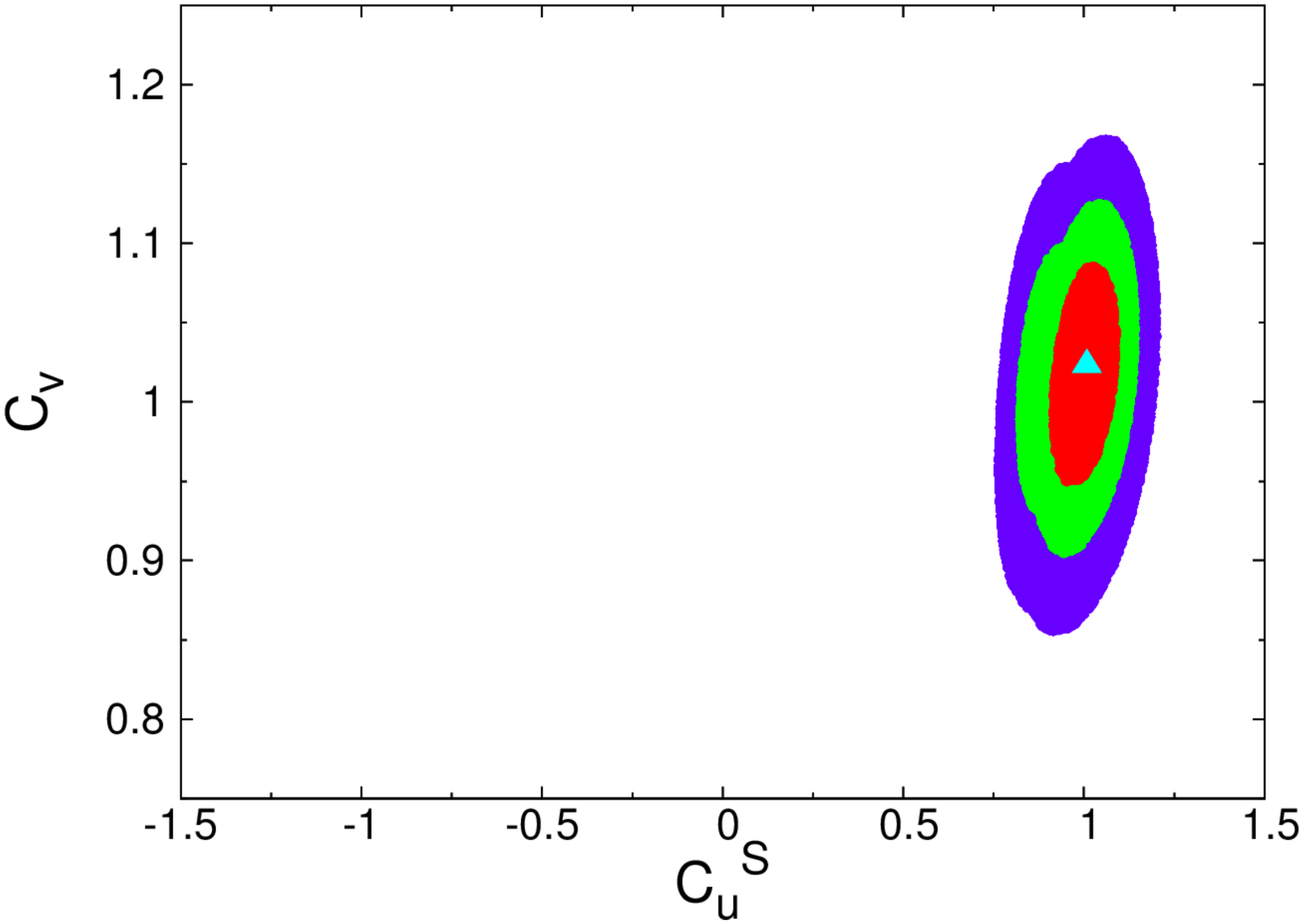}
\includegraphics[height=1.5in,angle=0]{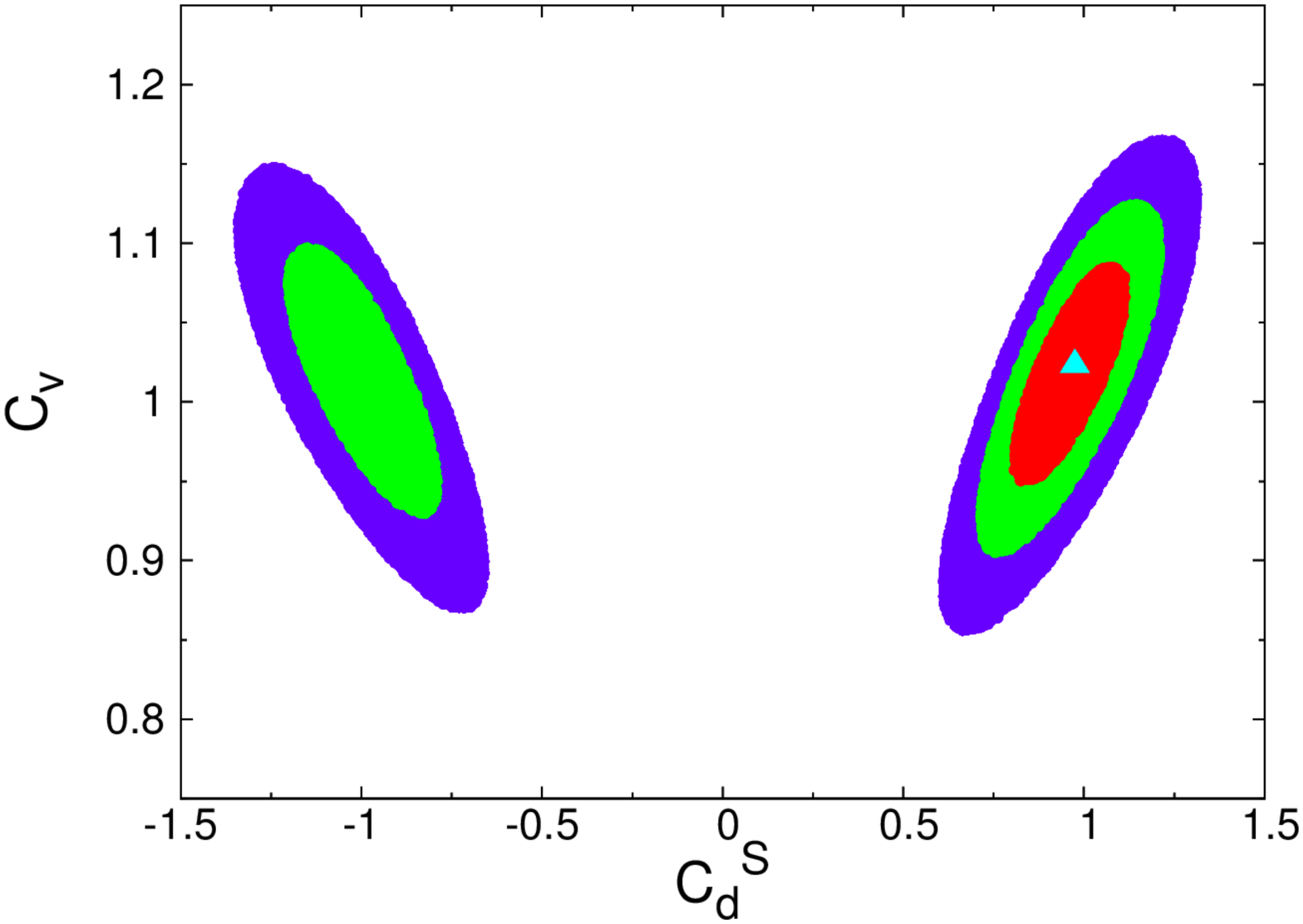}
\includegraphics[height=1.5in,angle=0]{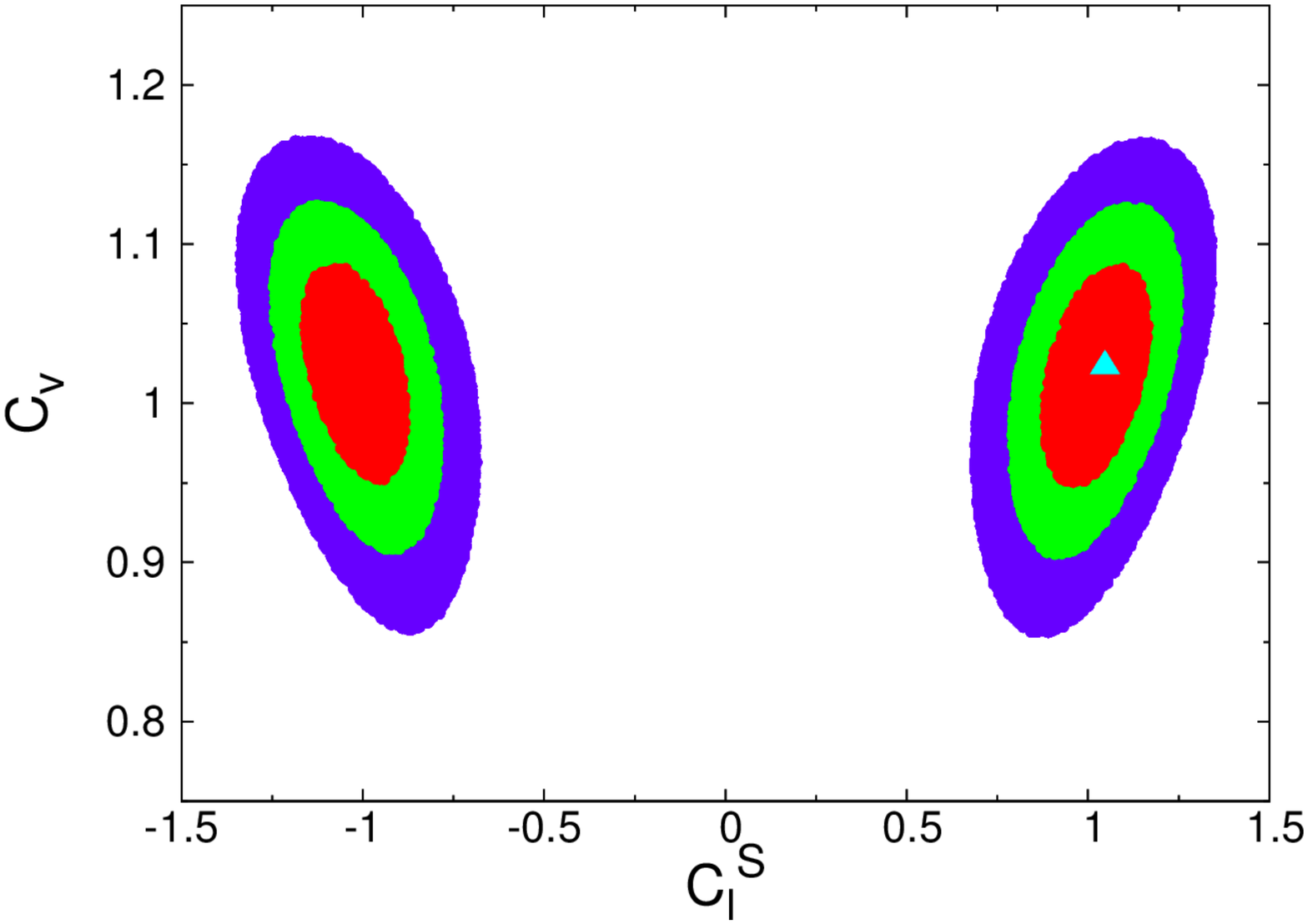}
\includegraphics[height=1.5in,angle=0]{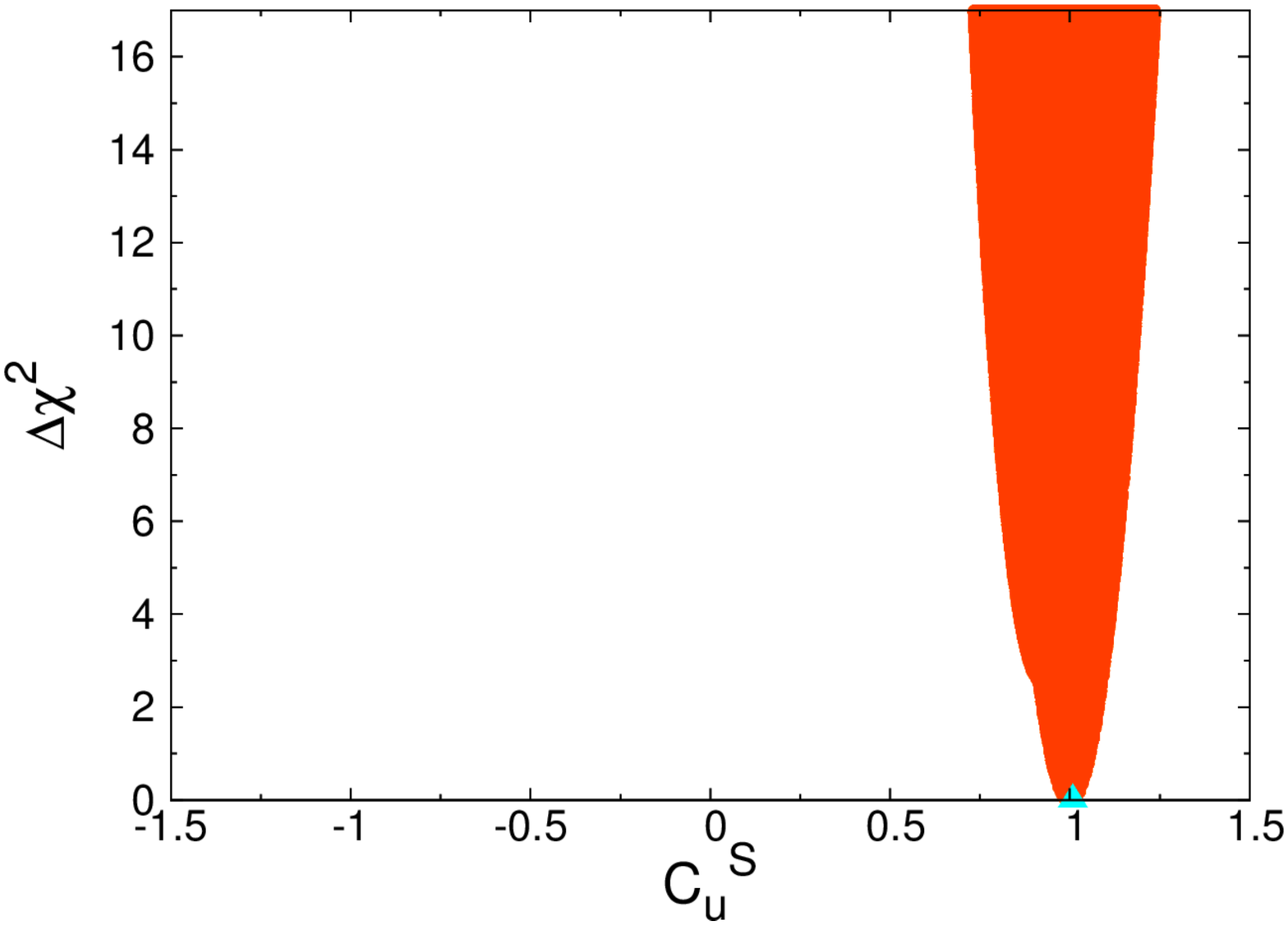}
\includegraphics[height=1.5in,angle=0]{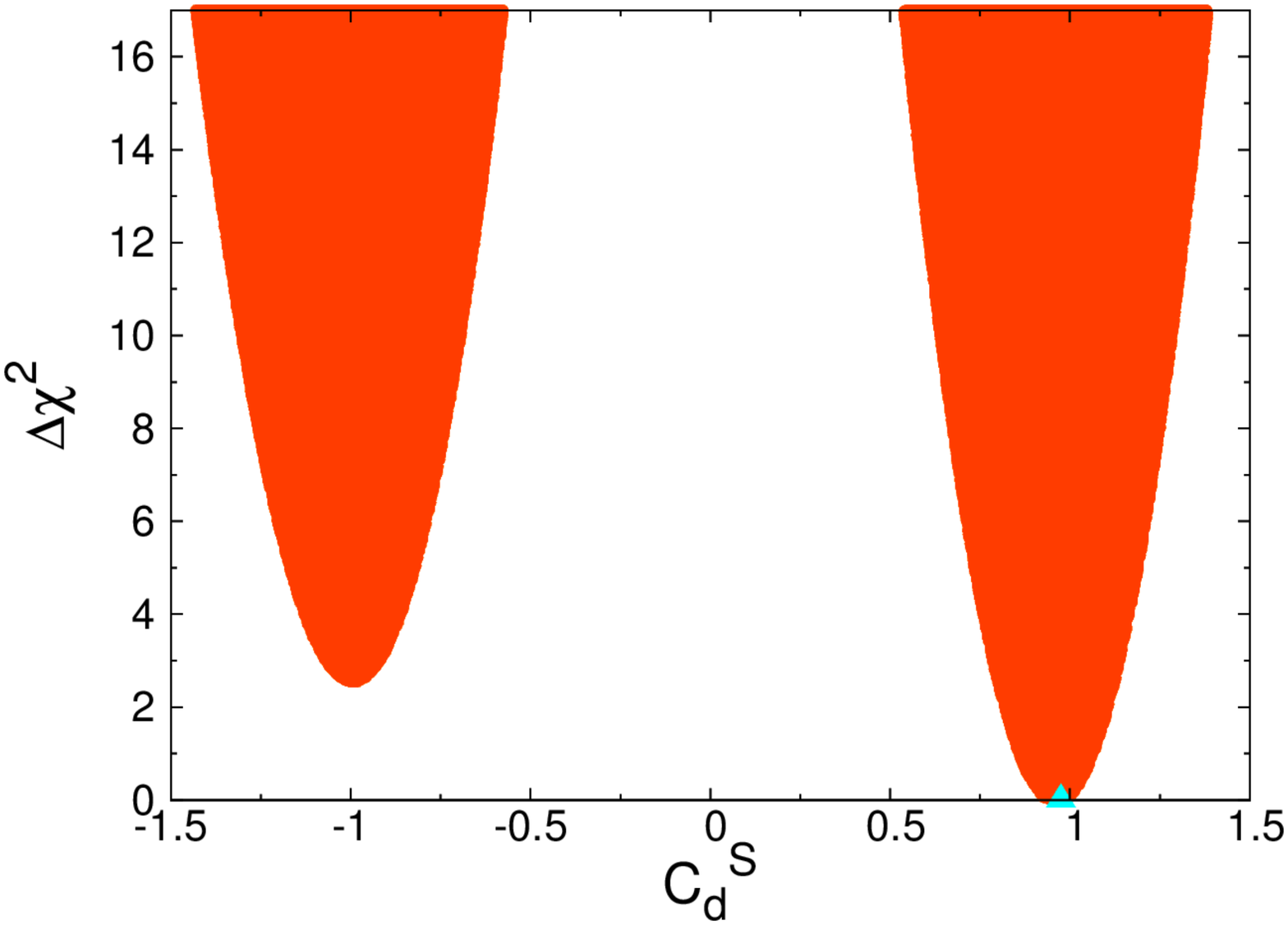}
\includegraphics[height=1.5in,angle=0]{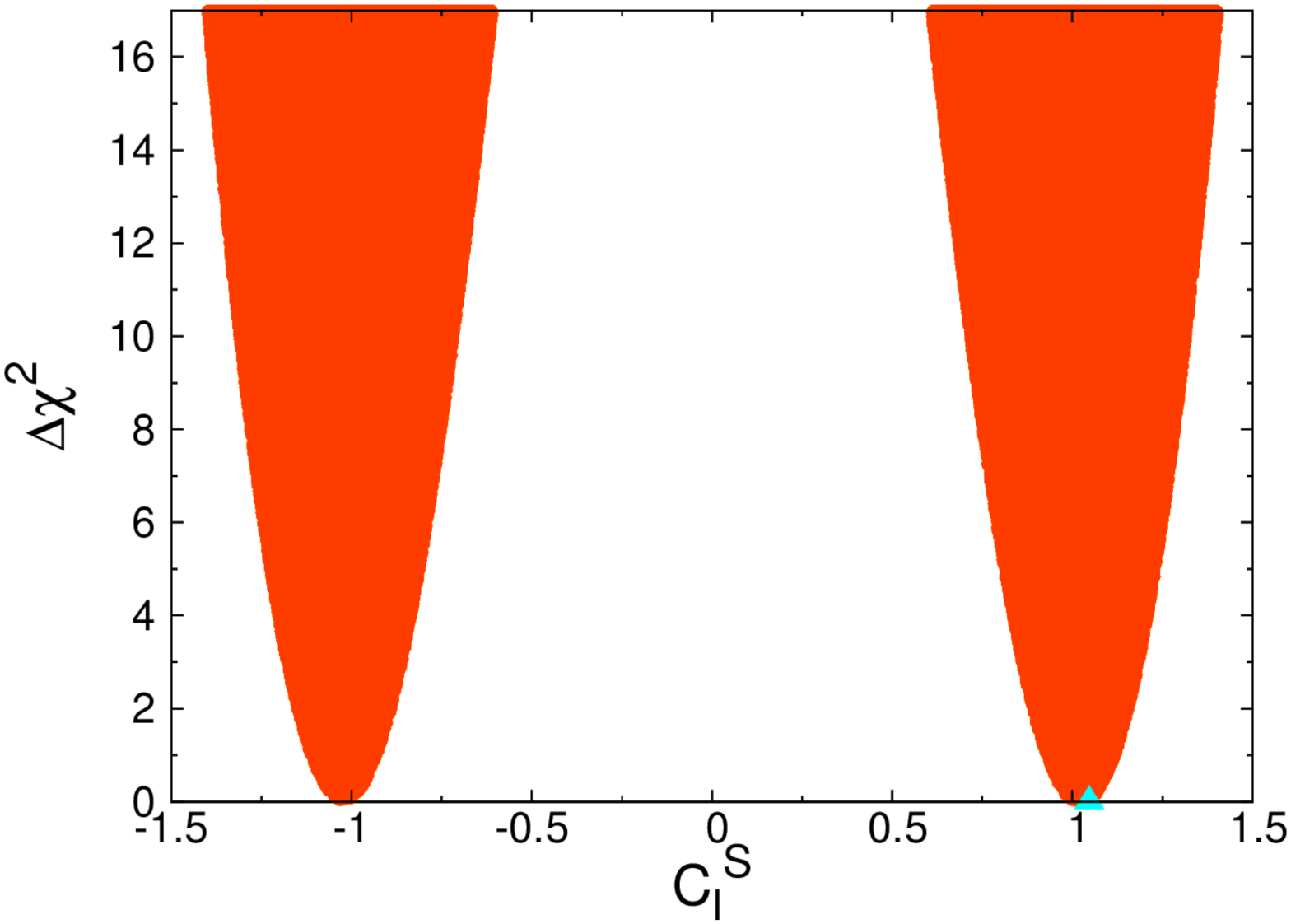}
\caption{\small \label{CPC4}
{\bf CPC4}: (Upper)
The confidence-level regions of the fit by varying $C_v$, $C_u^S$,
$C_d^S$, and $C_\ell^S$. The color code is the same
as Fig.~\ref{CPC2}. (Lower)
$\Delta \chi^2$ from the minimum versus Yukawa couplings. 
}
\end{figure}

\begin{figure}[t!]
\centering
\includegraphics[height=1.5in,angle=0]{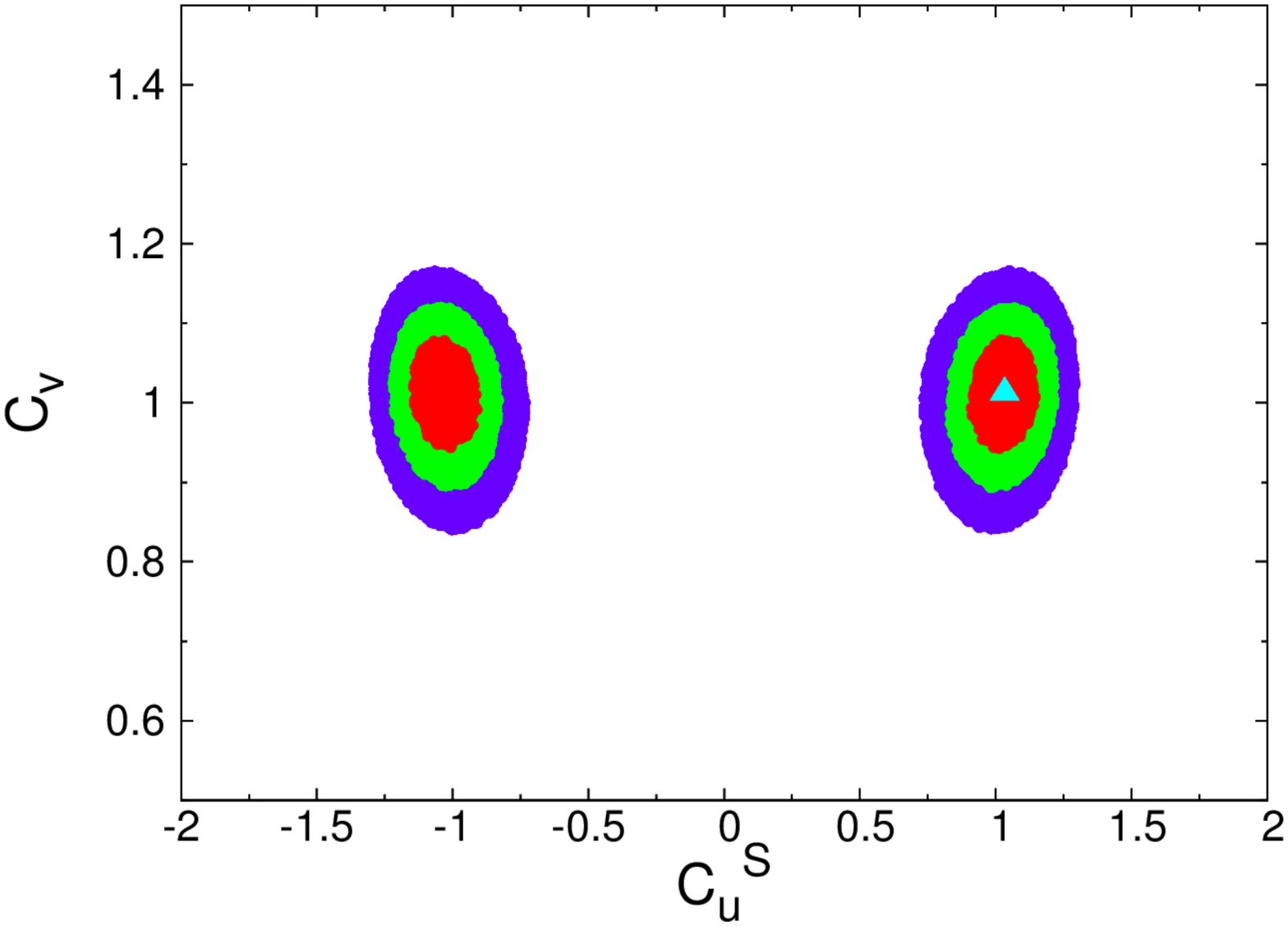}
\includegraphics[height=1.5in,angle=0]{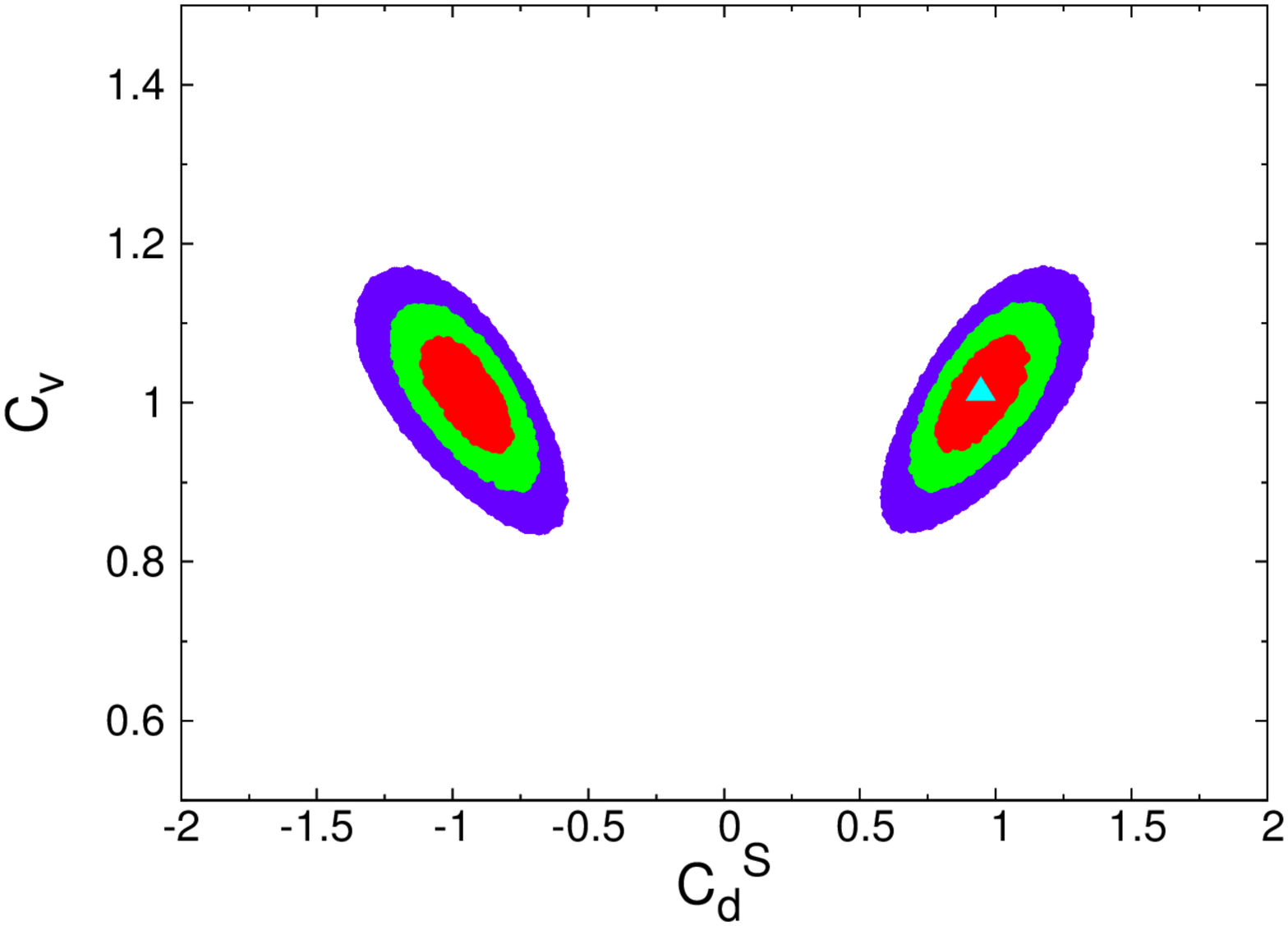}
\includegraphics[height=1.5in,angle=0]{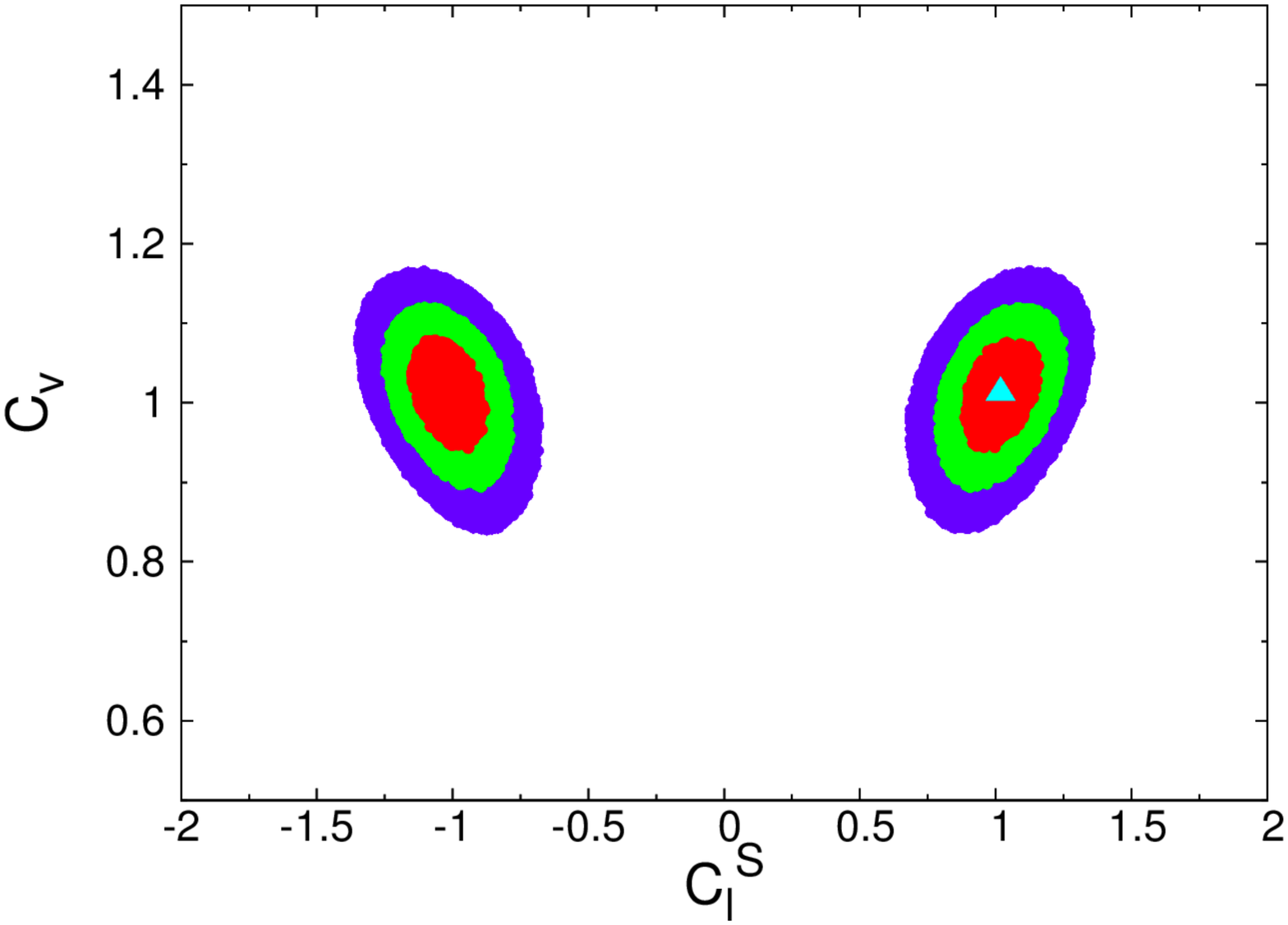}
\includegraphics[height=1.5in,angle=0]{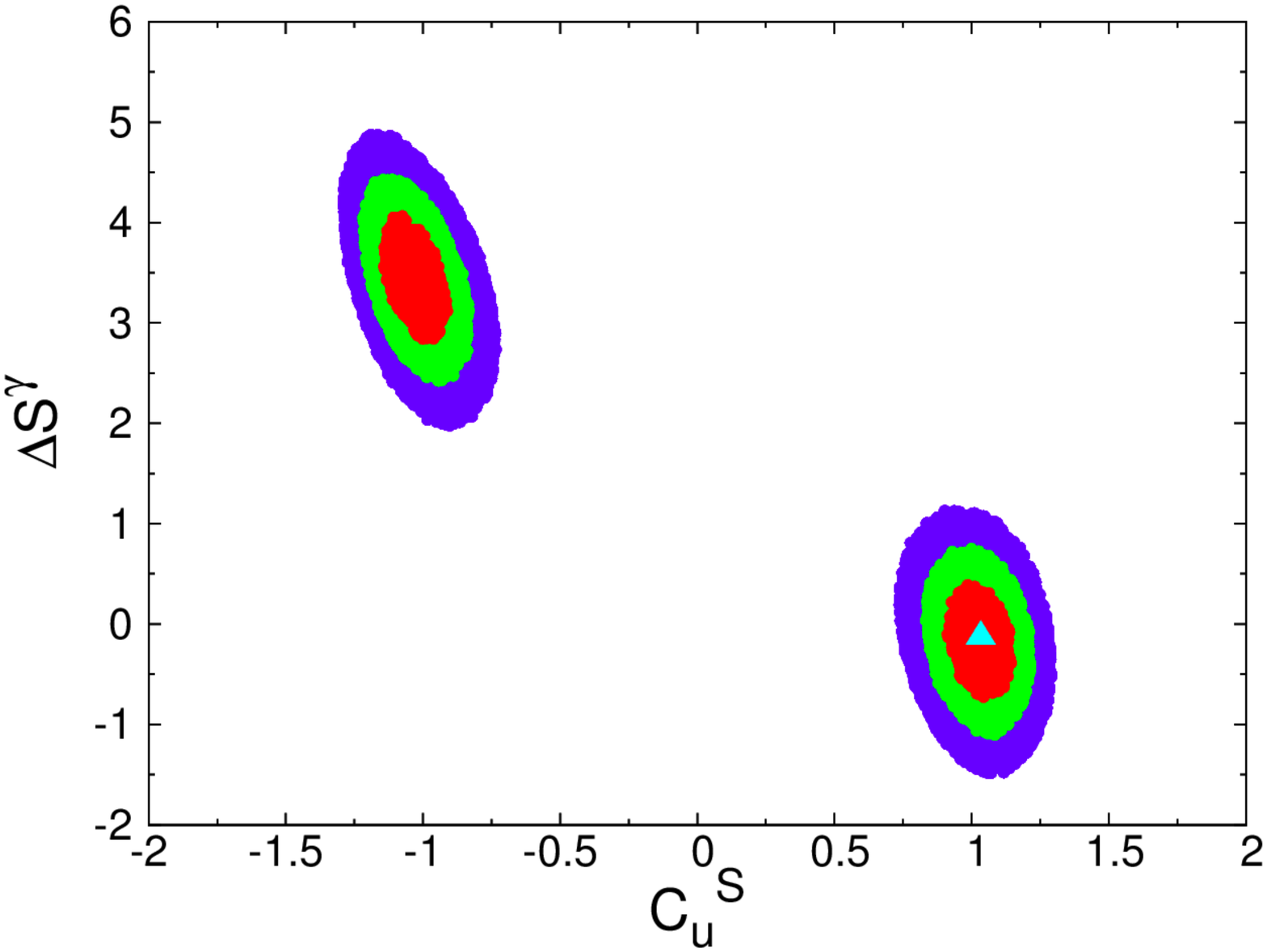}
\includegraphics[height=1.5in,angle=0]{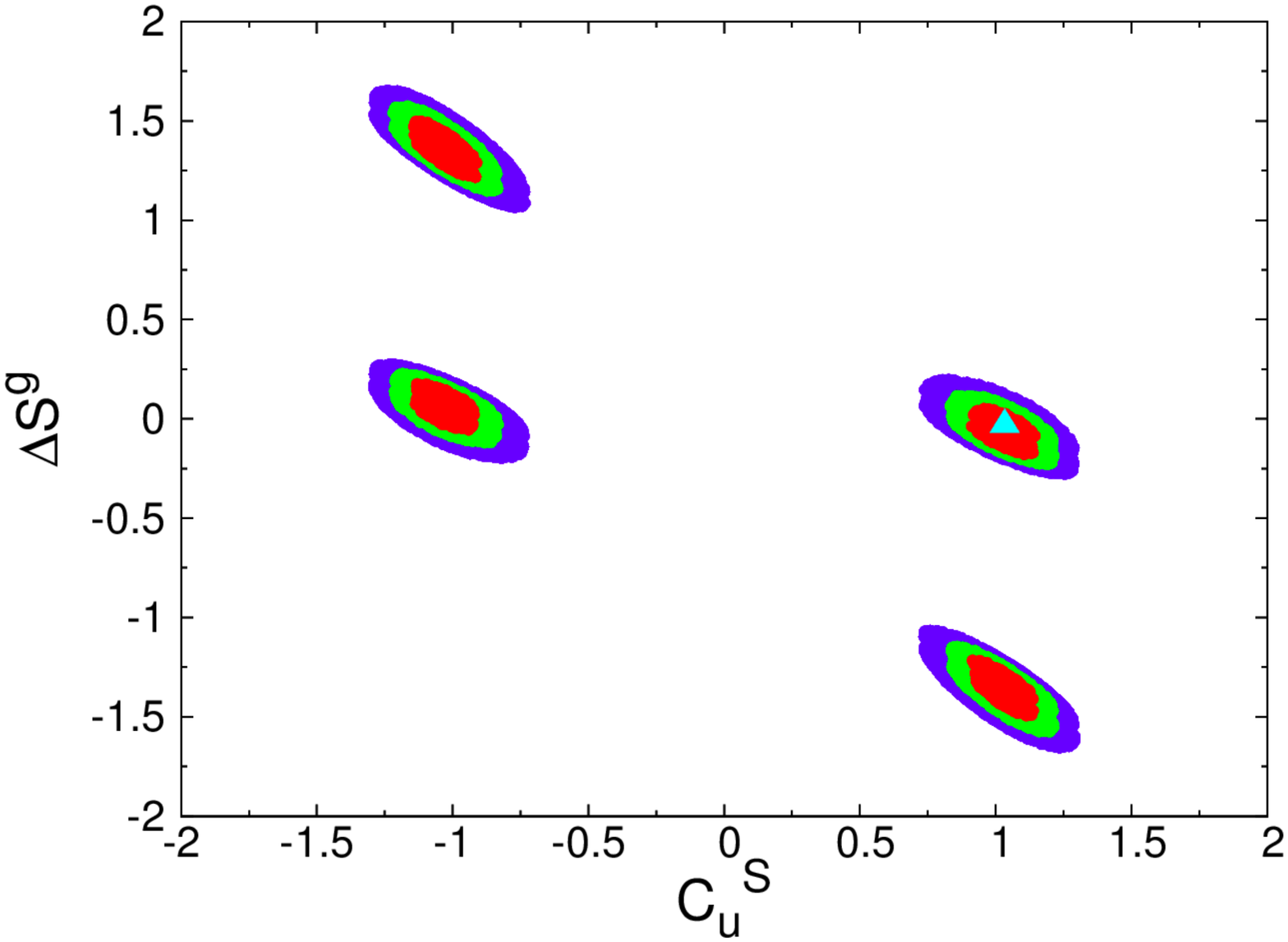}
\includegraphics[height=1.5in,angle=0]{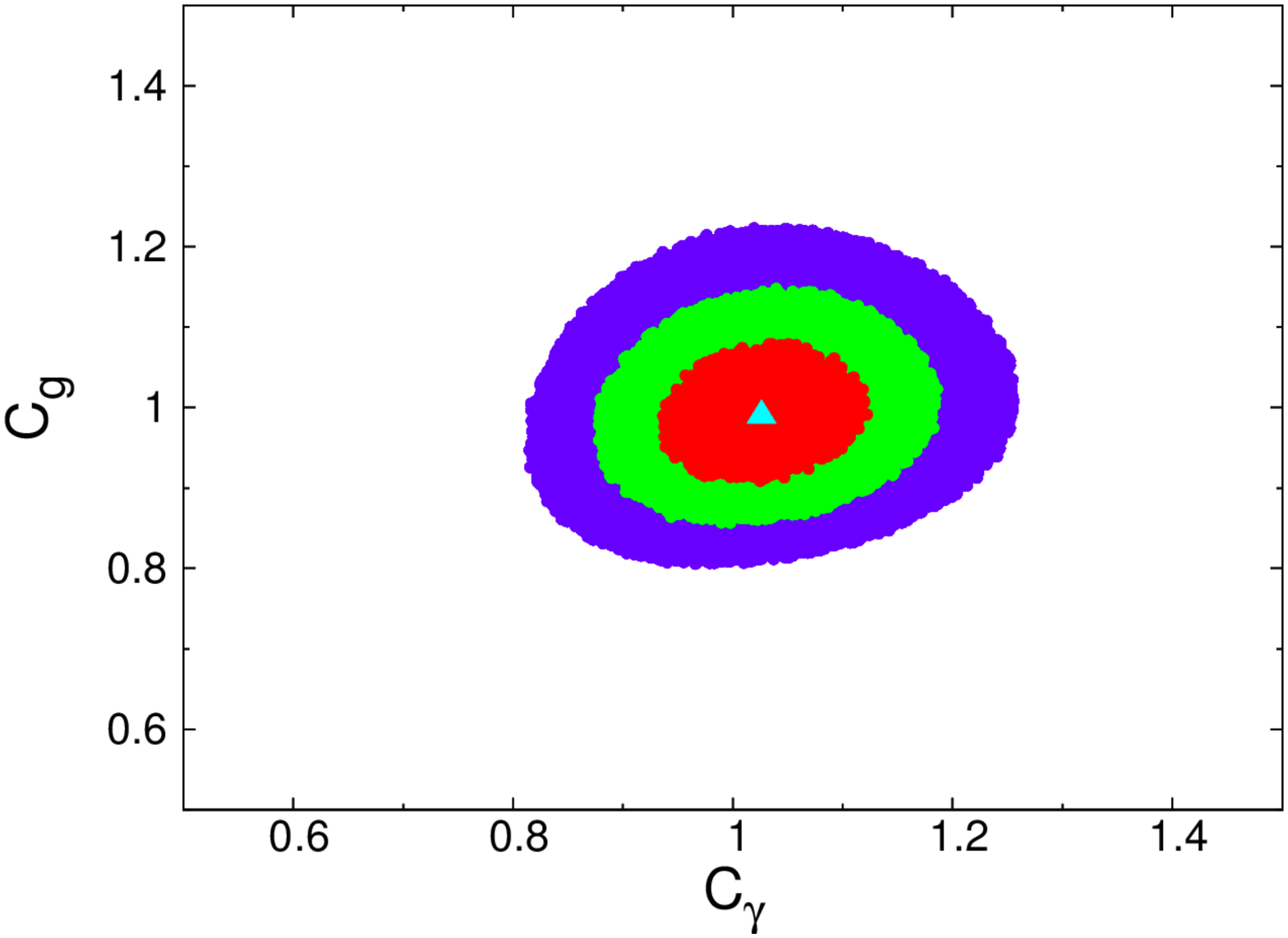}
\caption{\small \label{CPC6}
{\bf CPC6}:
The confidence-level regions of the fit by varying $\Delta S^\gamma$,
$\Delta S^g$, $C_v$, $C_u^S$, $C_d^S$, and $C_\ell^S$.
The color code is the same
as Fig.~\ref{CPC2}.
}
\end{figure}

The fitting results for {\bf CPC1} to {\bf CPC6} are shown in Table~\ref{CPC}. 
The corresponding figures for confidence regions are depicted
in Fig.~\ref{CPC1} to Fig.~\ref{CPC6}.
In the following, we are going through each fit one by one.

%
%
In {\bf CPC1}, the best-fit value for $\Delta \Gamma_{\rm tot}$ is 
\[
\Delta \Gamma_{\rm tot} = -0.285\,^{+0.18}_{-0.17} \; {\rm MeV}
\]
which is $1.6\,\sigma$ below zero. The $p$-value of this fit is 
$0.851$, which is indeed
better than the SM ($p$-value = 0.814).
This finding is consistent with the average signal
strength $\mu_{\rm All}= 1.10 \pm 0.05$. Nevertheless, we do not 
recall any new physics models that reduce the total decay width. From
the fit we can determine the upper limit for $\Delta \Gamma_{\rm tot}$.
The 95\% CL allowed range for 
$\Delta \Gamma_{\rm tot}=-0.285^{+0.38}_{-0.32}$, as shown in Fig.~\ref{CPC1}.
Assuming the fit is consistent with the SM, the 95\% CL 
upper limit for $\Delta \Gamma_{\rm tot} = 0.38$ MeV
(we simply take the central value equal to zero and use the upper error
as the upper limit), 
which translates to a branching ratio
\[
 B(H \to {\rm nonstandard}) < 8.4\% \,,
\]
which improves significantly from the previous value of 19\% \cite{update2014}.

In {\bf CPC2}, we vary $\Delta S^g$ and $\Delta S^\gamma$ -- the vertex factors
for $Hgg$ and $H\gamma\gamma$, respectively. This scenario accounts for
additional charged particles running in the loop of $H\gamma\gamma$ vertex
and additional colored particles running in the loop of $Hgg$ vertex.
The best-fit point $(\Delta S^\gamma, \Delta S^g) = (-0.226,0.016)$ shows
an increase of 3.4\% and 2.4\% in 
$|S^\gamma|$ and $|S^g|$,
respectively.
We note that the error of $\Delta S^g$ is now $\pm 0.025$,
which is numerically smaller than the SM bottom-quark contribution of $-0.037$
to the real part of $S^g$, see Eq.~(\ref{eq:hgg}), alerting that
we have reached the sensitivity to probe the sign
of the bottom-quark Yukawa coupling in gluon fusion.
The $p$-value of the best-fit point is about as the SM one.
In Fig.~\ref{CPC2}, we show the confidence-level regions of the fit 
for $\Delta \chi^2 \le 2.3$ (red), $5.99$ (green), and $11.83$ (blue) 
above the minimum, which correspond to 
confidence levels of 68.3\%, 95\%, and 99.7\%, respectively. 
The corresponding regions for $(C^\gamma, C^g)$ are also shown in the
right panel.

%
%
In {\bf CPC3},
$\Delta \Gamma_{\rm tot}$, $\Delta S^g$, and $\Delta S^\gamma$ are the
varying parameters. The best-fit point shows that the data prefer
modification of $\Delta \Gamma_{\rm tot}$ to accommodate the data 
rather than the other two parameters. It implies that
the excesses are seen in most channels, not just the diphoton channel.
Nevertheless, the $p$-value of this fit is very similar to {\bf CPC2} and the
SM. On the other hand, the better $p$-value of {\bf CPC1} indicates that 
the data prefer enhancement in all channels, instead of a particular one.
The confidence-level regions of the fit are shown in Fig.~\ref{CPC3}. 

%
The {\bf CPC4} fit allows $C_v, C_u^S,\, C_d^S, C_\ell^S$ to vary, and it shows
two most dramatic changes from previous results \cite{higgcision,update2014}.
(i) The ``island'' on the negative of $C_u^S$ in the $(C_u^S, C_v)$ plane
completely disappears, shown in the left panels of Fig.~\ref{CPC4}. 
(ii) The middle panels show that $C_d^S$ now prefers the positive sign
to the negative one.  It is more clear from the middle-lower panel
that the point $C_d^S = -1$ has $\Delta \chi^2 > 2$ above the minimum
at $C_d^S = +1$. This is the first time that the data prefer positive
bottom-Yukawa coupling to the negative one.  The key observation here
is that when we change the sign of bottom-Yukawa coupling, the
vertex factor $S^\gamma$ only changes by $0.03/6.64 = 0.0045$, which is too small
compared with experimental uncertainty. On the other hand, the
vertex factor $S^g$ changes by $0.074/0.651 = 0.11$, which now becomes 
comparable to experimental uncertainty. This is the reason why 
the positive bottom-Yukawa is more preferred in the scenario with
$\Delta S^g = 0$.
Yet, the current data precision still do not show any preference 
for the sign of tau-Yukawa coupling,
as shown in the right panels of Fig.~\ref{CPC4}.

%
%
{\bf CPC6} is the most general scenario that we consider. Now confidence regions,
as in the upper panels of Fig.~\ref{CPC6}, show that both signs ($\pm 1$) of 
top-Yukawa $C_u^S$, bottom-Yukawa $C_d^S$,
and tau-Yukawa $C_\ell^S$
are equally good in describing
the data, because of the compensations from $\Delta S^g$ and $\Delta S^\gamma$.
For the positive sign of $C_u^S$, there are 4 possible combinations of $C_d^S$ and
$C_\ell^S$ with $\Delta S^\gamma \sim 0$
\footnote{In this work, we neglect the other possibility 
of $\Delta S^\gamma \sim 13\,(17)$ for positive (negative) $C_u^S$.},
see the lower-left panel of Fig.~\ref{CPC6}.
Together with the two minima at $\Delta S^g=-0.03\,(-0.10)$ and $-1.32\,(-1.39)$
for $C_d^S\sim 1\,(-1)$ as shown in lower-middle panel of Fig.~\ref{CPC6}, 
one has 8 minima. Similarly, for the negative sign of
$C_u^S$, there are also 8 minima with $\Delta S^\gamma \sim 3.4$. 
In total, therefore, there
are 16 degenerate minima in the {\bf CPC6} fit.
In Table~\ref{CPC}, we only show the minimum at $C_{u,d,\ell}^{S} \sim 1$
and $\Delta S^{\gamma ,g}\sim 0$.
A substantial improvement from previous results is that 
the confidence-level regions shown in Fig.~\ref{CPC6} are now 
well separated islands, while in previous
results \cite{update2014} those islands are ``connected''. For example,
in the plane of $(C_u^S, C_v)$, the negative and positive islands of $C_u^S$
are now separated but they were connected in previous results. It means
that previously $C_u^S = 0$ was allowed but not in the current data.

Before moving to {\bf CPCN} fits,
we note that the negative top-quark Yukawa coupling is allowed only in the
presence of non-zero $\Delta S^\gamma$
which can offset the flipped top-quark contribution to $S^\gamma$.
The required tuning is now $\delta(\Delta S^\gamma) \simeq \pm 0.4$ 
at 1 $\sigma$ level, which is
about 10\% of the change in $\Delta S^\gamma$ due to the negative top-quark
Yukawa coupling.
The tuning will be more and more severe as more data accumulate.

\begin{table}[t!]
\caption{\small \label{CPCN}
({\bf CPCN})
The best-fitted values in various CP conserving fits and the corresponding 
chi-square per degree of freedom and goodness of fit.
The $p$-value for each fit hypothesis against the SM null
hypothesis is also shown.
For the SM, we obtain $\chi^2=53.81$, $\chi^2/dof=53.81/64$,
and so the goodness of fit $=0.814$. 
}
\begin{ruledtabular}
\begin{tabular}{c|cc|cccc}
Cases & {\bf CPCN2} & {\bf CPCN3} & \multicolumn{4}{c}{\bf CPCN4} \\
\hline
      & Vary $C^S_u,C_v$ & {Vary $C^S_u,C_v$} & \multicolumn{4}{c}{Vary $C^S_u,C_v$} \\
Parameters &  & {$\Delta S^\gamma$} & \multicolumn{4}{c}{$\Delta S^\gamma$, $\Delta S^g$}   \\
\hline
\multicolumn{7}{c}{After ICHEP 2018}\\
\hline
$C^S_u$           & $1.017^{+0.039}_{-0.037}$ &  $1.016^{+0.039}_{-0.038}$ 
                  & $1.042^{+0.077}_{-0.081}$ & $1.042^{+0.078}_{-0.081}$
                  & $-1.042^{+0.081}_{-0.078}$ & $-1.042^{+0.081}_{-0.078}$ 
                   \\
$C^S_d$           & 1 & 1 & 1 & 1 & 1 & 1  \\
$C^S_\ell$        & 1 & 1 & 1 & 1 & 1 & 1 \\
$C_v$           & $1.030^{+0.028}_{-0.028}$ & $1.025^{+0.034}_{-0.035}$ 
                  & $1.027^{+0.034}_{-0.036}$ & $1.027^{+0.034}_{-0.036}$
                  & $1.028^{+0.034}_{-0.036}$ & $1.028^{+0.034}_{-0.036}$ 
                   \\
$\Delta S^\gamma$ & 0     & $-0.090^{+0.36}_{-0.36}$ 
                  & $-0.129^{+0.37}_{-0.37}$   & $-0.129^{+0.37}_{-0.37}$
                  & $3.524^{+0.41}_{-0.42}$  & $3.523^{+0.41}_{-0.42}$ 
                    \\
$\Delta S^g$      & 0     & 0  
                  & $-0.021^{+0.057}_{-0.055}$  & $-1.34^{+0.066}_{-0.065}$
                  & $0.095^{+0.055}_{-0.057}$  & $1.414^{+0.066}_{-0.066}$ 
                       \\
$\Delta \Gamma_{\rm tot}$ (MeV) & 0 & 0 
                                & 0 & 0  & 0 & 0 \\
\hline
$\chi^2/dof$ & 51.16/62 & {51.10/61} & \multicolumn{4}{c}{50.96/60}  \\
goodness of fit    &  0.835   &   0.813    &  \multicolumn{4}{c}{0.791}    \\
$p$-value    &  0.266 & 0.439  &  \multicolumn{4}{c}{0.583}    \\
\end{tabular}
\end{ruledtabular}
\end{table}
\subsubsection{{\bf CPCN2} to {\bf CPCN4}}
\begin{figure}[h!]
\centering
\includegraphics[height=2.8in,angle=0]{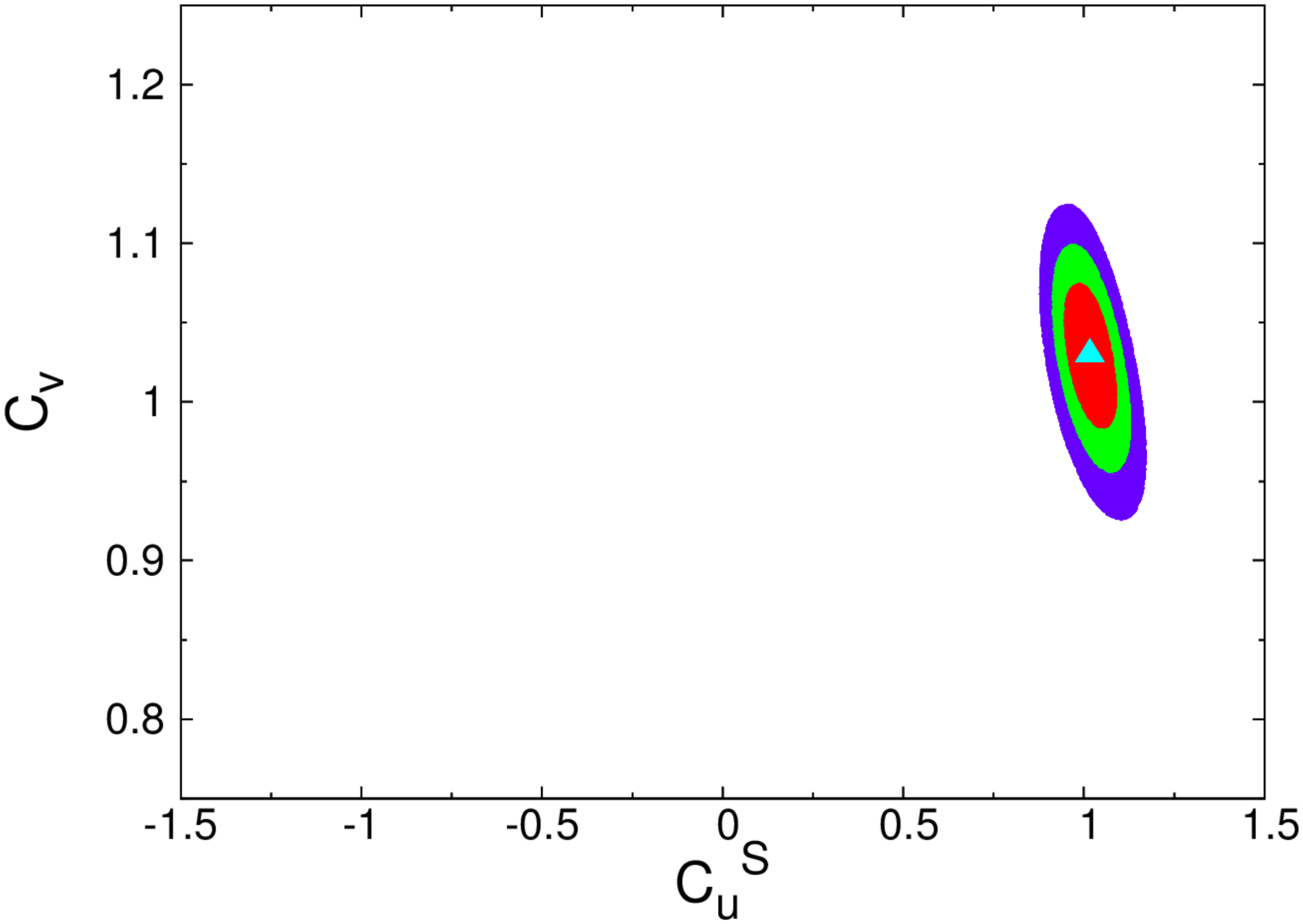}
\caption{\small \label{CPCN2} 
{\bf CPCN2}:
The confidence-level regions of the fit by varying 
$C_u^S$ and $C_v$. The color code is the same as in Fig.~\ref{CPC2}.
}
\end{figure}

\begin{figure}[t!]
\centering
\includegraphics[height=1.8in,angle=0]{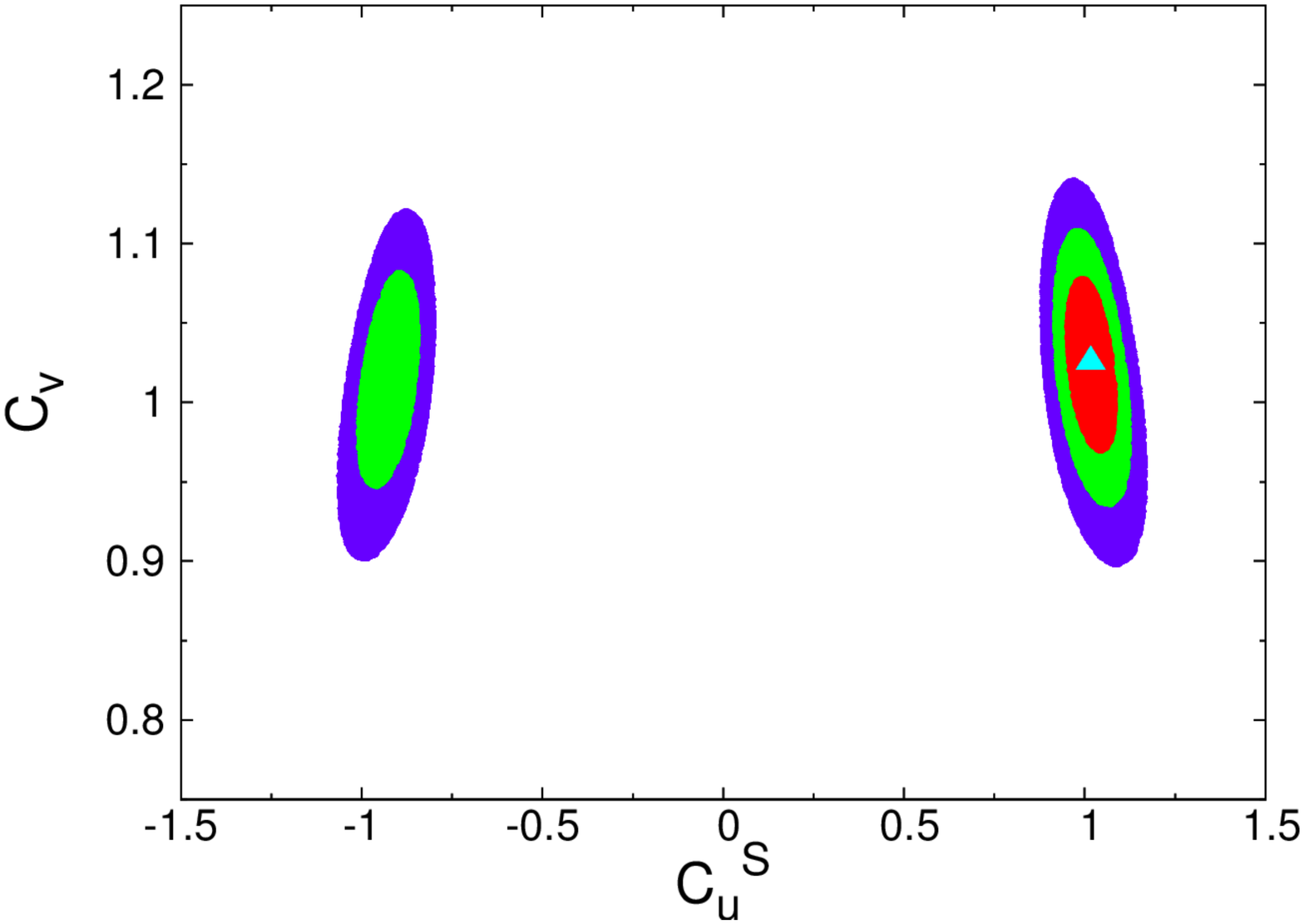}
\includegraphics[height=1.8in,angle=0]{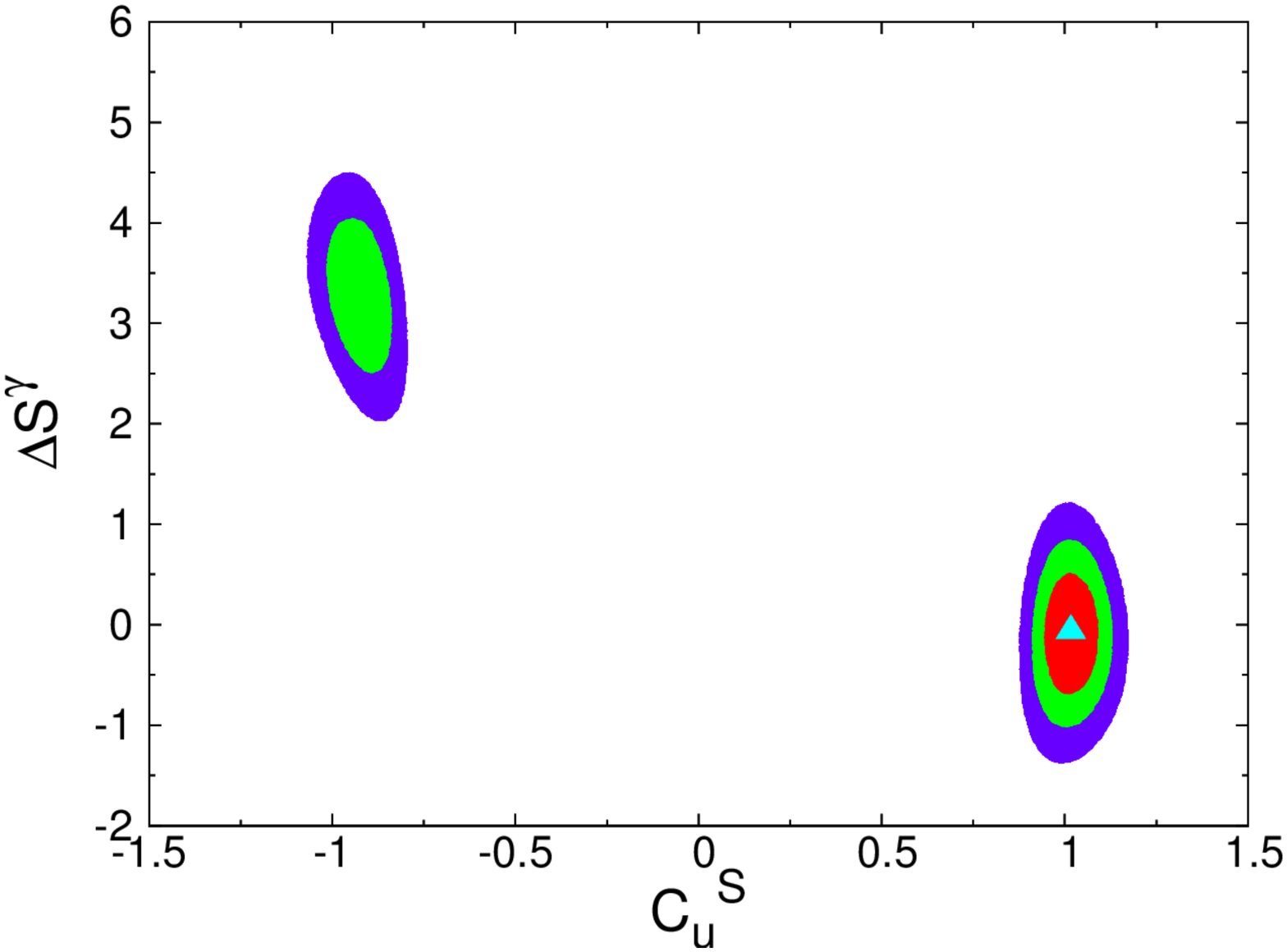}
\includegraphics[height=1.8in,angle=0]{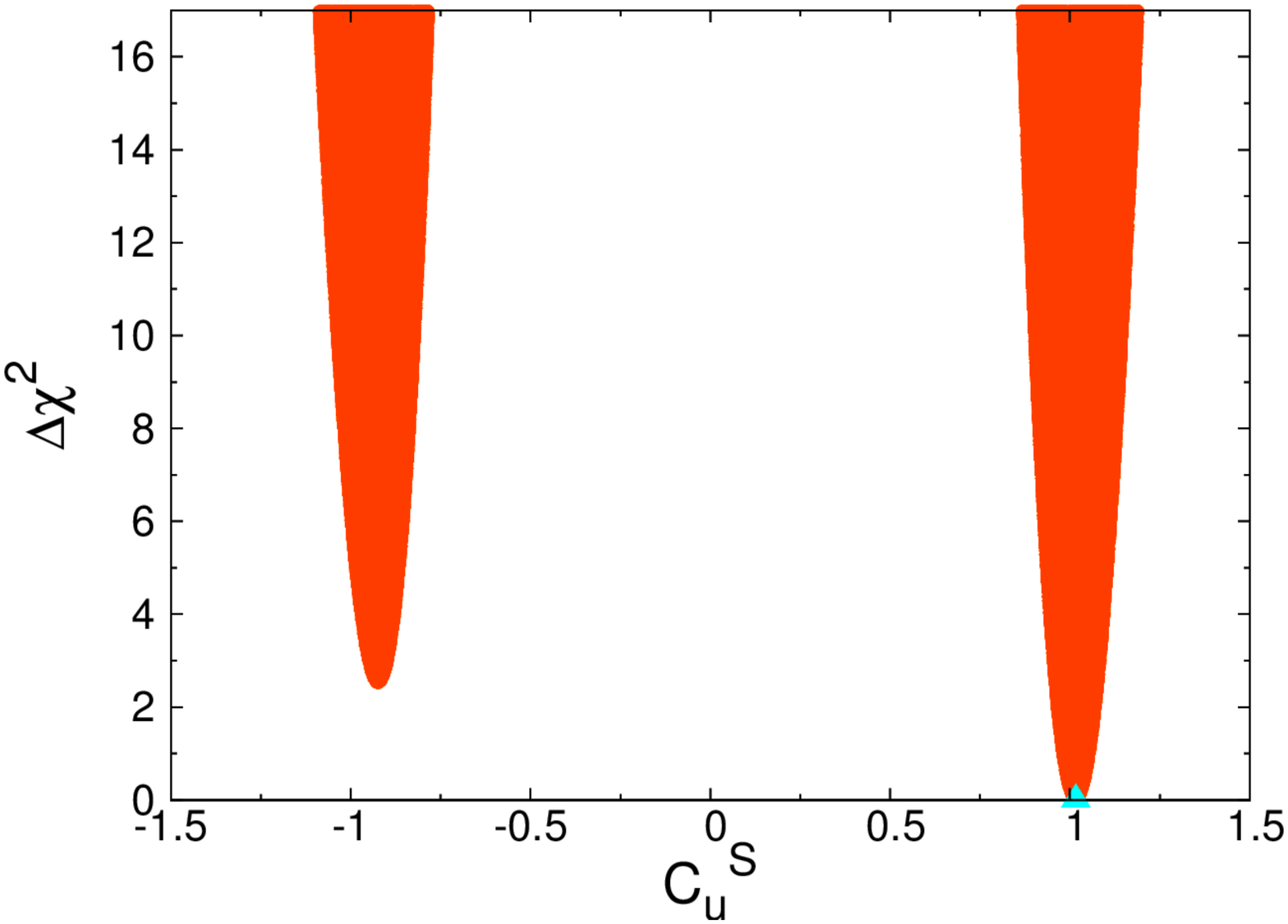}
\includegraphics[height=1.8in,angle=0]{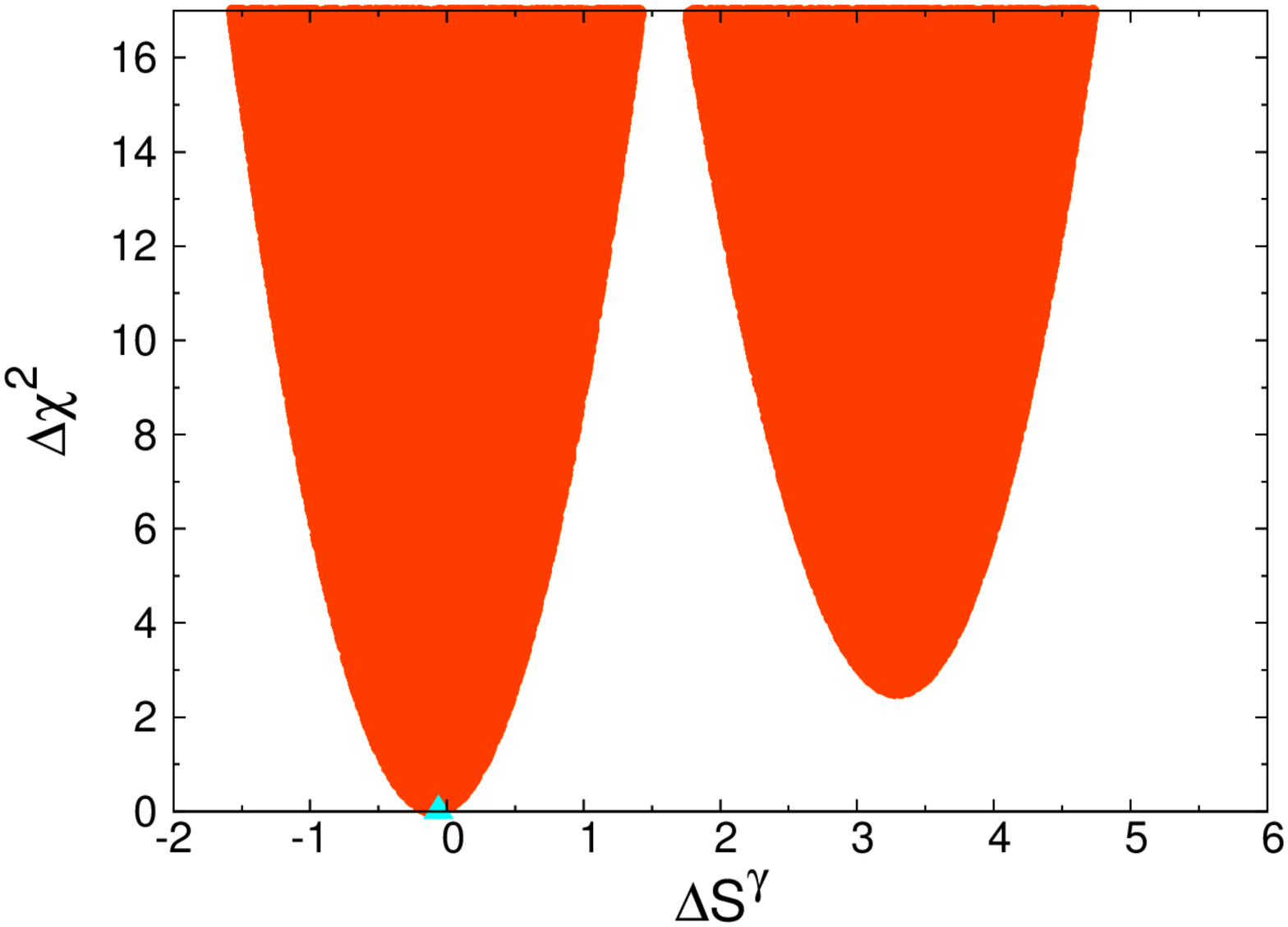}
\caption{\small \label{CPCN3}
{\bf CPCN3}: (Upper)
The confidence-level regions of the fit by varying
$\Delta S^\gamma$, $C_u^S$, and $C_v$.
The color code is the same as in Fig.~\ref{CPC2}.
(Lower) 
$\Delta \chi^2$ versus $C_u^S$ (left) and 
$\Delta \chi^2$ versus $\Delta S^\gamma$ (right).
}
\end{figure}

\begin{figure}[t!]
\centering
\includegraphics[height=1.5in,angle=0]{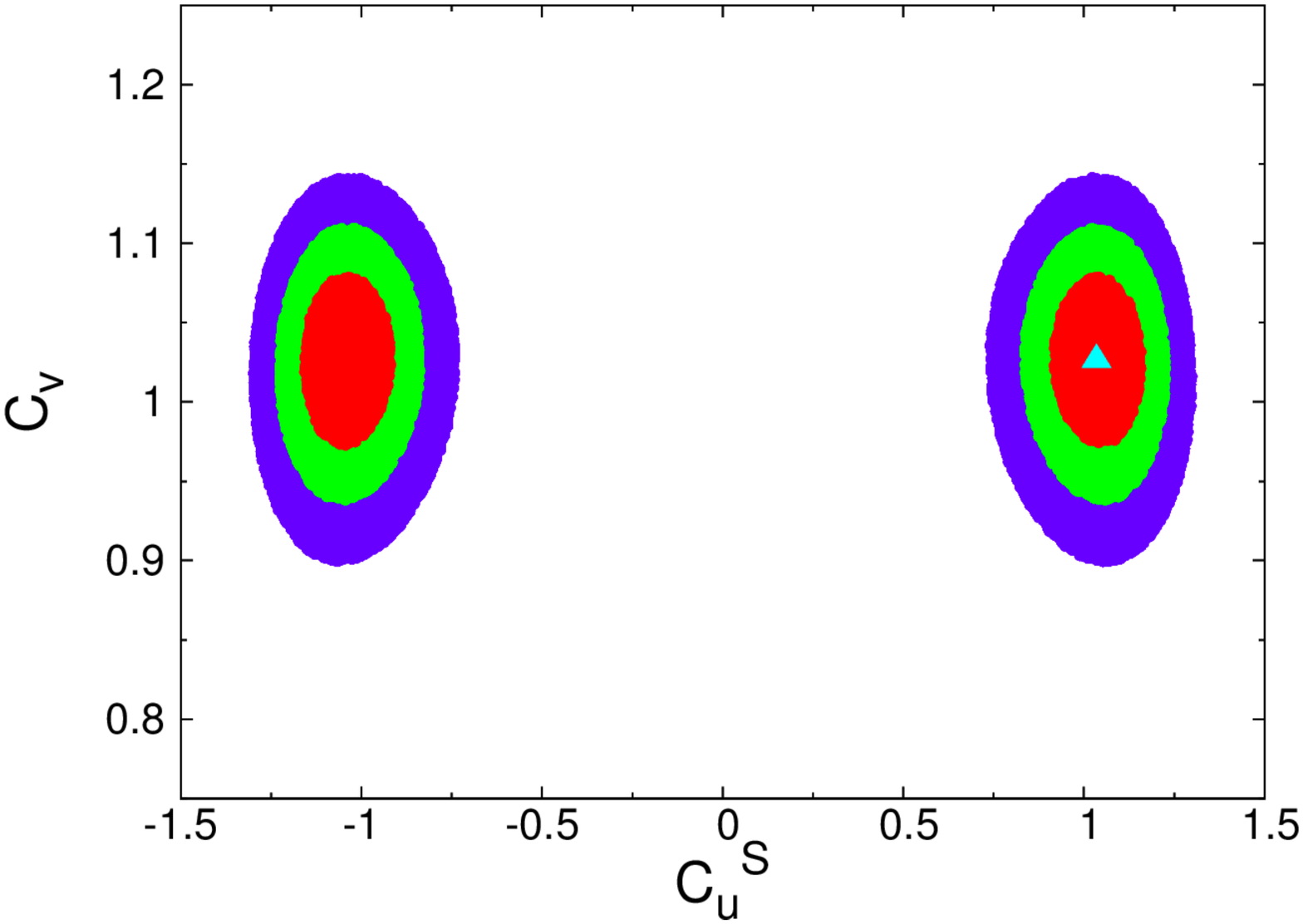}
\includegraphics[height=1.5in,angle=0]{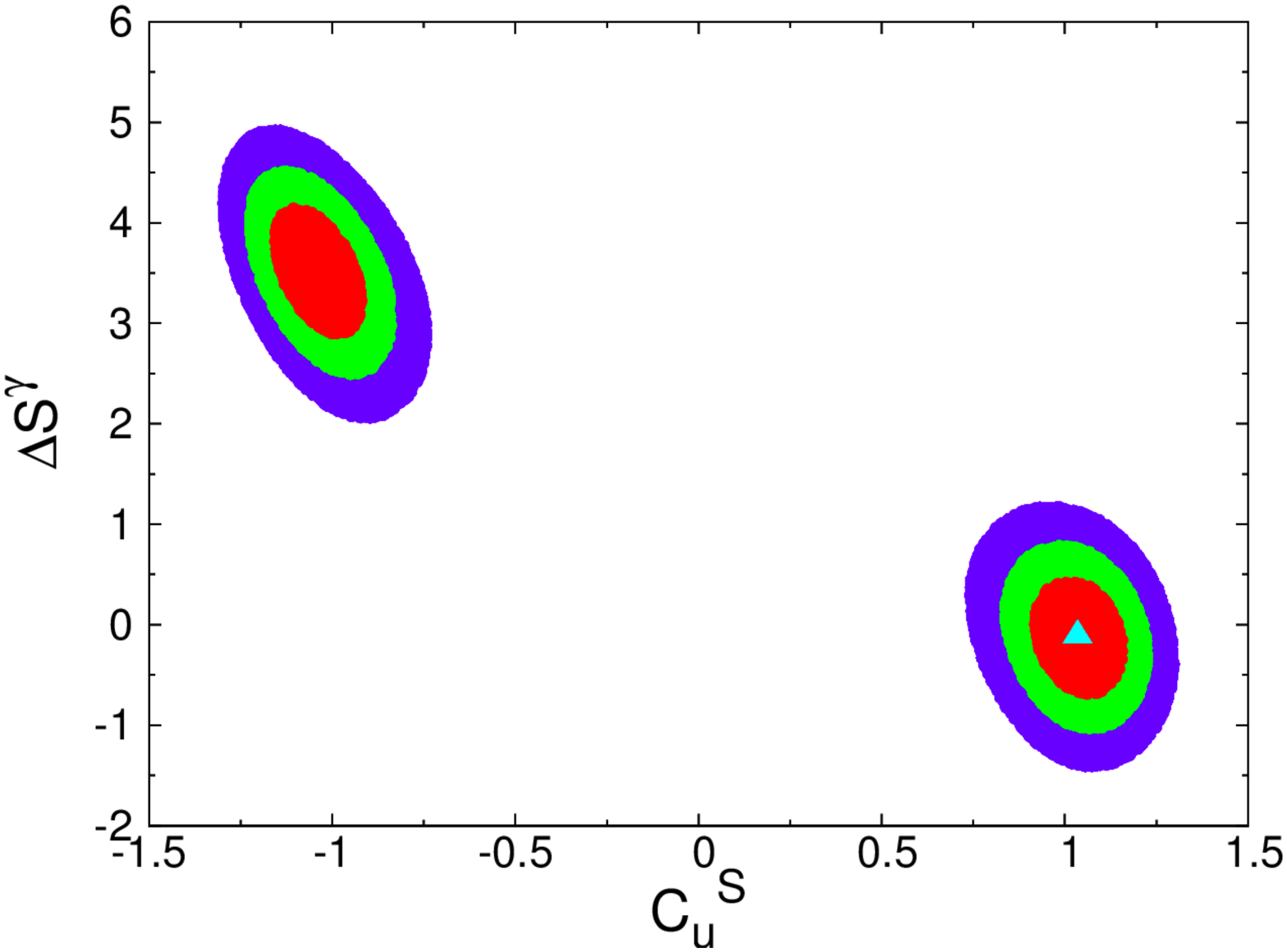}
\includegraphics[height=1.5in,angle=0]{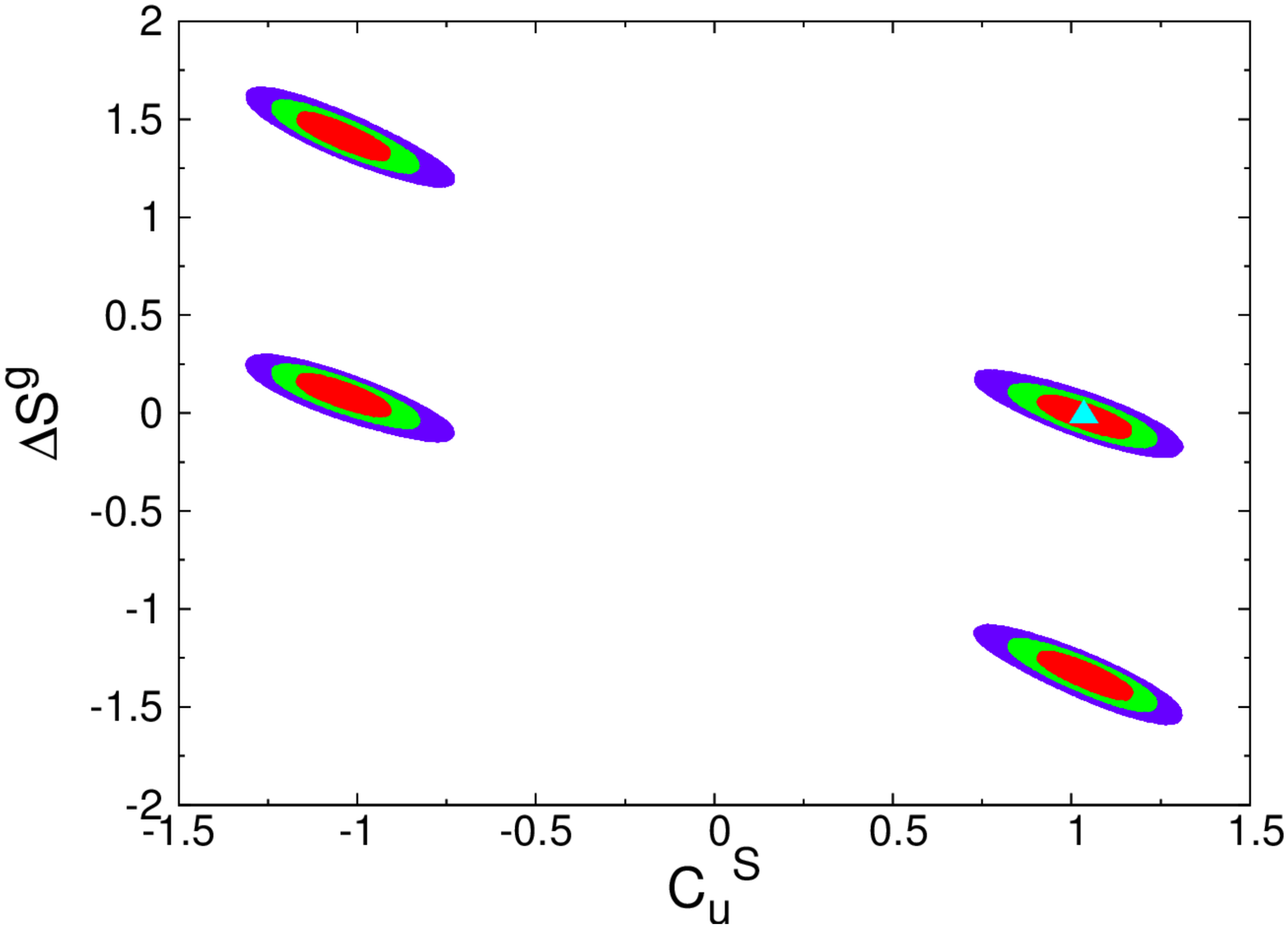}
\includegraphics[height=1.5in,angle=0]{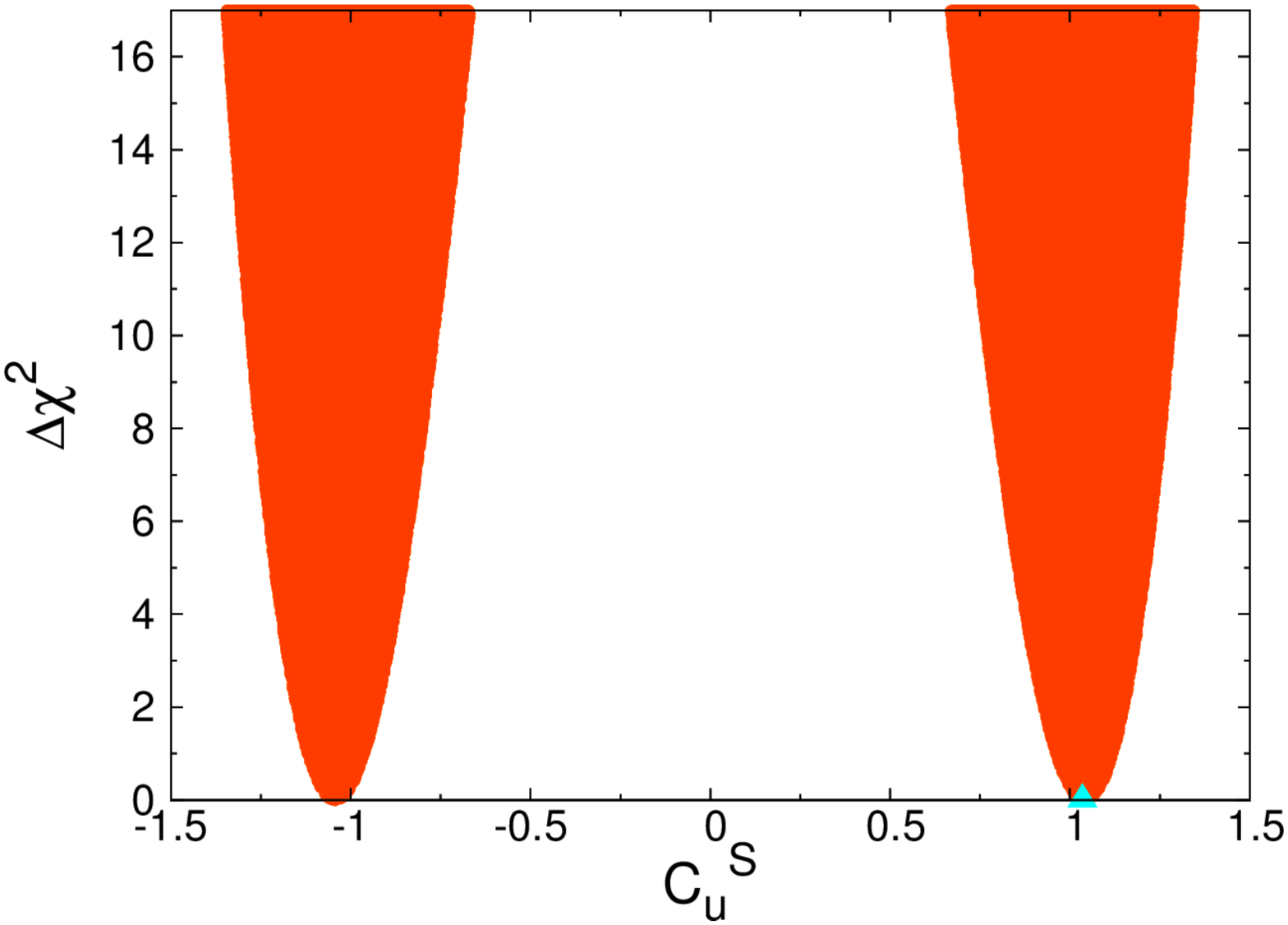}
\includegraphics[height=1.5in,angle=0]{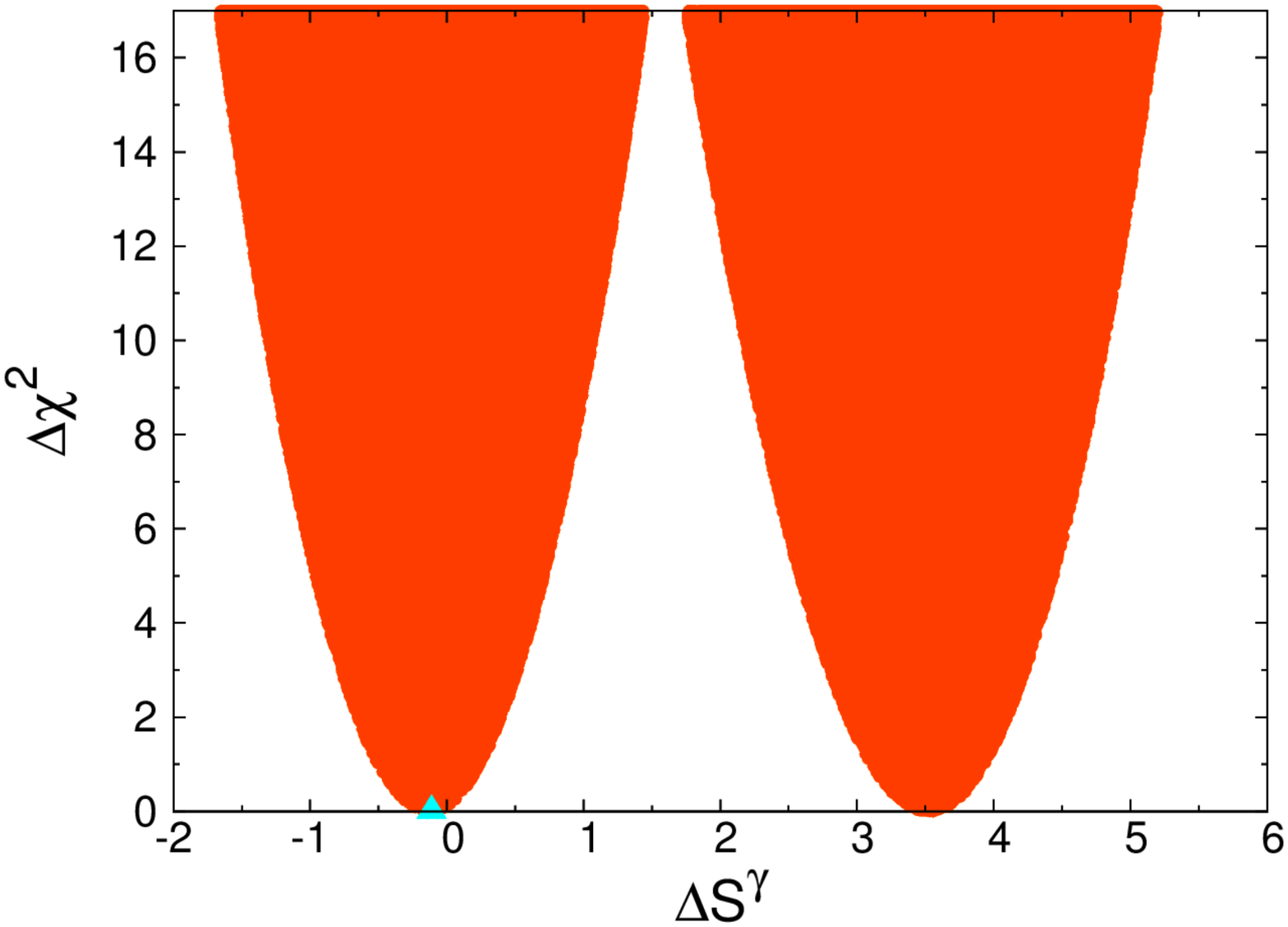}
\includegraphics[height=1.5in,angle=0]{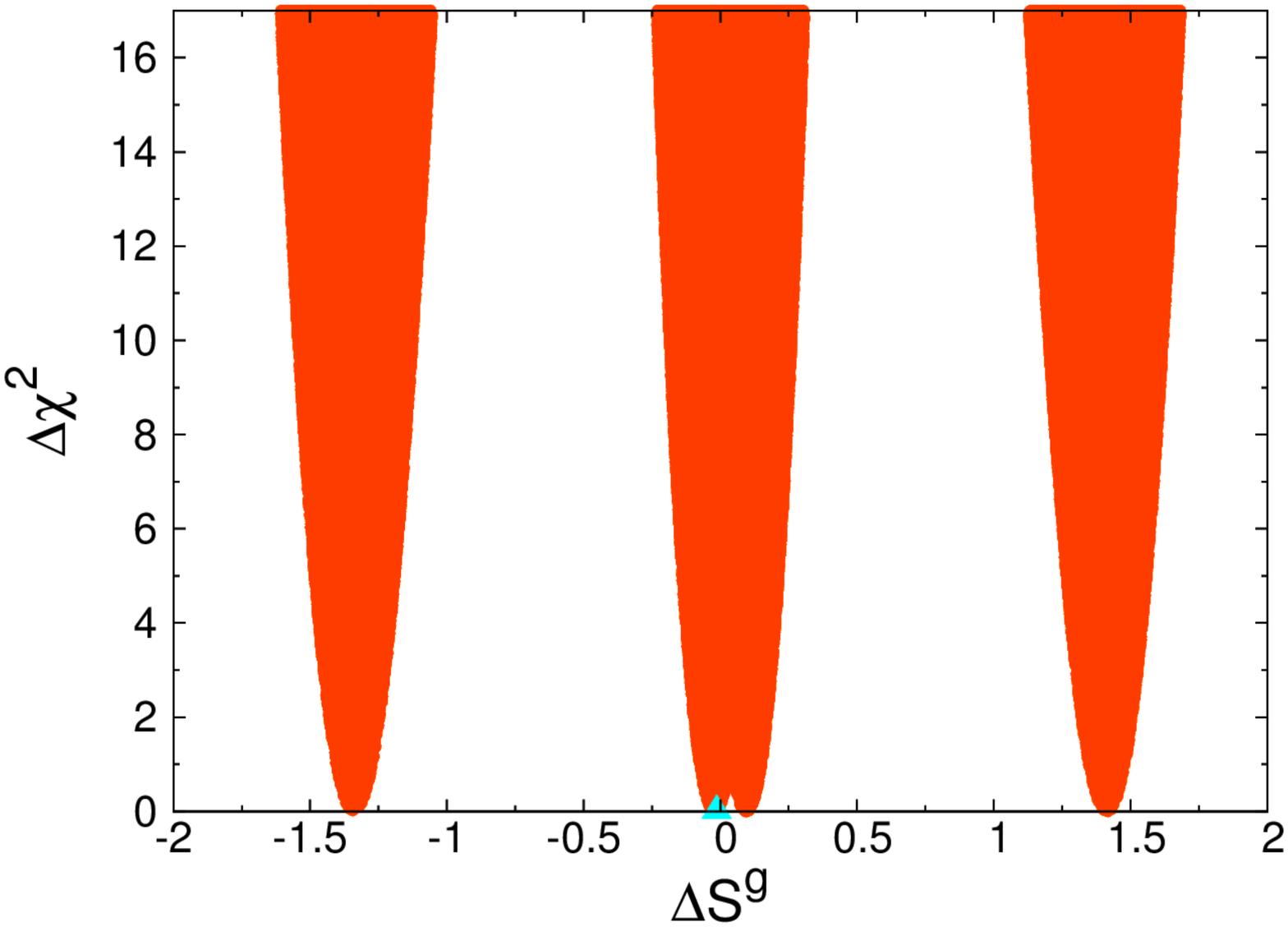}
\caption{\small \label{CPCN4}
{\bf CPCN4}: (Upper)
The confidence-level regions of the fit by varying
$\Delta S^g$, $\Delta S^\gamma$, $C_u^S$, and $C_v$.
The color code is the same as in Fig.~\ref{CPC2}.
(Lower) 
$\Delta \chi^2$ versus $C_u^S$ (left),
$\Delta \chi^2$ versus $\Delta S^\gamma$ (middle),  and 
$\Delta \chi^2$ versus $\Delta S^g$ (right).
}
\end{figure}

We can see in {\bf CPC2} to {\bf CPC6}, the bottom-Yukawa and tau-Yukawa couplings
are not very sensitive to the overall fits, though the bottom-Yukawa
shows slight preference on the positive side in {\bf CPC4}. Here we 
attempt to use the more effective parameters in the fits. 
The best-fits points and their $p$-values are shown in Table~\ref{CPCN},
and their corresponding figures in Fig.~\ref{CPCN2} to \ref{CPCN4}.

%
%
In {\bf CPCN2}, we vary only $C_v$ and $C_u^S$.
This fit offers a slightly better $p$-value than the SM. 
While in {\bf CPCN3},
we also vary $\Delta S^\gamma$ in addition to $C_v$ and $C_u^S$. We find
that it has little improvement over the {\bf CPCN2} 
in terms of total $\chi^2$ but, with one less 
degree of freedom, the $p$-value indeed decreases. As shown in 
Fig.~\ref{CPCN3}, there are two minima: 
the minimum near $(C_u^S,\Delta S^\gamma)=(1,0)$ provides a 
better solution by $\Delta\chi^2 \approx 2$ than the other one 
near $(C_u^S,\Delta S^\gamma)=(-1,3.2)$.
The $\Delta S^\gamma$ can compensate the flip in sign of $C_u^S$ in 
the vertex factor $S^\gamma$.  However, when
$C_u^S$ flips the sign, $|S^g|$ increases by more than 
10\% leading to a worse fit.

%
%
In {\bf CPCN4}, we vary the four most efficient fitting parameters
$C_v$, $C_u^S$, $\Delta S^\gamma$, and $\Delta S^g$. Therefore, in contrast 
to {\bf CPCN3}, the $\Delta S^g$ here can compensate the sign change in 
$C_u^S$, such that there are four minima in this fit with the 
same $p$-value, as shown in Table~\ref{CPCN} and Fig.~\ref{CPCN4}.

%
\begin{table}[t!]
\caption{\small \label{CPCX}
({\bf CPCX})
The best-fitted values in various CP conserving fits and the corresponding
chi-square per degree of freedom and goodness of fit.
The $p$-value for each fit hypothesis against the SM null
hypothesis is also shown.
}
\begin{ruledtabular}
\begin{tabular}{c|c|cc||c|cc}
Cases & {\bf CPCX2} & \multicolumn{2}{c||}{\bf CPCX3} & Cases &
\multicolumn{2}{c}{\bf CPCX4} \\
\hline
      & Vary $C_v,\Delta\Gamma_{\rm tot}$ & \multicolumn{2}{c||}{Vary $C^S_u,C_v$}
& & \multicolumn{2}{c}{Vary $C^S_u,C_w$}  \\
Parameters &  & \multicolumn{2}{c||}{$\Delta S^g$} & Parameters &
\multicolumn{2}{c}{$C_z$, $\Delta S^g$}   \\
\hline
\multicolumn{7}{c}{After ICHEP 2018}\\ 
\hline   
$C^S_u$           & 1 & $1.04^{+0.08}_{-0.08}$ & $1.04^{+0.08}_{-0.08}$
& $C^S_u$ & $1.045^{+0.078}_{-0.081}$ & $1.045^{+0.078}_{-0.081}$  \\
$C^S_d$           & 1 & 1 & 1
& $C^S_d$           & 1 & 1  \\
$C^S_\ell$        & 1 & 1 & 1  
& $C^S_\ell$        & 1 & 1 \\
$C_v$           & $1.020^{+0.051}_{-0.049}$ & $1.03^{+0.03}_{-0.03}$ &
$1.03^{+0.03}_{-0.03}$
& $C_w$           & $1.040^{+0.033}_{-0.034}$ & $1.040^{+0.032}_{-0.034}$ \\
- &&&  
& $C_z$           & $1.015^{+0.048}_{-0.049}$ & $1.015^{+0.048}_{-0.049}$\\
$\Delta S^\gamma$ & 0     & 0 & 0  
& $\Delta S^\gamma$ & 0 & 0 \\
$\Delta S^g$      & 0     & $-0.02^{+0.06}_{-0.05}$ & $-1.34^{+0.07}_{-0.06}$
& $\Delta S^g$      & $-0.020^{+0.056}_{-0.054}$ & $-1.345^{+0.067}_{-0.067}$ \\
$\Delta \Gamma_{\rm tot}$ (MeV) & $-0.134^{+0.43}_{-0.36}$ & 0 & 0
& $\Delta \Gamma_{\rm tot}$ (MeV) & 0 & 0 \\ 
\hline 
$\chi^2/dof$ & 51.25/62 & \multicolumn{2}{c||}{51.08/61}
& $\chi^2/dof$ & \multicolumn{2}{c}{50.84/60} \\
goodness of fit    &   0.833  &   \multicolumn{2}{c||}{0.813}
& goodness of fit & \multicolumn{2}{c}{0.820} \\
$p$-value    &   0.278  &   \multicolumn{2}{c||}{0.435}
& $p$-value & \multicolumn{2}{c}{0.5631} \\
\end{tabular}
\end{ruledtabular}
\end{table}
\subsubsection{{\bf CPCX2} to {\bf CPCX4}}
\begin{figure}[t!]
\centering
\includegraphics[height=3.0in,angle=0]{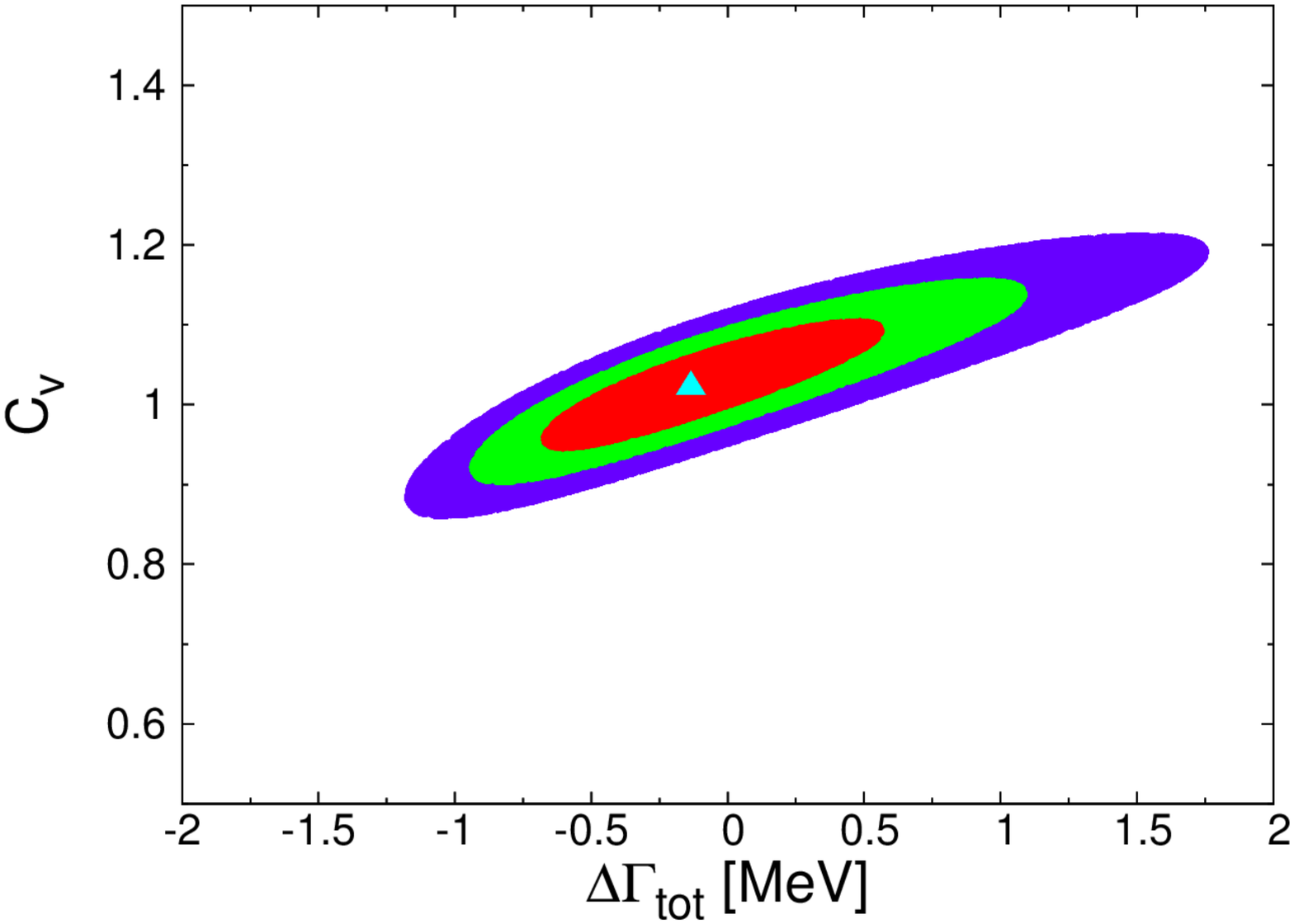}
\caption{\small \label{CPCX2} 
{\bf CPCX2}:
The confidence-level regions of the fit by varying 
$C_v$ and $\Delta \Gamma_{\rm tot}$.
The color code is the same as in Fig.~\ref{CPC2}.
}
\end{figure}

\begin{figure}[t!]
\centering
\includegraphics[height=2.0in,angle=0]{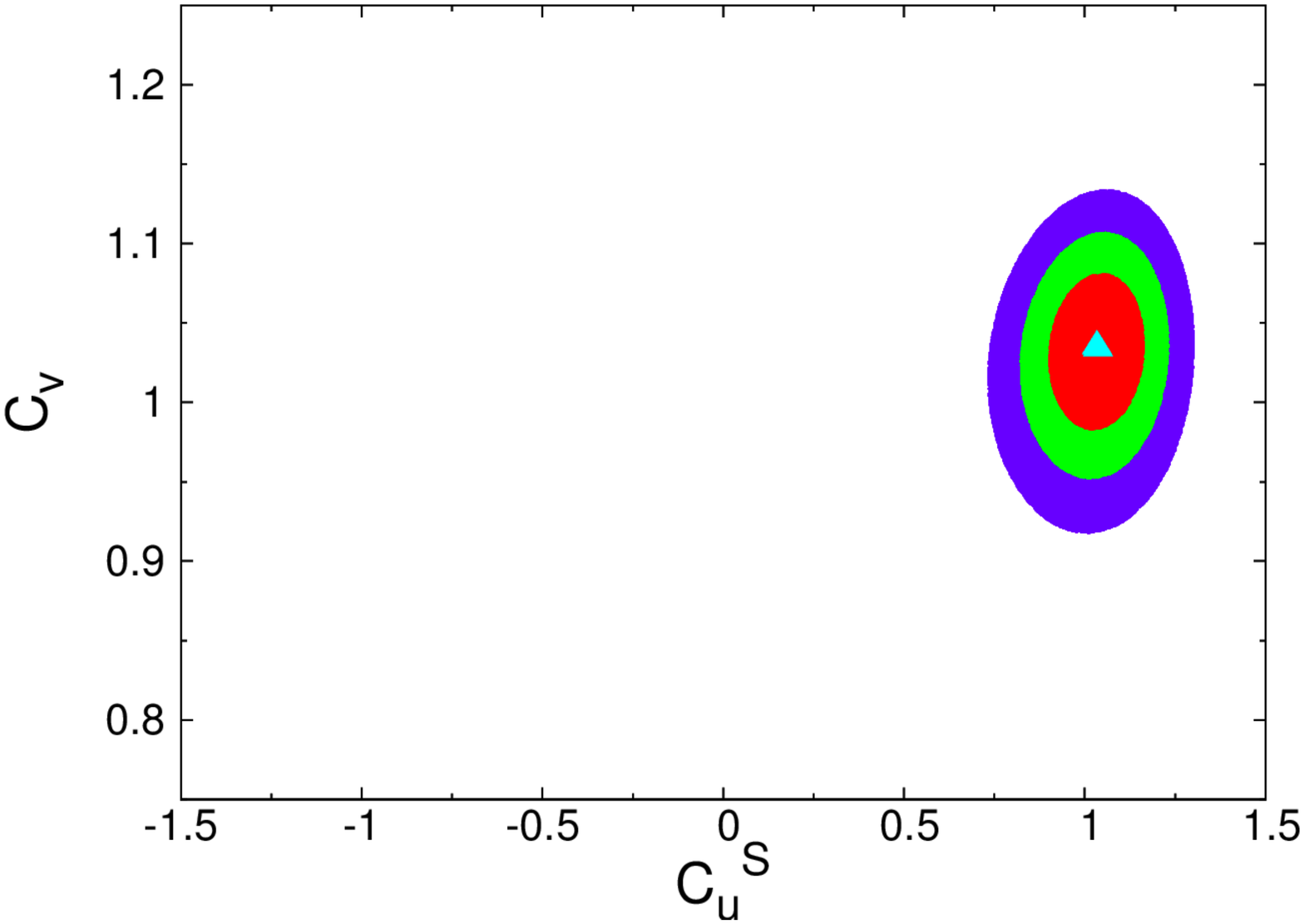}
\includegraphics[height=2.0in,angle=0]{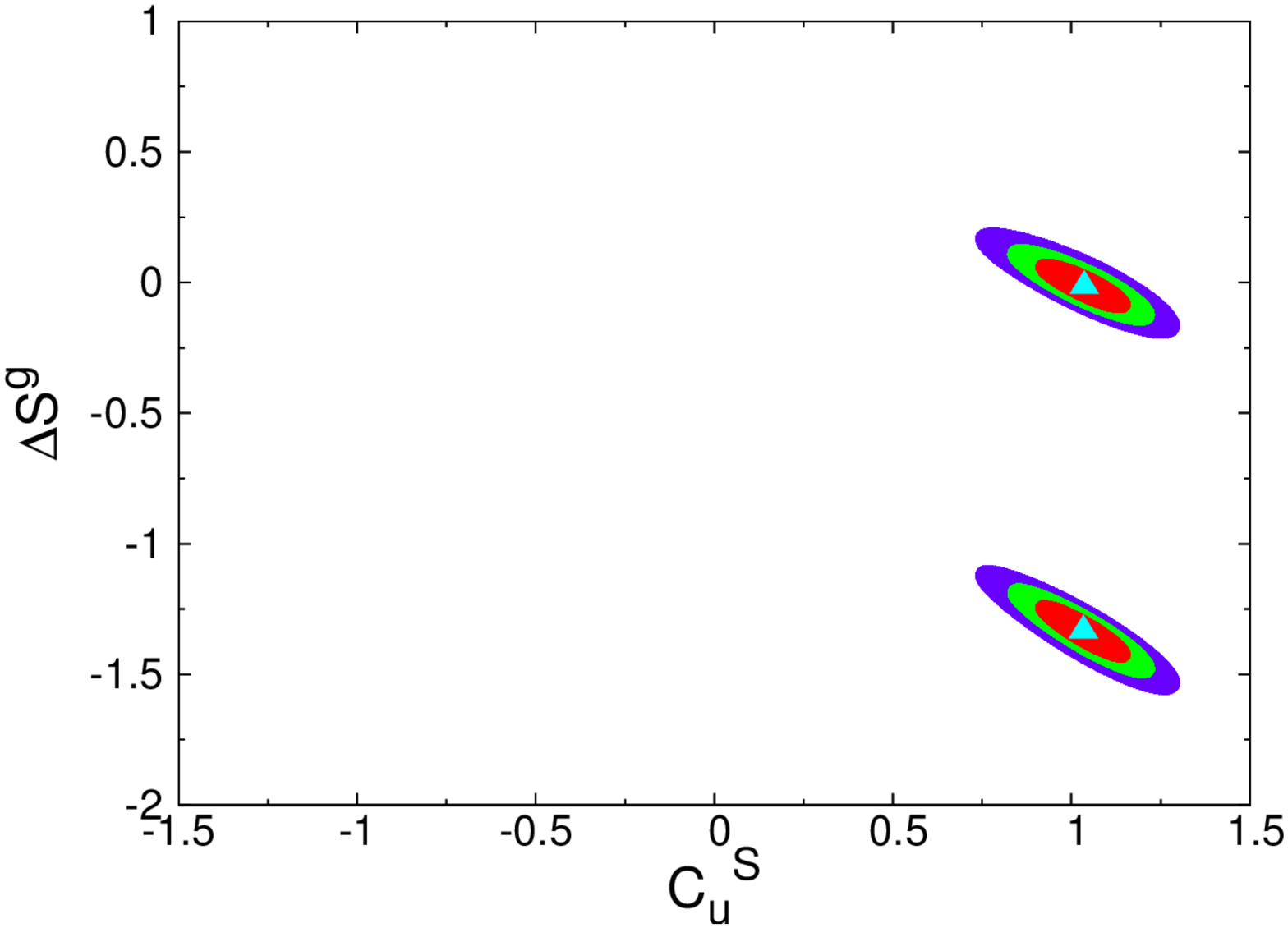}
\caption{\small \label{CPCX3} 
{\bf CPCX3}:
The confidence-level regions of the fit by varying 
$C_v$, $C_u^S$ and $\Delta S^g$.
The color code is the same as in Fig.~\ref{CPC2}.
}
\end{figure}

\begin{figure}[t!]
\centering
\includegraphics[height=1.5in,angle=0]{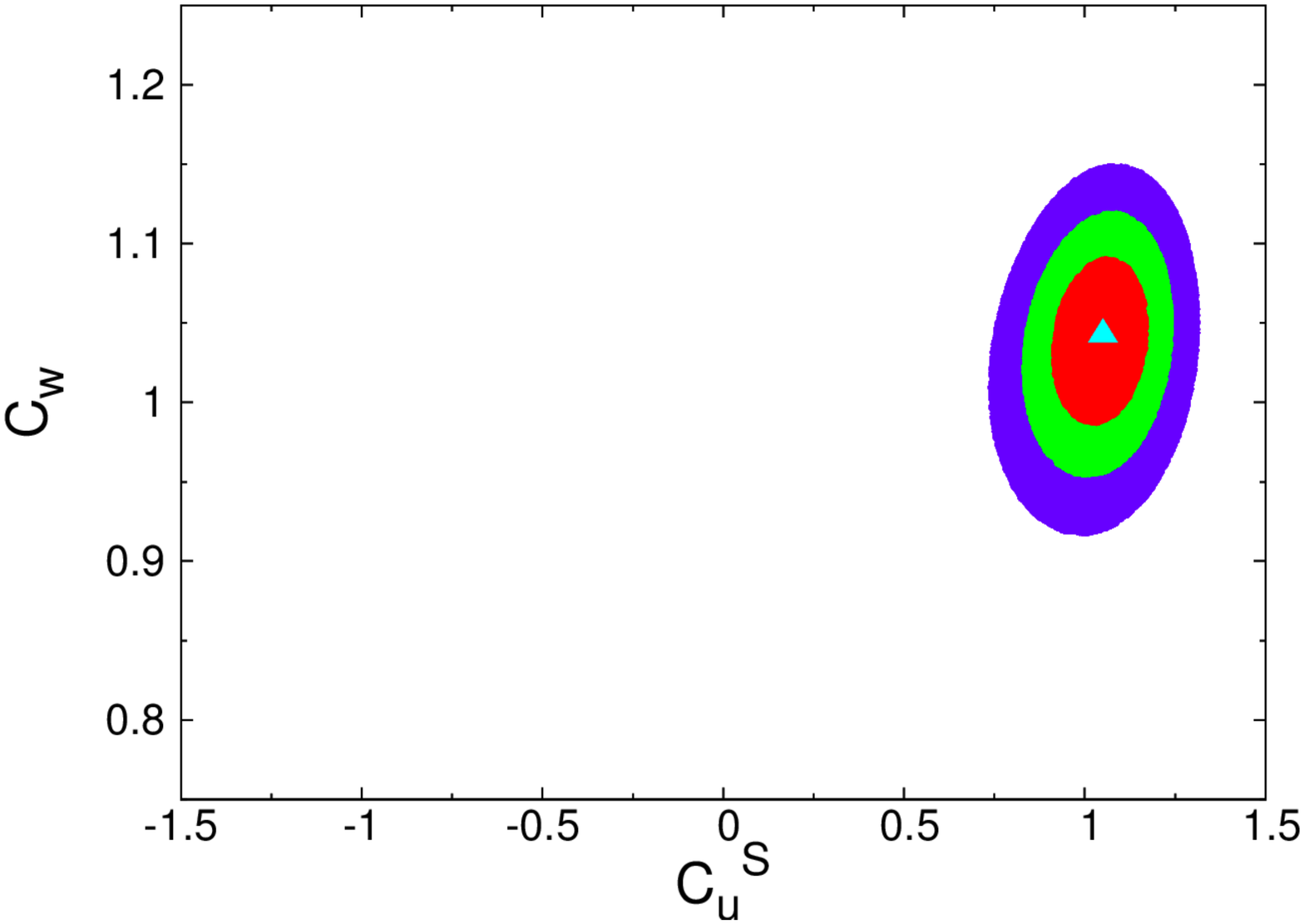}
\includegraphics[height=1.5in,angle=0]{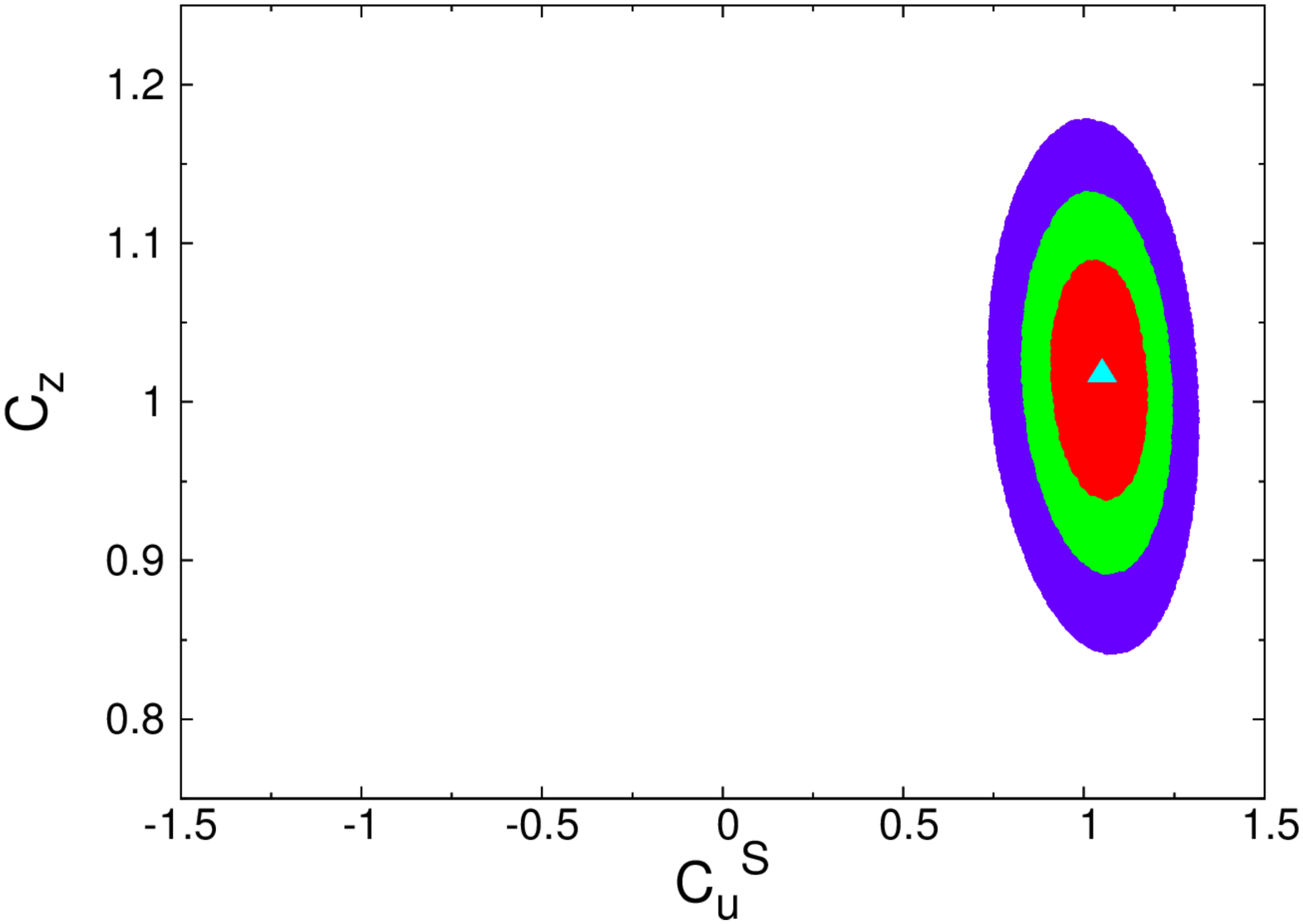}
\includegraphics[height=1.5in,angle=0]{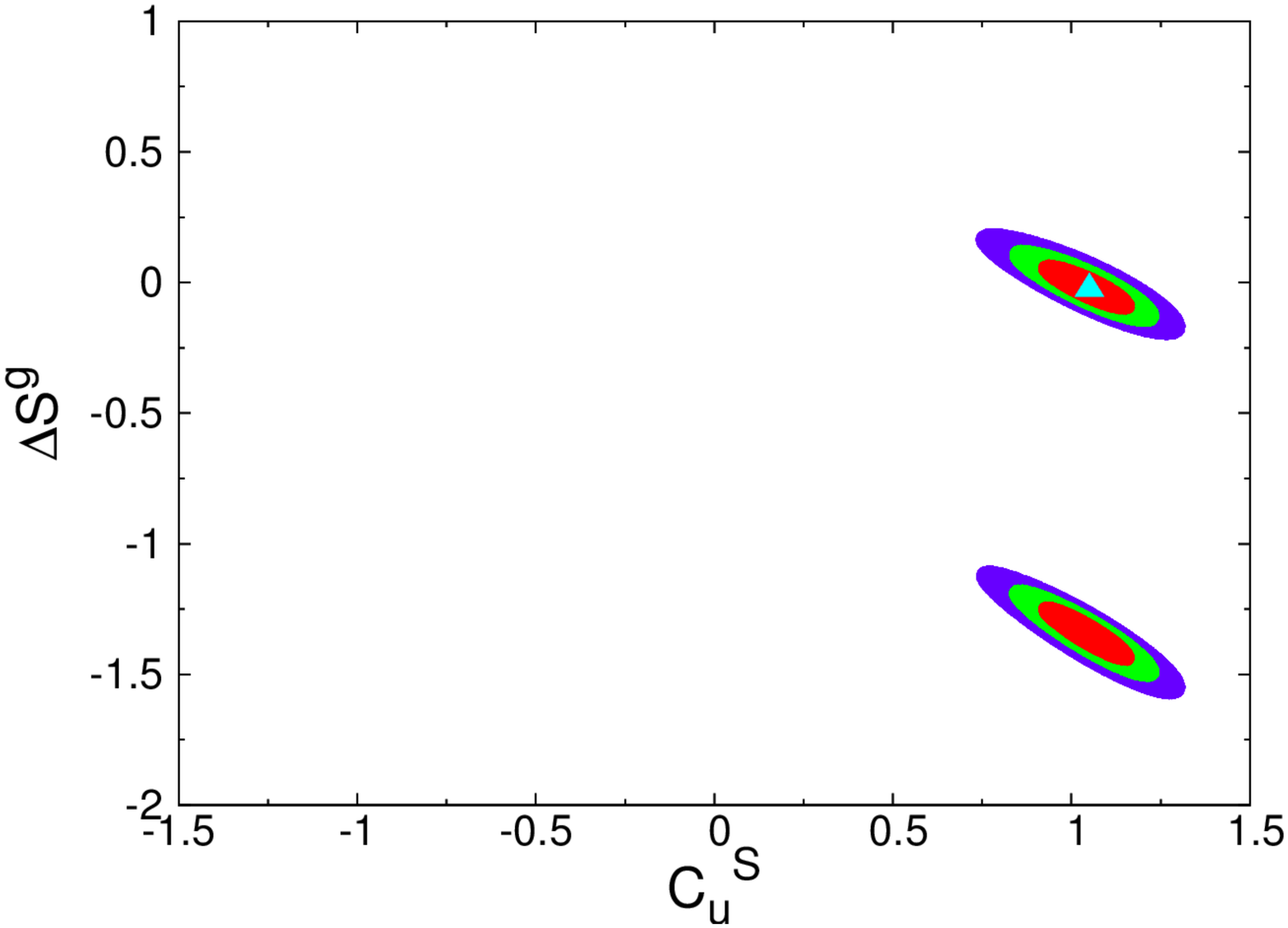}
\includegraphics[height=1.5in,angle=0]{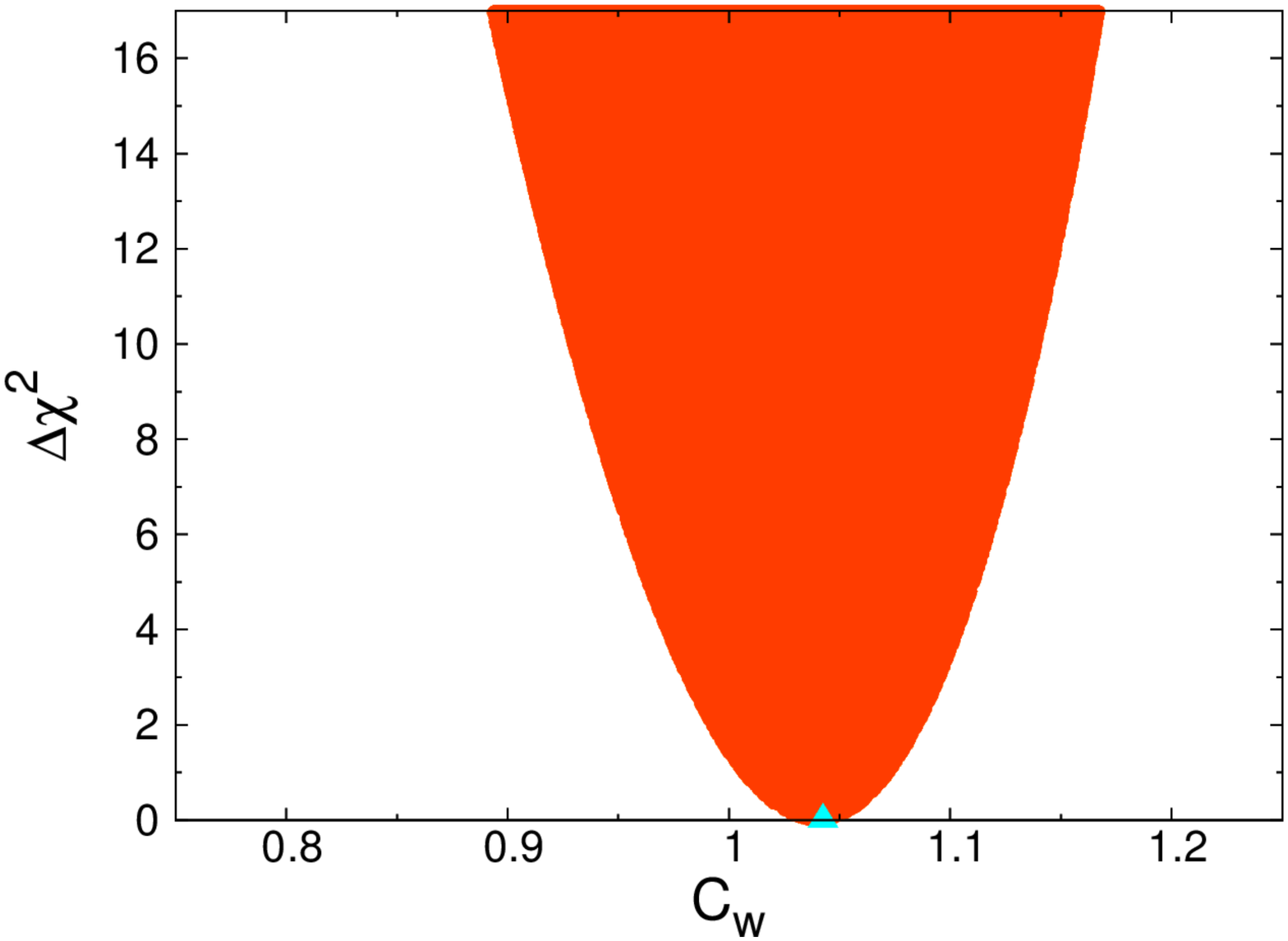}
\includegraphics[height=1.5in,angle=0]{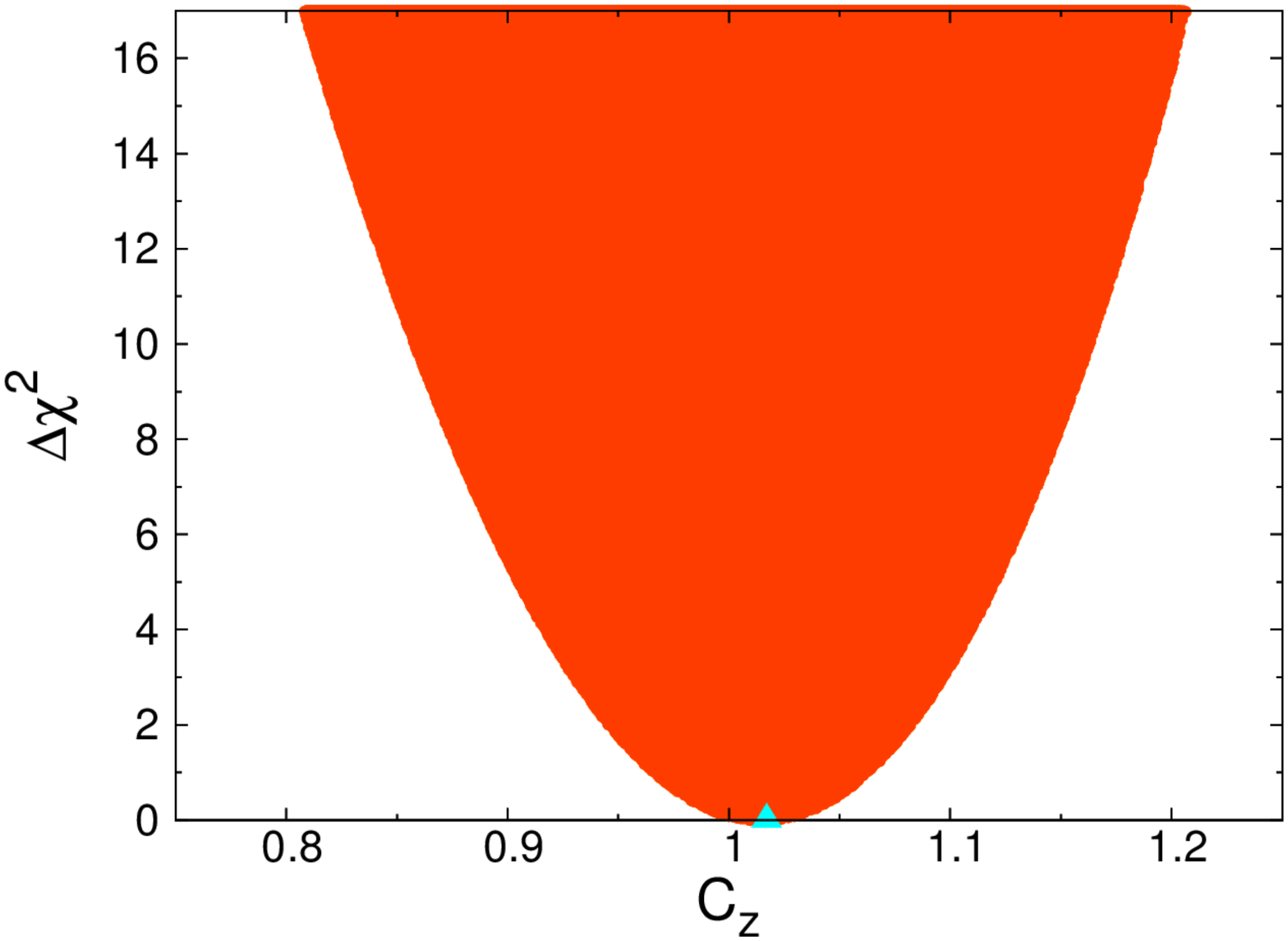}
\includegraphics[height=1.5in,angle=0]{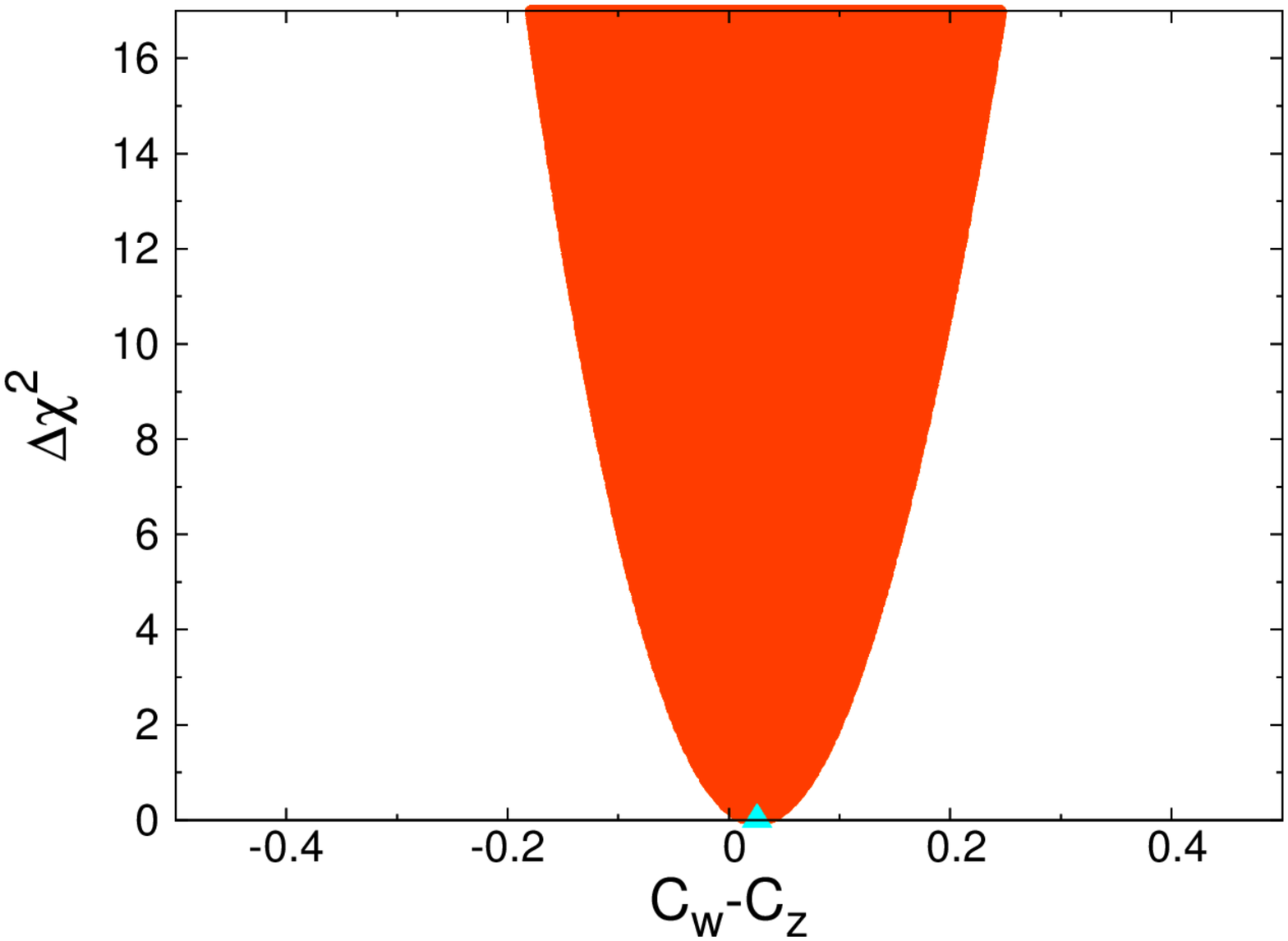}
\caption{\small \label{CPCX4} 
{\bf CPCX4}: (Upper)
The confidence-level regions of the fit by varying 
$C_w$, $C_z$, $\Delta S^g$, and $C_u^S$. 
The color code is the same as in Fig.~\ref{CPC2}.
(Lower) 
$\Delta \chi^2$ versus $C_w$ (left),
$\Delta \chi^2$ versus $C_z$ (middle), and
$\Delta \chi^2$ versus $C_w-C_z$ (right).
}
\end{figure}

We perform some more fits, which were not considered in our previous works.
The best-fit points for {\bf CPCX2} to {\bf CPCX4} are shown in Table~\ref{CPCX}
and the corresponding figures in Fig.~\ref{CPCX2} to Fig.~\ref{CPCX4}.

%
%
The {\bf CPCX2} fit 
involves $C_v$ and $\Delta \Gamma_{\rm tot}$. Both parameters shift
from the corresponding SM values in order to enhance the signal strengths.
The confidence-level regions are shown in Fig.~\ref{CPCX2}. 

%
%
In addition to $C_v$ and $C_u^S$ (similar to {\bf CPCN2}), 
the {\bf CPCX3} fit also varies $\Delta S^g$. 
The result is very similar to {\bf CPCN2},
but $\Delta S^g$ has two solutions with the same $p$-values: see
Fig.~\ref{CPCX3}.

%
%
In the {\bf CPCX4} fit, we relax the requirement of $C_w = C_z$ because we can
see from the 13 TeV data that the signal strengths for
$H \to W W^*$ are generically larger than those for $H \to ZZ^*$, 
see Table~\ref{all13}. 
The result is shown in Table~\ref{CPCX}. The best-fitted values
for $C_w$ and $C_z$ are within $1\sigma$ and $C_w > C_z$ as demanded
by the data. Again there are two solutions for $\Delta S^g$: see
Fig.~\ref{CPCX4}. 
We note that, compared to $C_z$ which
is only constrained by $H \to Z Z^*$ decay, 
$C_w$ is constrained by both VBF and WH production as well as
$H\to\gamma\gamma$ and $H \to W W^*$ decays.
This leads to the narrower $\Delta\chi^2$ curve in
$C_w$ than in $C_z$, as shown in lower
frames of Fig.~\ref{CPCX4}.

\begin{table}[thb!]
\caption{\small \label{CPV}
({\bf CPV})
  The best-fitted values in various CP violating fits and the
  corresponding chi-square per degree of freedom and goodness of fit.
The $p$-value for each fit hypothesis against the SM null
hypothesis is also shown.
}
\begin{ruledtabular}
\begin{tabular}{c|cc|c|c|cc}
Cases & \multicolumn{2}{c}{\bf CPV2} & {\bf CPV3} 
      & {\bf CPV4} & \multicolumn{2}{c}{\bf CPVN3} \\
\hline
      & \multicolumn{2}{c}{Vary $C^S_u,C^P_u$} & {Vary $C^S_u,C^P_u$}  
      & {Vary $\Delta S^\gamma,\Delta S^g$} & \multicolumn{2}{c}{Vary $C^S_u,C^P_u$}  \\
Parameters &  &  &  {$C_v$} 
           & {$\Delta P^\gamma$, $\Delta P^g$} & 
  \multicolumn{2}{c}{$\Delta \Gamma_{\rm tot}$}    \\
\hline
\multicolumn{7}{c}{After ICHEP 2018}\\
\hline
$C^S_u$           & $1.00^{+0.07}_{-0.11}$ & $1.00^{+0.07}_{-0.11}$
                  & $1.02^{+0.04}_{-0.10}$ & 1 
                  & $0.99^{+0.07}_{-0.10}$ & $0.99^{+0.07}_{-0.10}$  \\
$C^S_d$           & 1 & 1 & 1 & 1 & 1 & 1  \\
$C^S_\ell$        & 1 & 1 & 1 & 1 & 1 & 1   \\
$C_v$           & 1 & 1 & $1.03^{+0.03}_{-0.03}$ & 1 & 1 & 1  \\
$\Delta S^\gamma$ & 0 & 0 & 0 & $0.26^{+13.56}_{-0.81}$ & 0 & 0    \\
$\Delta S^g$      & 0 & 0 & 0 & $0.016^{+0.025}_{-}$  & 0 & 0  \\
$\Delta \Gamma_{\rm tot}$ (MeV) & 0  & 0 & 0 & 0 
                  & $-0.27^{+0.34}_{-0.28}$ & $-0.27^{+0.34}_{-0.28}$    \\
\hline
$C^P_u$           & $0.19^{+0.14}_{-0.52}$ & $-0.19^{+0.52}_{-0.14}$ 
                  & $0.00^{+0.28}_{-0.28}$ & 0 
                  & $0.11^{+0.19}_{-0.41}$ & $-0.11^{+0.41}_{-0.19}$  \\
$\Delta P^\gamma$      & 0 & 0 & 0 & $-2.54^{+9.72}_{-4.65}$ & 0 & 0  \\
$\Delta P^g$           & 0 & 0 & 0 & $0.00^{+0.69}_{-0.69}$ & 0 & 0 \\
\hline
$\chi^2/dof$ & \multicolumn{2}{c}{52.07/62} & 51.16/61 
             & 51.87/60 & \multicolumn{2}{c}{51.42/61} \\
goodness of fit    &  \multicolumn{2}{c}{0.812}   &   0.811 
             & 0.763 &  \multicolumn{2}{c}{0.804}  \\
$p$-value    &  \multicolumn{2}{c}{0.419}   &   0.449 
             & 0.747 &  \multicolumn{2}{c}{0.495}  \\
\end{tabular}
\end{ruledtabular}
\end{table}
\subsection{CP Violating fits}
\begin{figure}[t!]
\centering
\includegraphics[height=2.5in,angle=0]{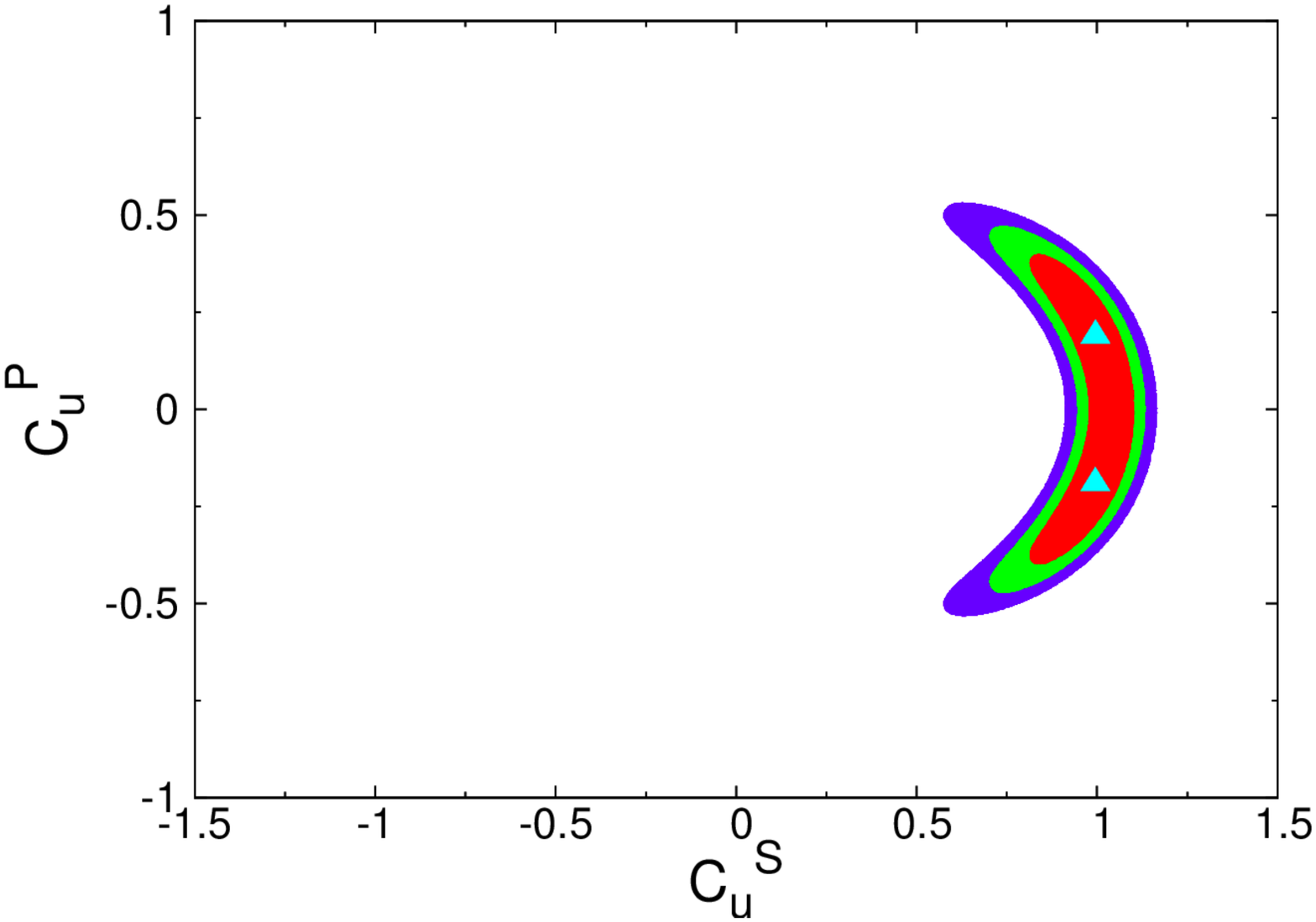}
\caption{\small \label{CPV2} 
{\bf CPV2}:
The confidence-level regions of the fit by varying 
$C_u^S$ and $C_u^P$.
The color code is the same as in Fig.~\ref{CPC2}.
}
\end{figure}

\begin{figure}[t!]
\centering
\includegraphics[height=1.5in,angle=0]{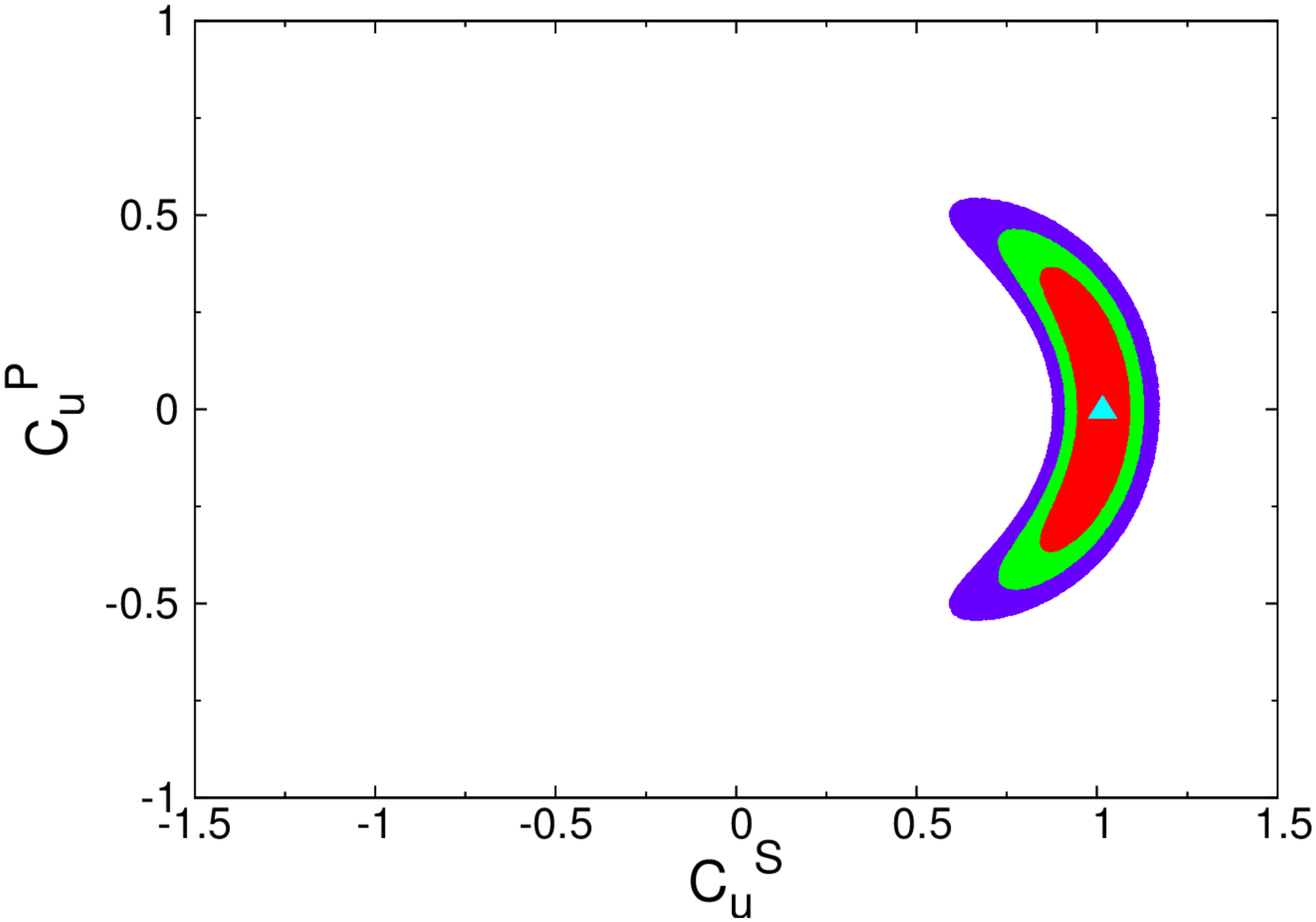}
\includegraphics[height=1.5in,angle=0]{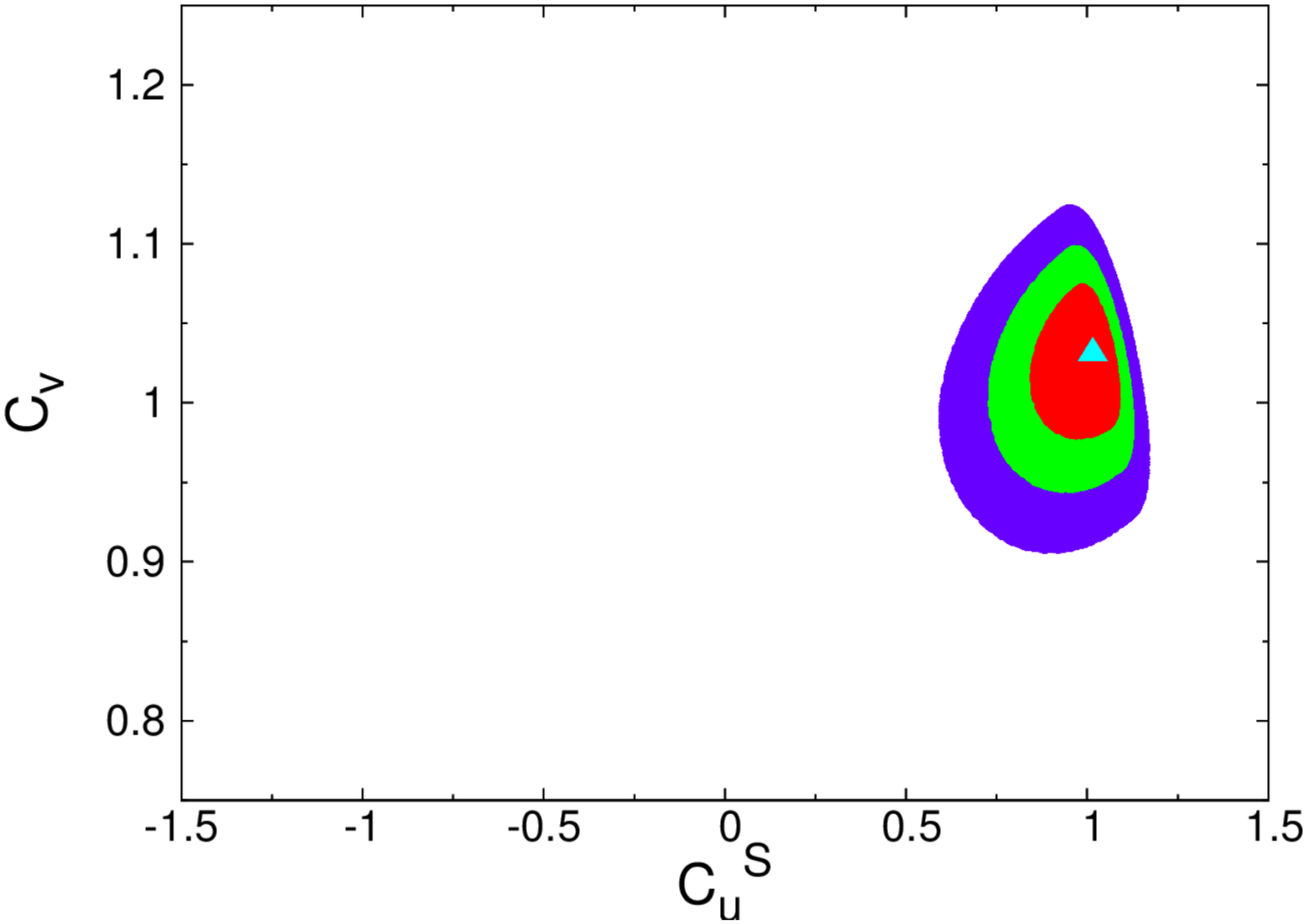}
\includegraphics[height=1.5in,angle=0]{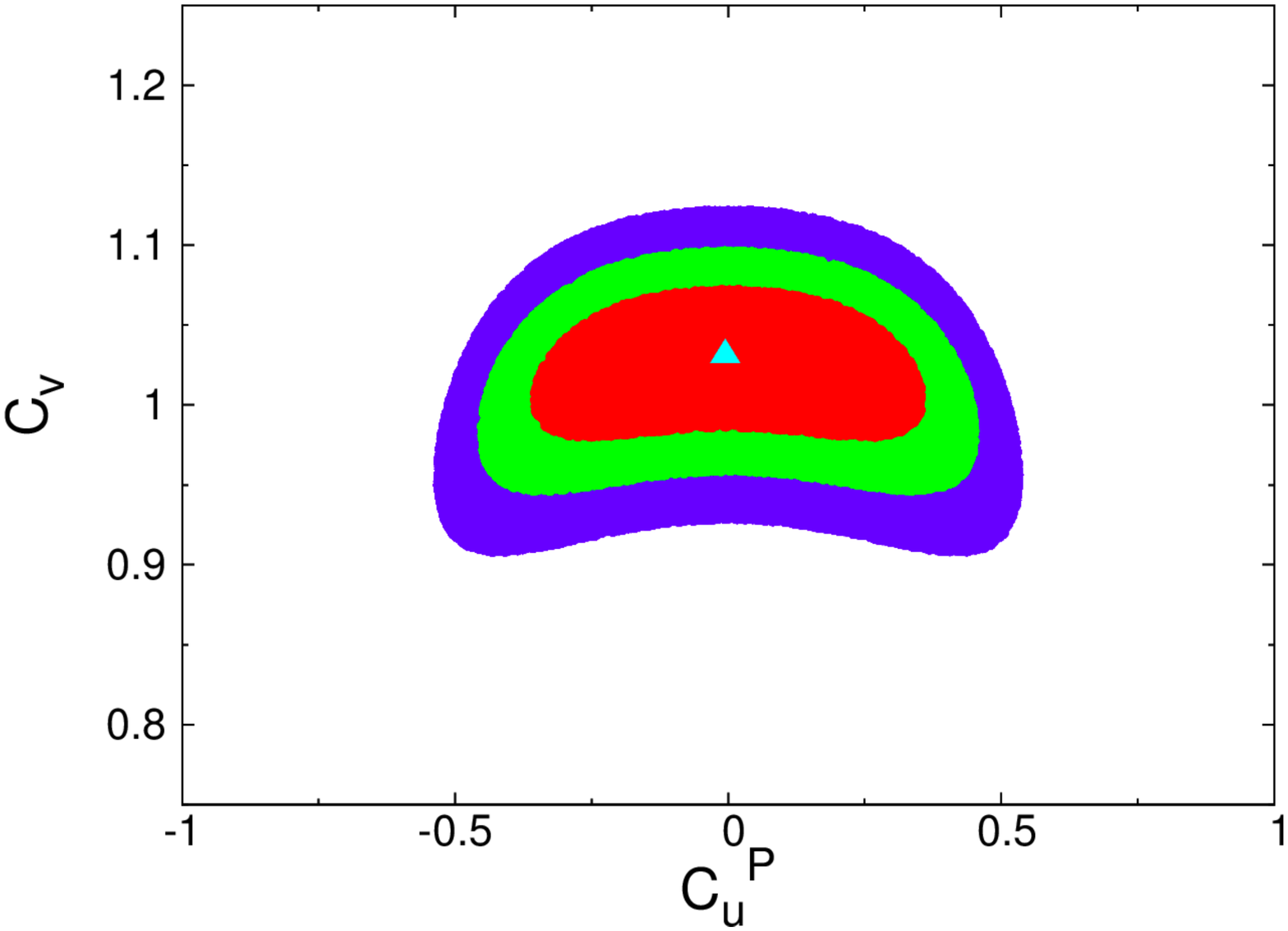}
\caption{\small \label{CPV3} 
{\bf CPV3}:
The confidence-level regions of the fit by varying 
$C_u^S$,  $C_u^P$, and $C_v$.
The color code is the same as in Fig.~\ref{CPC2}.
}
\end{figure}

\begin{figure}[t!]
\centering
\includegraphics[height=1.5in,angle=0]{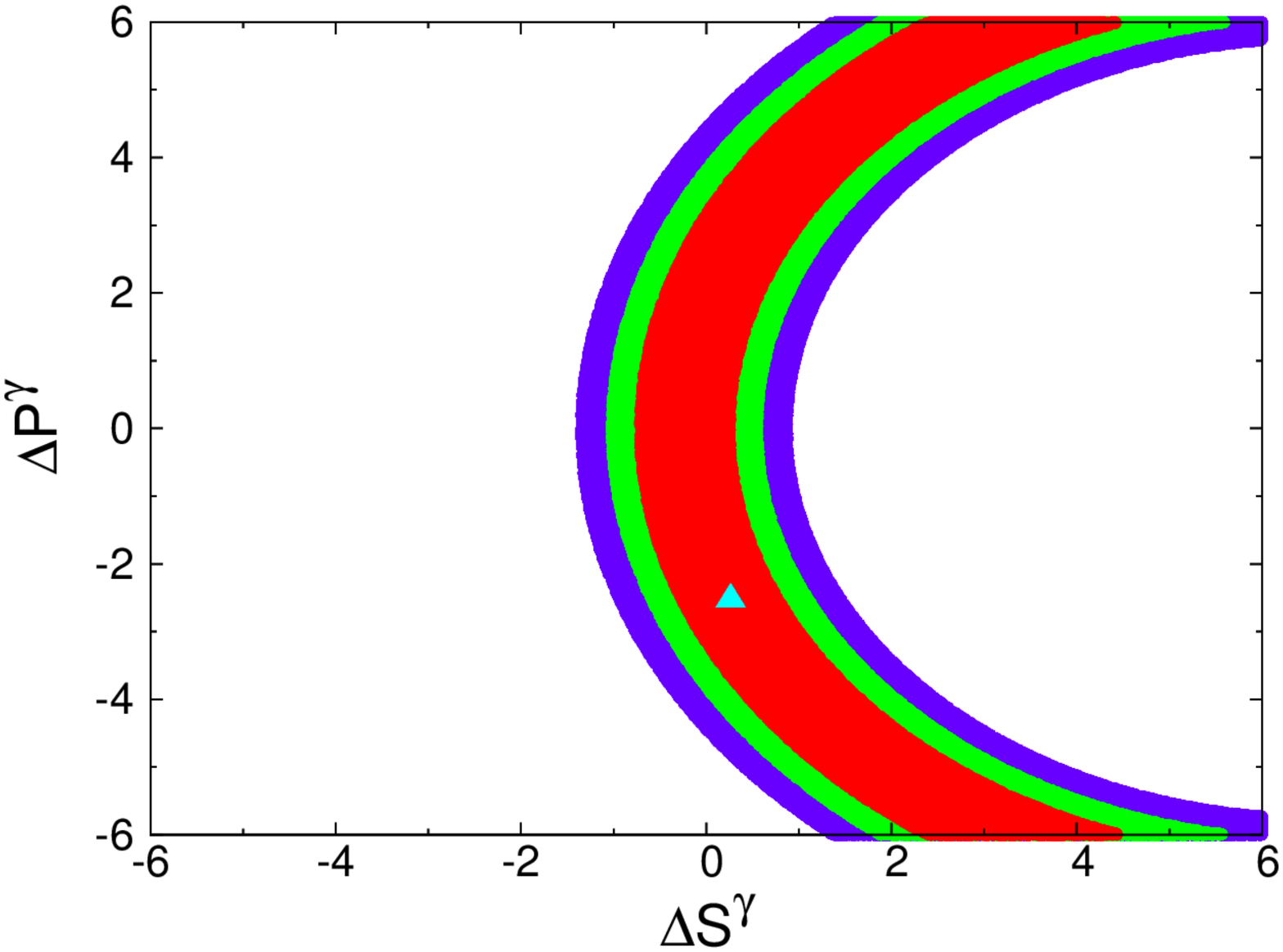}
\includegraphics[height=1.5in,angle=0]{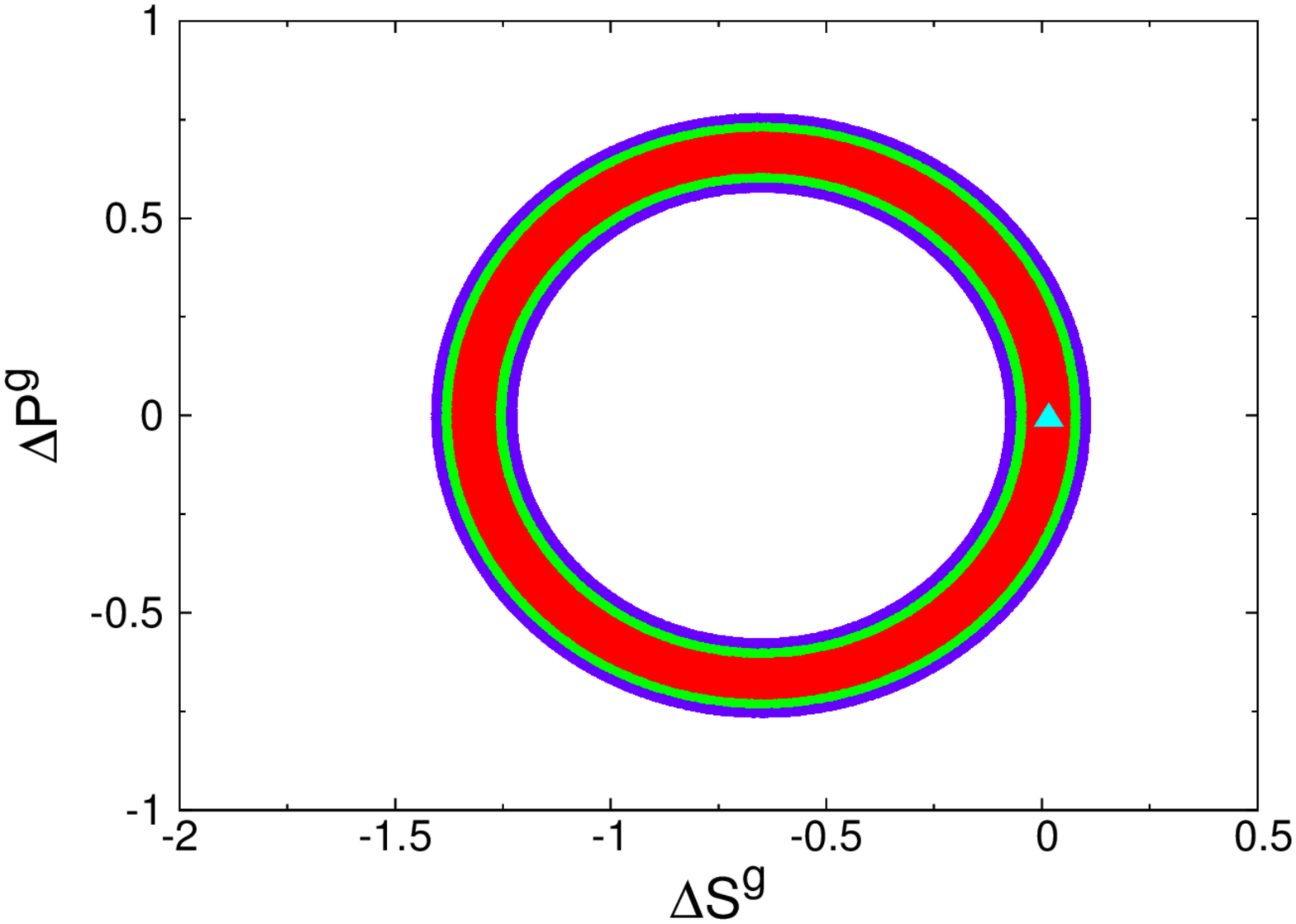}
\includegraphics[height=1.5in,angle=0]{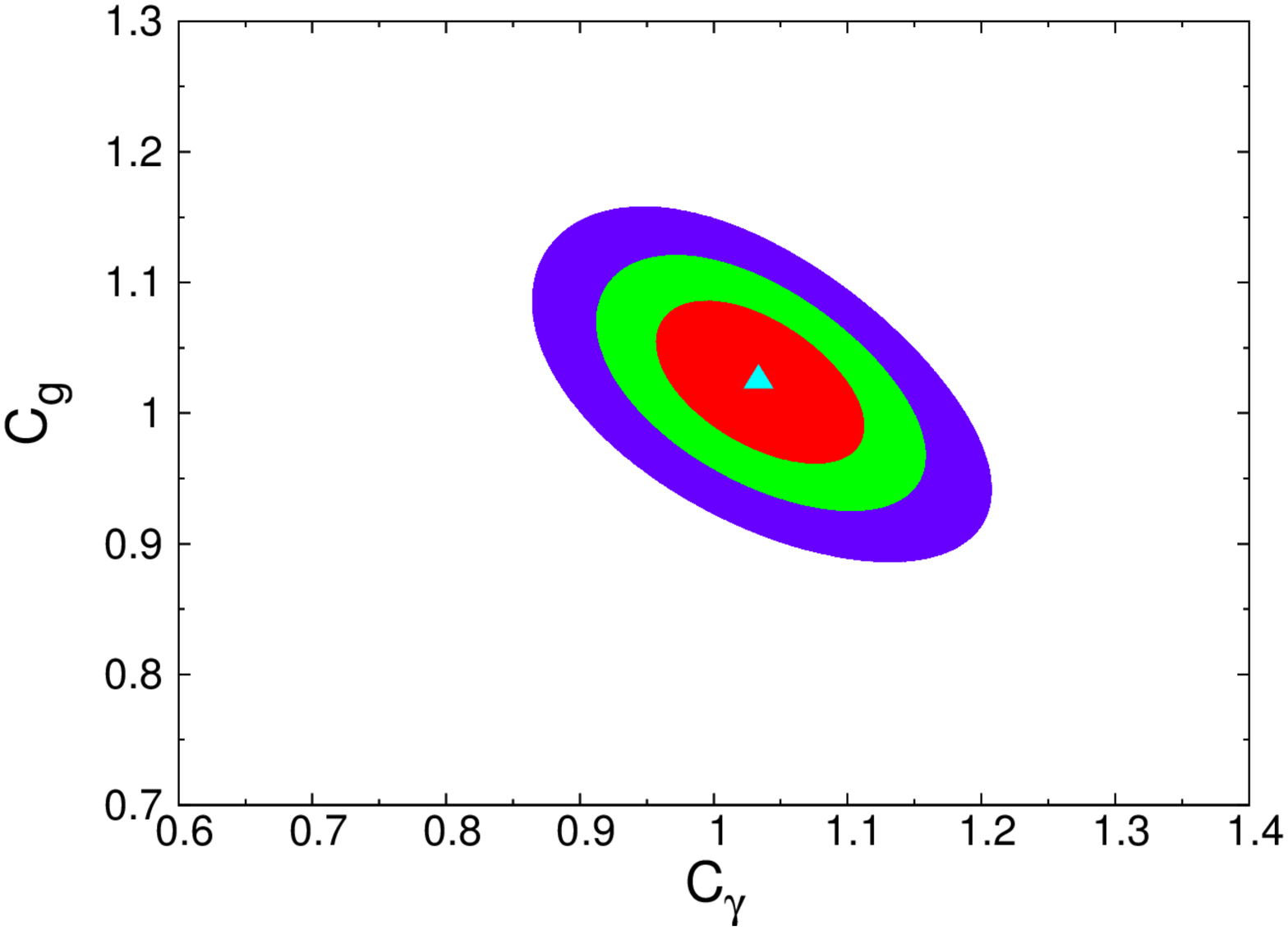}
\caption{\small \label{CPV4} 
{\bf CPV4}:
The confidence-level regions of the fit by varying 
$\Delta S^\gamma$, $\Delta S^g$, $\Delta P^\gamma$, and $\Delta P^g$.
The color code is the same as in Fig.~\ref{CPC2}.
}
\end{figure}

\begin{figure}[t!]
\centering
\includegraphics[height=1.5in,angle=0]{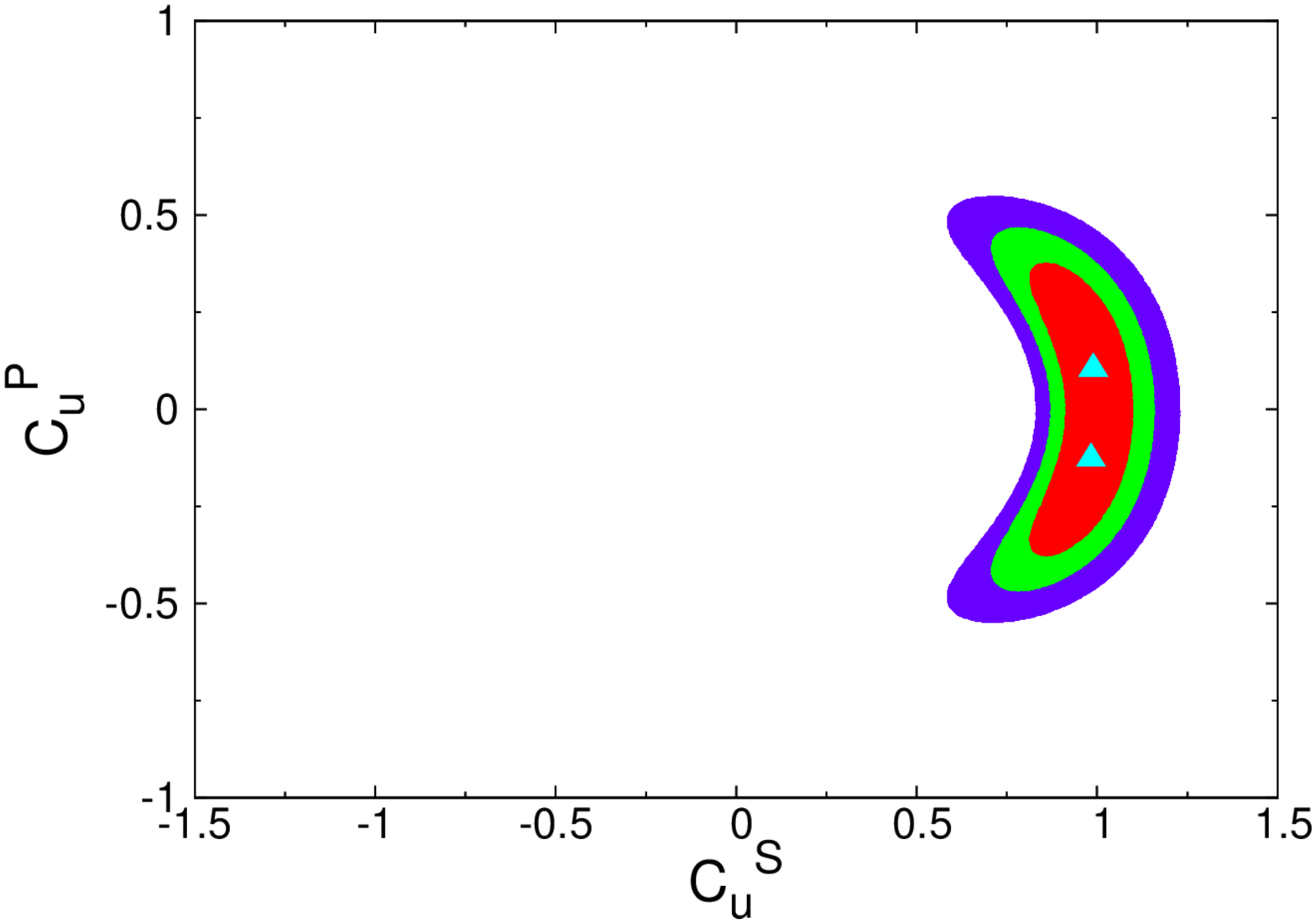}
\includegraphics[height=1.5in,angle=0]{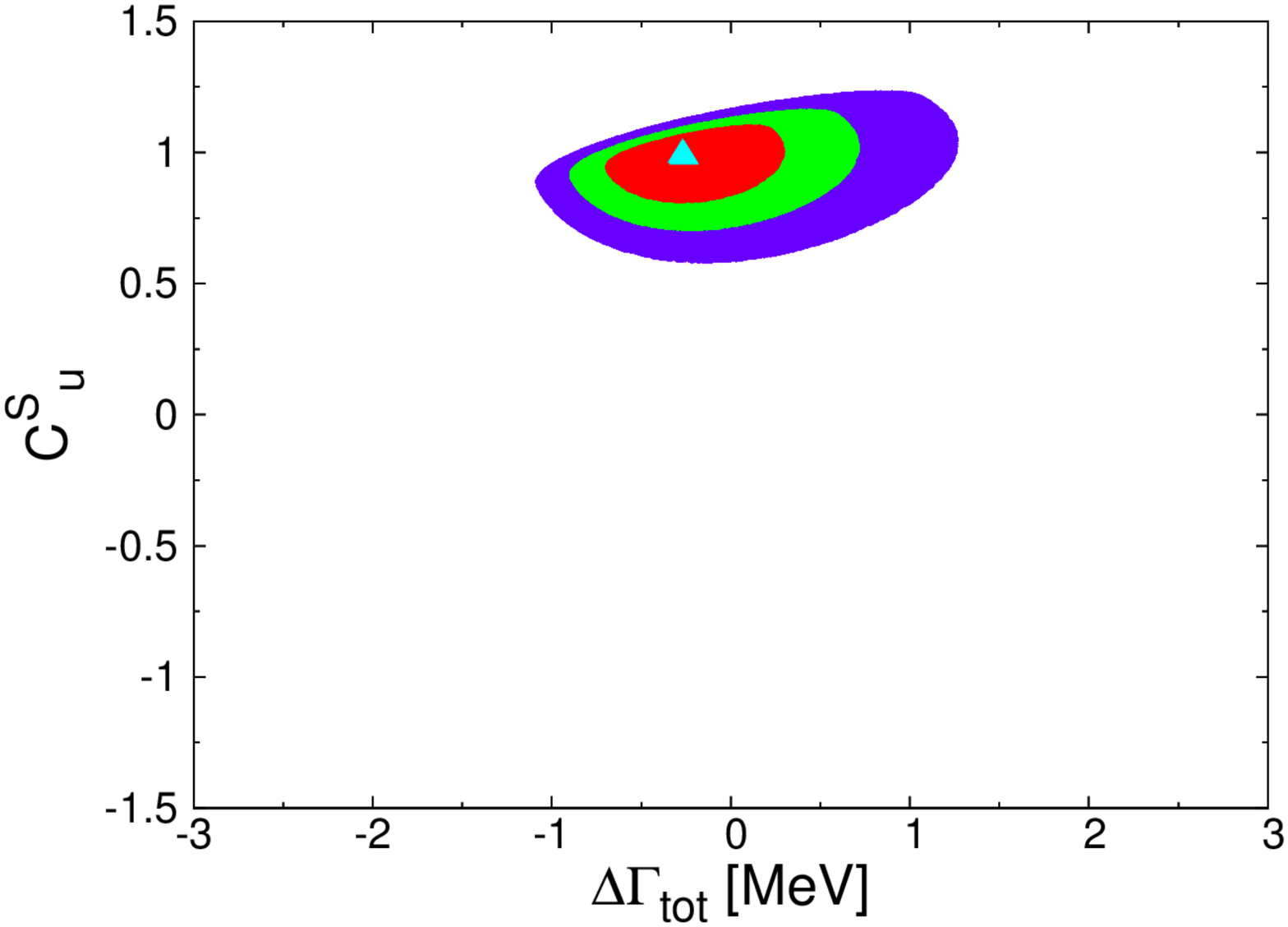}
\includegraphics[height=1.5in,angle=0]{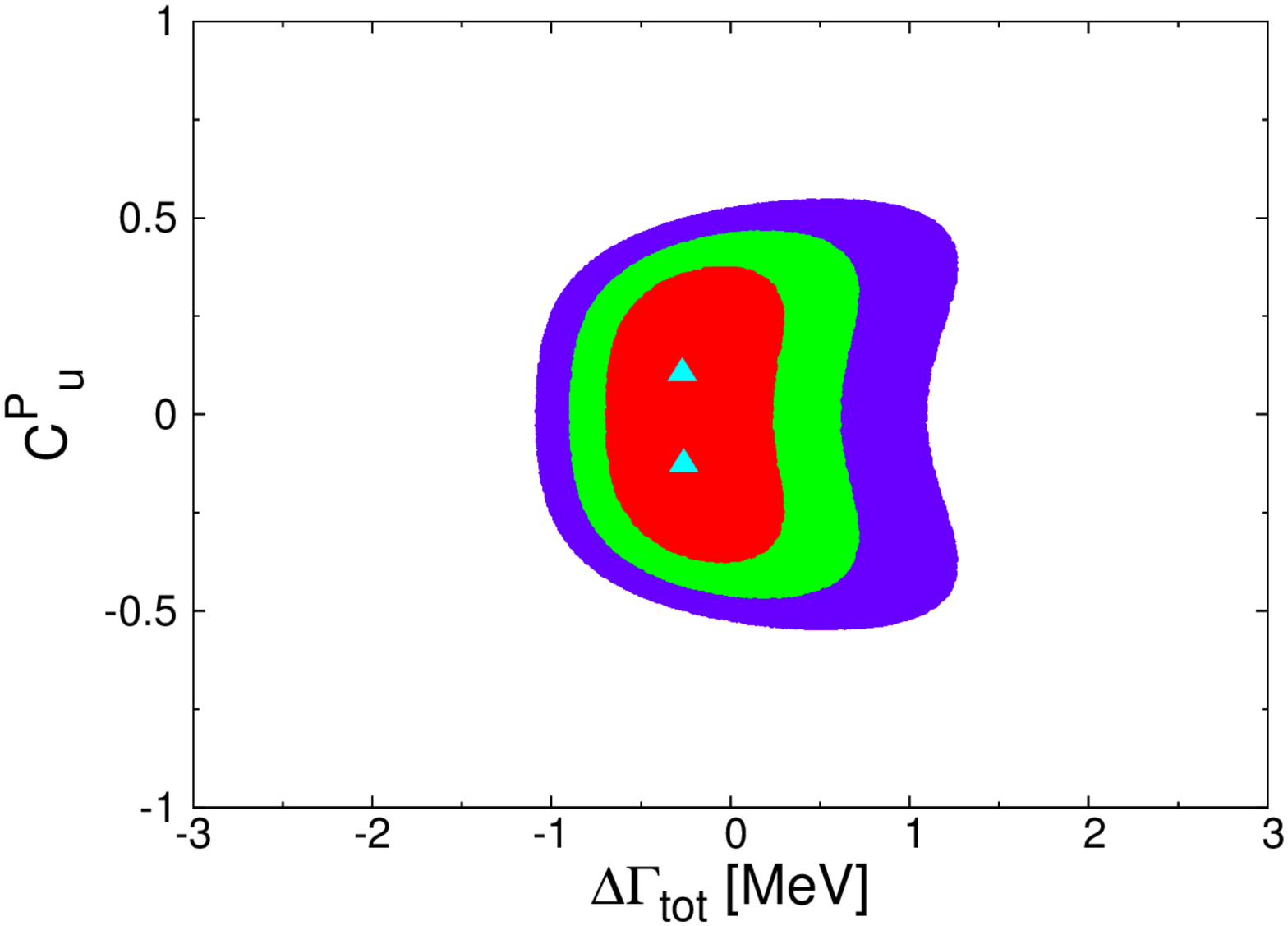}
\caption{\small \label{CPVN3} 
{\bf CPVN3}:
The confidence-level regions of the fit by varying 
$C_u^S$,  $C_u^P$, and $\Delta \Gamma_{\rm tot}$.
The color code is the same as in Fig.~\ref{CPC2}.
}
\end{figure}

For the CP-violating fits, we consider the following 4 scenarios:
\begin{itemize} 
\item {\bf CPV2}: vary $C^S_u,~C^P_u$.
\item {\bf CPV3}: vary $C^S_u,~C^P_u,~C_v$.
\item {\bf CPV4}: vary $\Delta S^\gamma,~\Delta S^g
,~\Delta P^\gamma,~\Delta P^g$.
\item {\bf CPVN3}: vary $C^S_u,~C^P_u,~\Delta\Gamma_{\rm tot}$.
\end{itemize}
The current Higgs boson data ruled out a pure pseudoscalar
\cite{Aad:2013xqa,Khachatryan:2014kca},
but the data cannot rule out a mixed state~\cite{Aad:2016nal}. 
Noting that the CP-odd coupling to gauge bosons only arises from loop 
corrections, we only
allow the top-quark Yukawa coupling and the vertex factors for
$Hgg$ and $H\gamma\gamma$ to develop sizeable CP-odd couplings.
Therefore, CP violation is signaled by
the simultaneous existence of $C_u^S$ and $C_u^P$ 
as in {\bf CPV2}, {\bf CPV3}, and {\bf CPVN3}
or of $\Delta S^{\gamma\,,g}$ and $\Delta P^{\gamma\,,g}$ in {\bf CPV4}.
The results for {\bf CPV2} to {\bf CPVN3} are shown in Table~\ref{CPV} and the
corresponding figures in Fig.~\ref{CPV2} to Fig.~\ref{CPVN3}.

%
The simplest choice {\bf CPV2} happens in the
coexistence of CP-even and CP-odd top-Yukawa couplings: 
$C_u^S$ and $C_u^P$. 
Since the signal strengths are CP-even quantities, in general,
they do not contain any CP-odd products of 
$C_u^S\times C_u^P$ and $S^{g,\gamma}\times P^{g,\gamma}$ even though
the products are non-vanishing.
This is why the confidence-level regions appear like a circle
or an arc of a circle
 in the planes of $(C_u^S\,,C_u^P)$,
$(\Delta S^\gamma\,, \Delta P^\gamma)$, and
$(\Delta S^g\,, \Delta P^g)$.
In Fig.~\ref{CPV2}, we show two best-fit points with equal
$p$-value for {\bf CPV2}, 
indeed the arc joining these two points essentially has 
the same $p$-value.

%
We vary $C_u^S$, $C_u^P$, and $C_v$ in {\bf CPV3} fit. 
The confidence-level regions, shown in Fig.~\ref{CPV3},
shrink a lot from previous results \cite{update2014}. Previously, the blue 
region forms a closed ellipse, but now all regions hardly form a closed
ellipse, showing the data are getting much more stringent than before.

%
We vary $\Delta S^g$, $\Delta S^\gamma$,  $\Delta P^g$, and $\Delta P^\gamma$
in {\bf CPV4}.  As explained in our previous work \cite{higgcision}, the
solutions to $\Delta S^g$ and $\Delta P^g$, as well as to
$\Delta S^\gamma$ and $\Delta P^\gamma$ appear to be an ellipse. It is 
quite clear in Fig.~\ref{CPV4}. The best-fit points are in fact an arc 
inside the red region that passed through the triangle. 
Note that
we are not considering the scenarios with too large values of
$|\Delta S^\gamma|$ in our fits.
Otherwise, the left frame of Fig.~\ref{CPV4}
may complete an ellipse as in the middle frame.

%
In the {\bf CPVN3} fit, 
we try a different combination of parameters: 
$C_u^S$, $C_u^P$, and $\Delta \Gamma_{\rm tot}$.
With the help of $\Delta \Gamma_{\rm tot}$ the ``banana'' shaped
regions originally in CPV2 now become fattened.

\begin{figure}[t!]
\centering
\includegraphics[height=1.5in,angle=0]{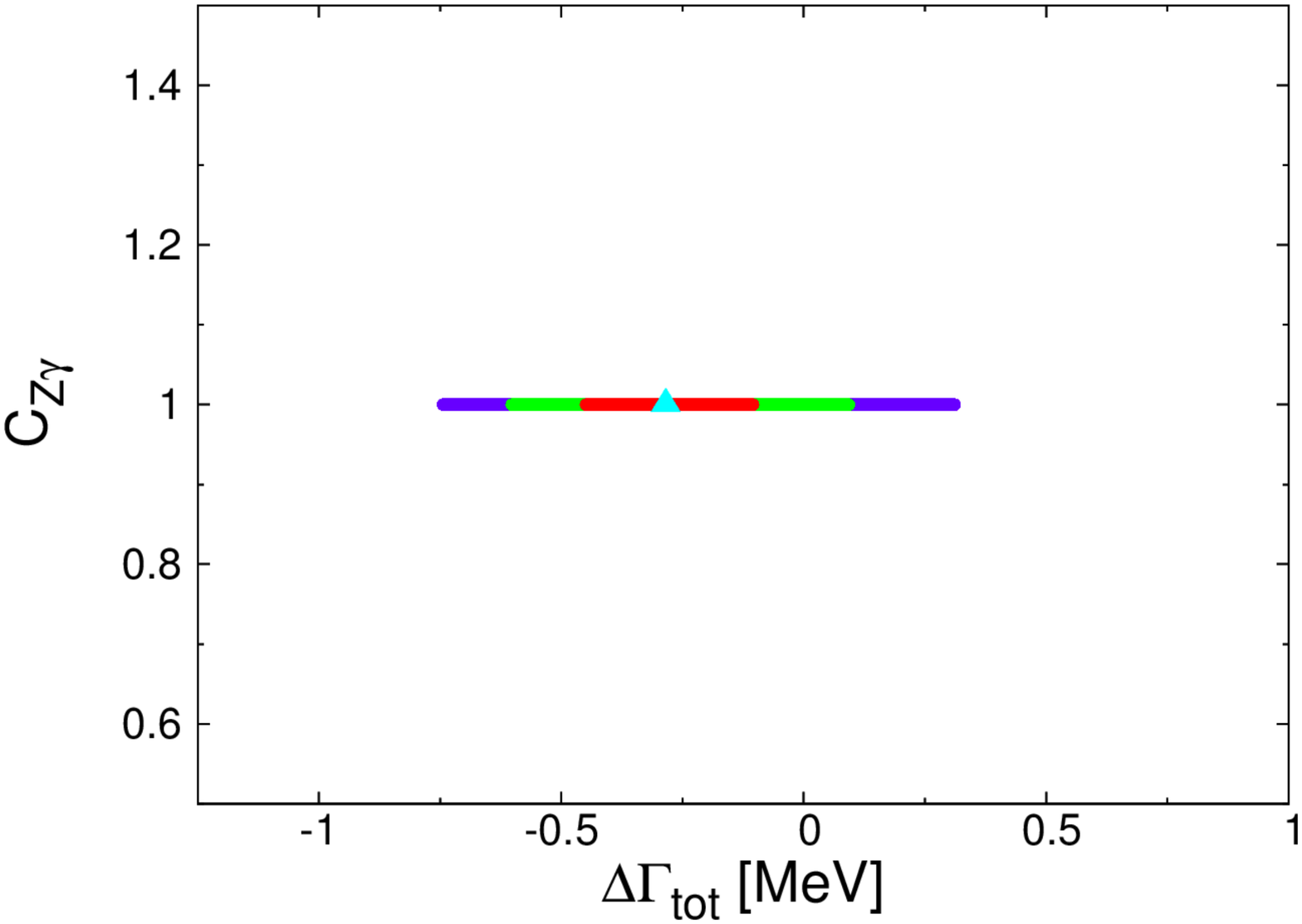}
\includegraphics[height=1.5in,angle=0]{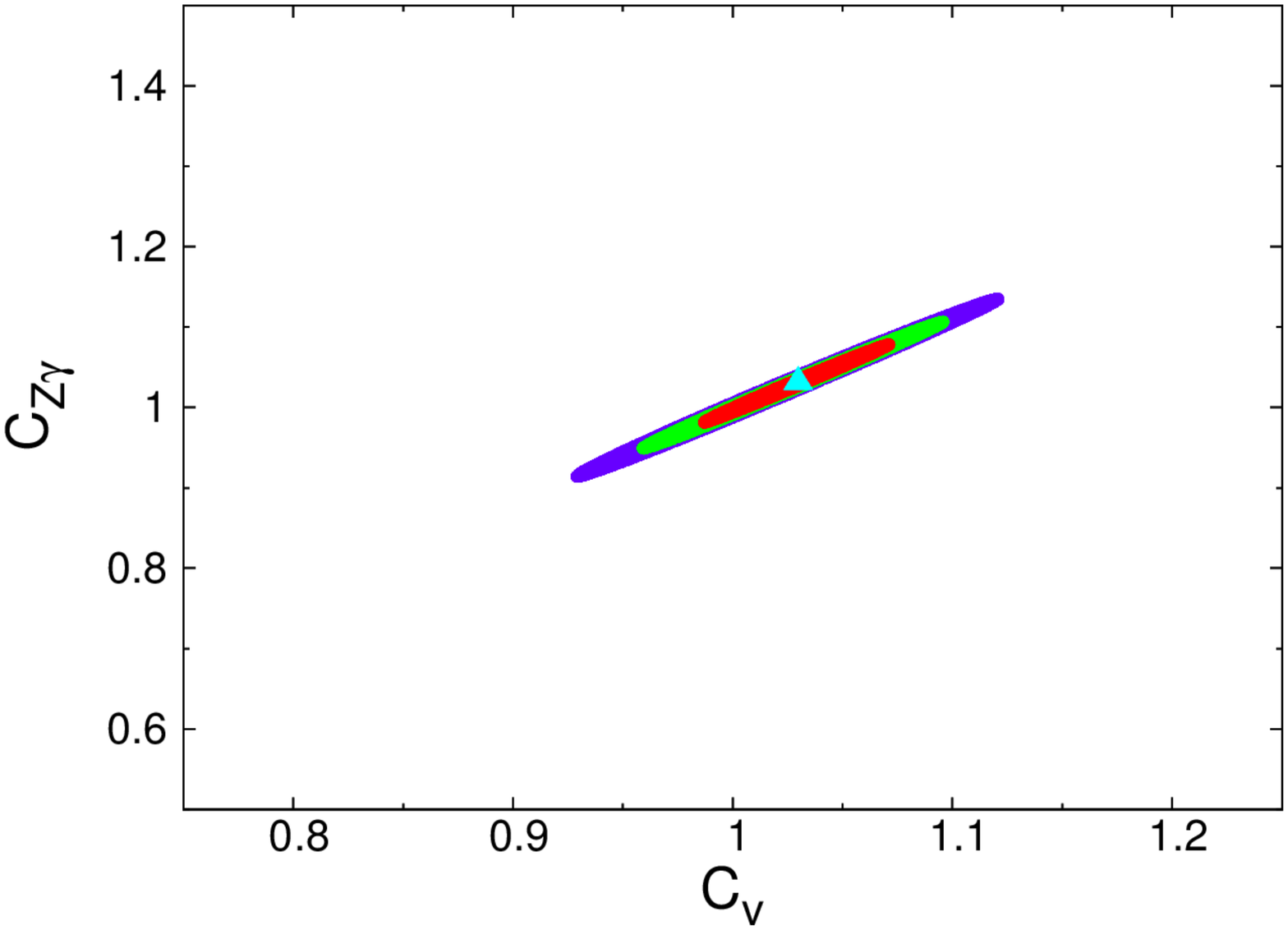}
\includegraphics[height=1.5in,angle=0]{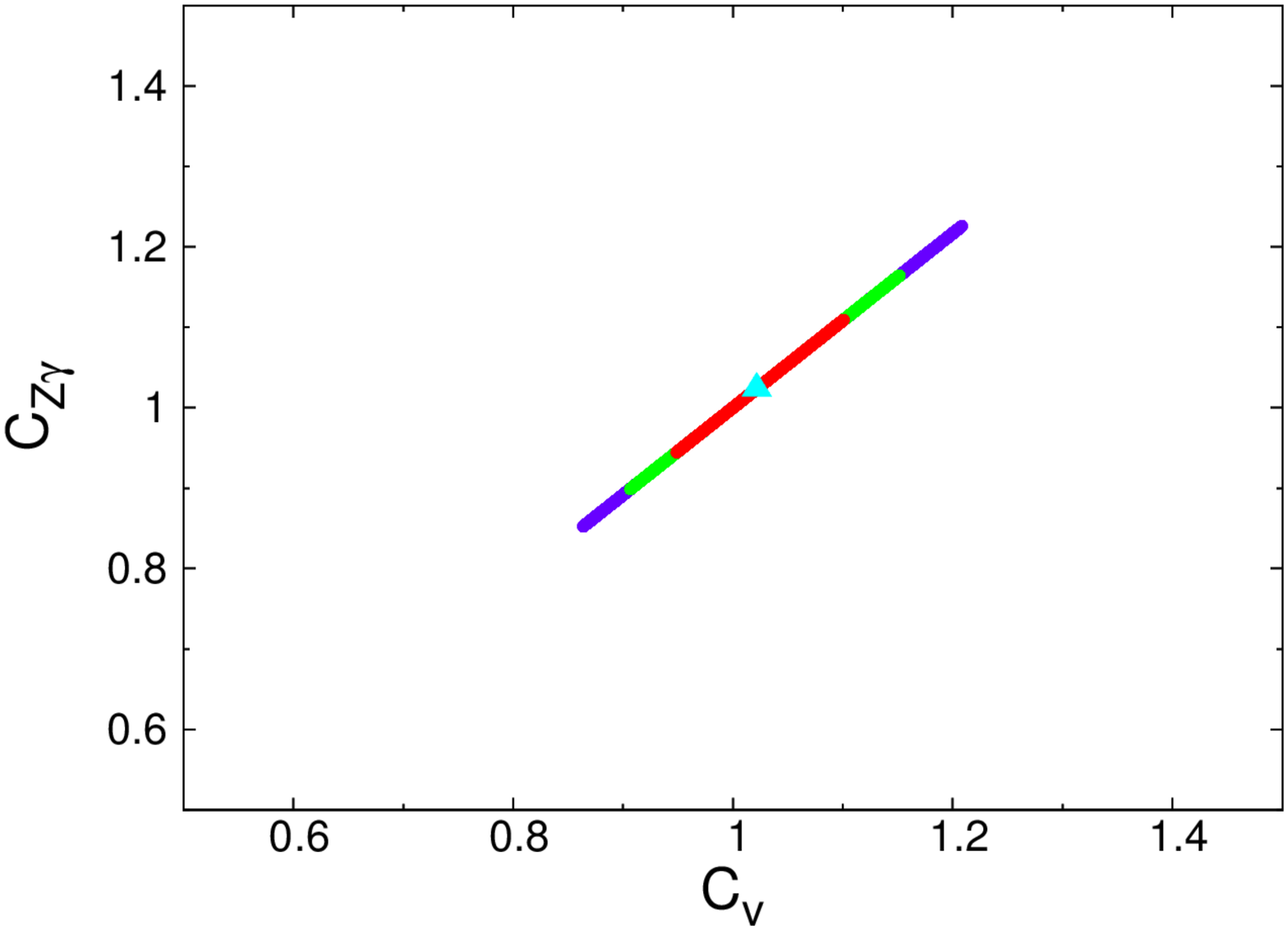}
\includegraphics[height=1.5in,angle=0]{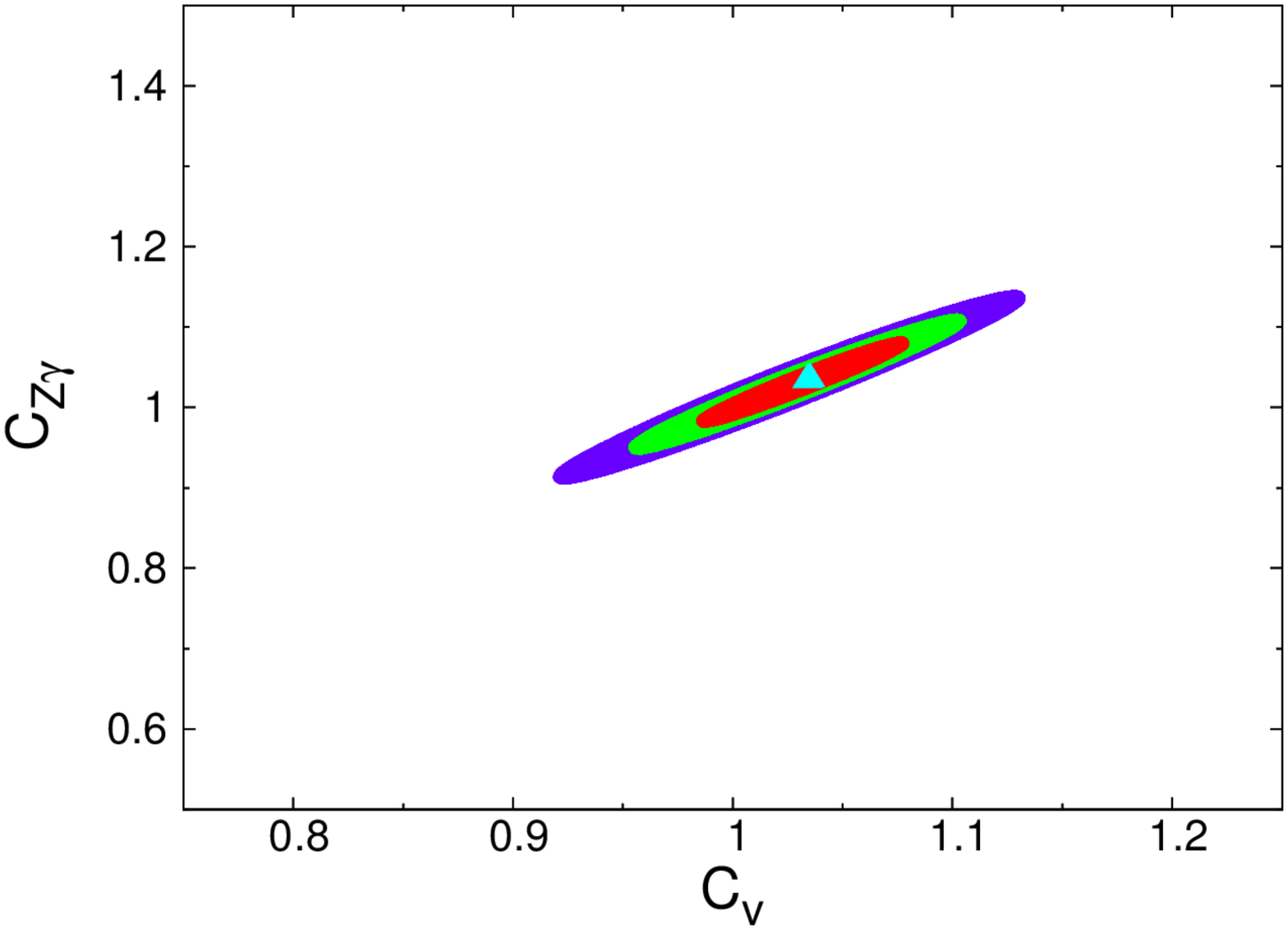}
\includegraphics[height=1.5in,angle=0]{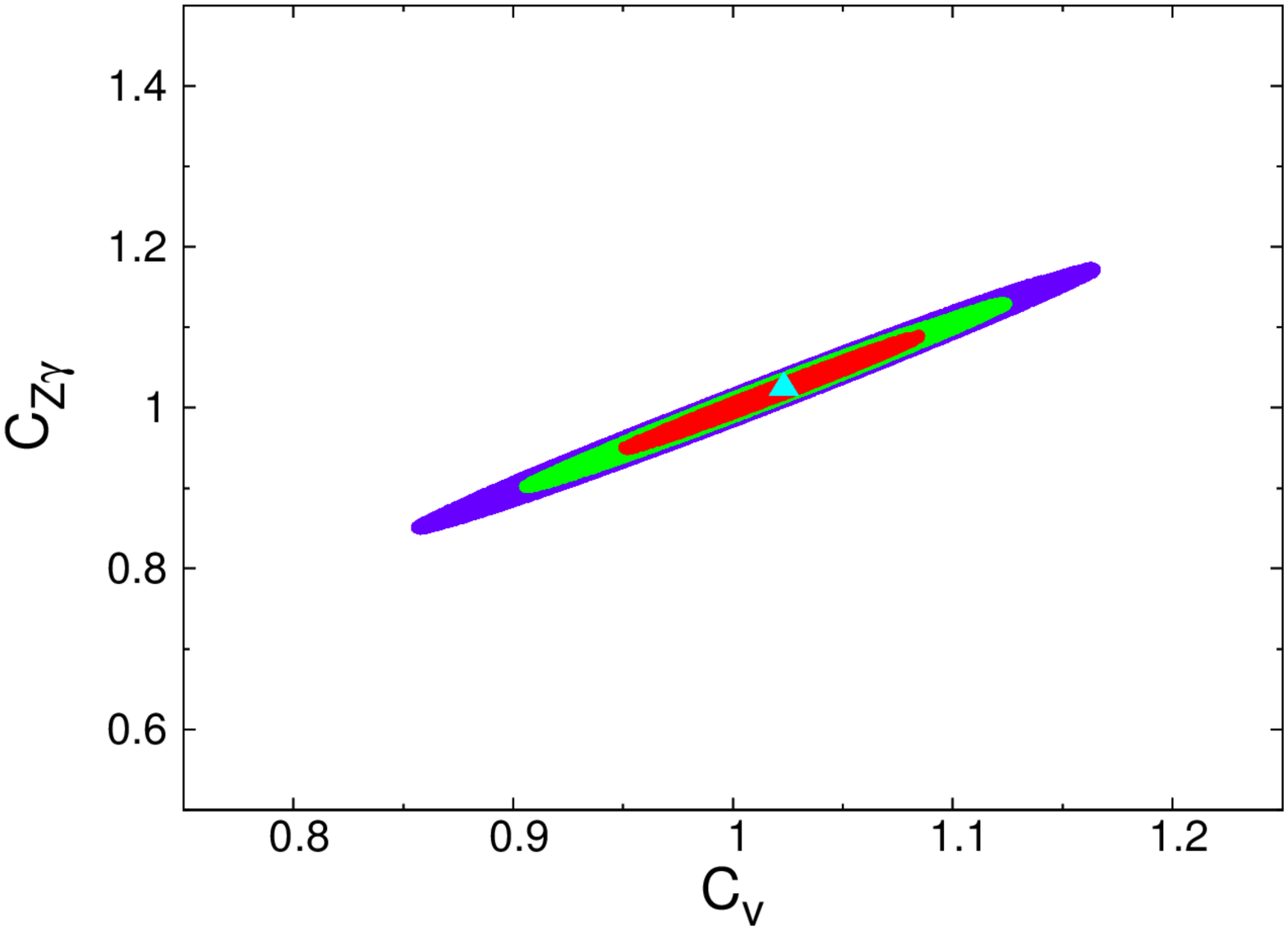}
\includegraphics[height=1.5in,angle=0]{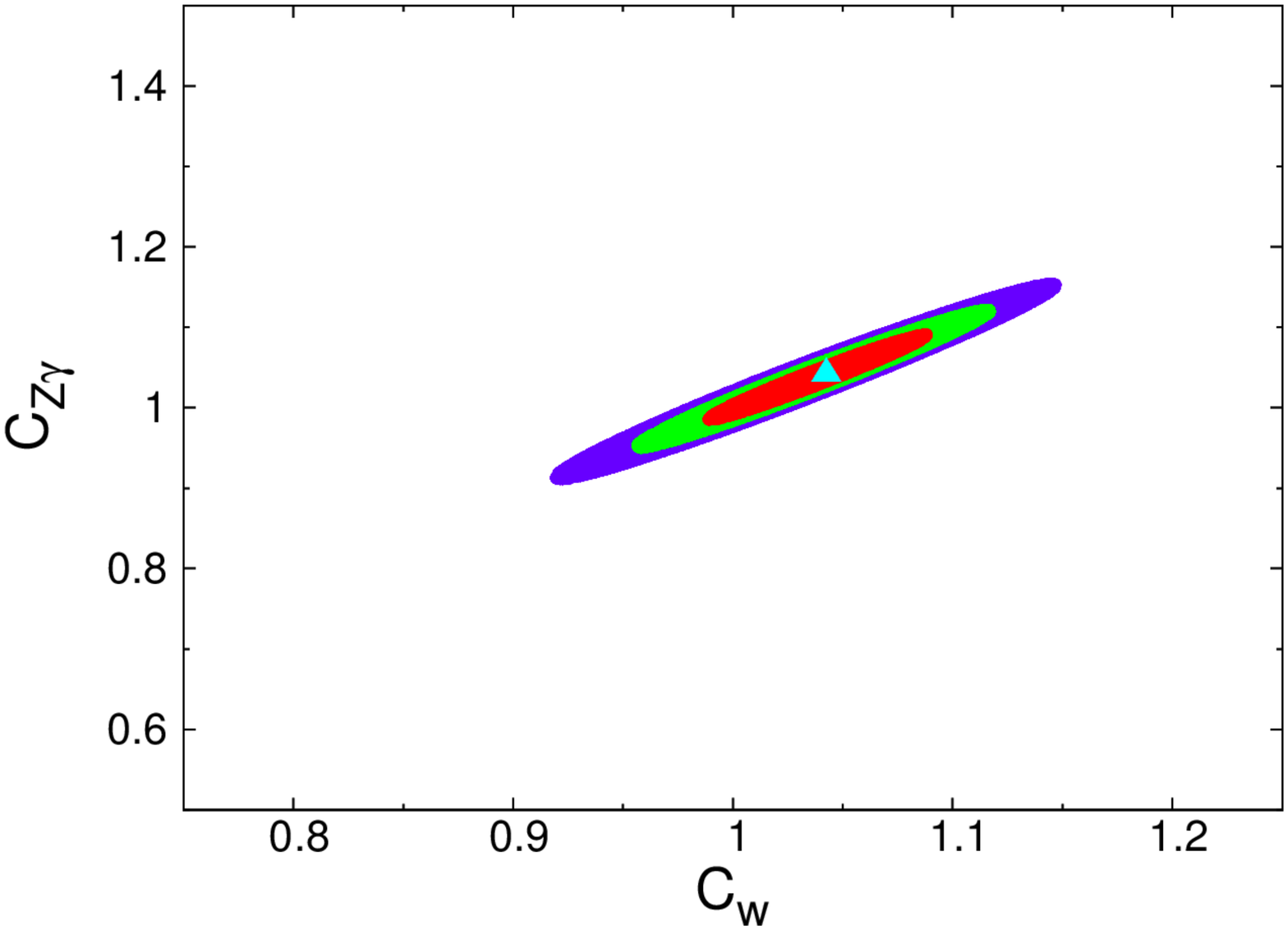}
\includegraphics[height=1.5in,angle=0]{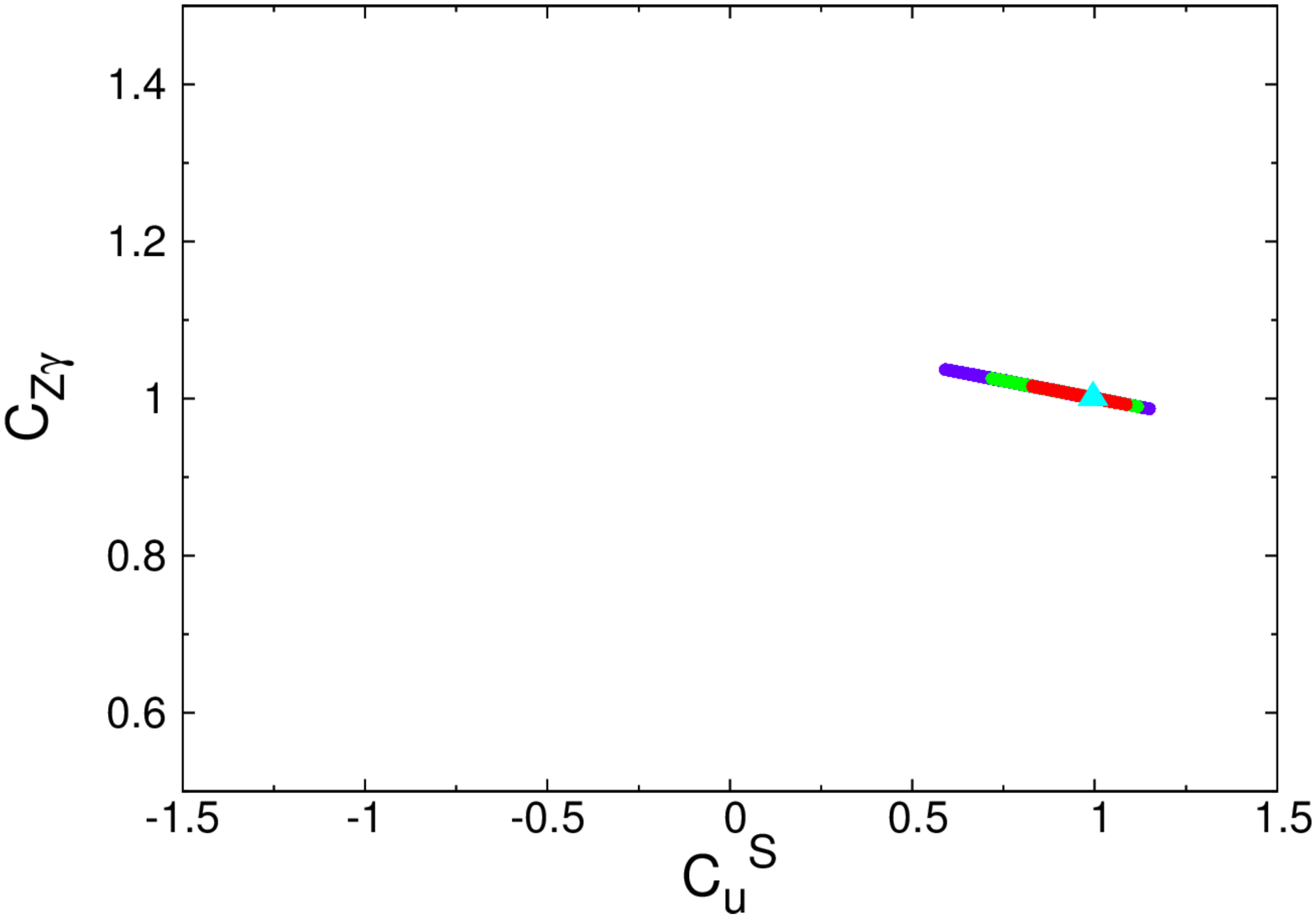}
\includegraphics[height=1.5in,angle=0]{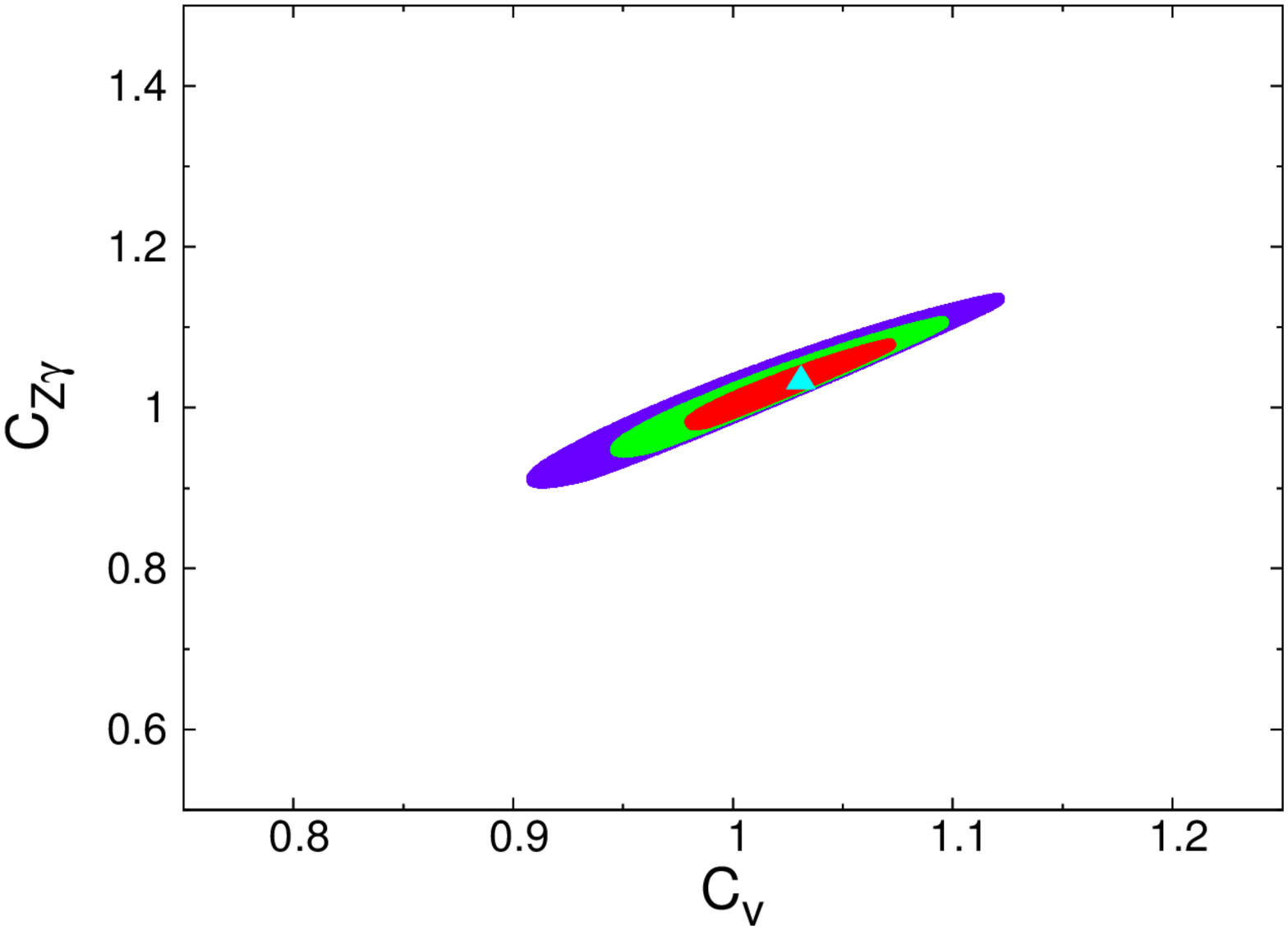}
\includegraphics[height=1.5in,angle=0]{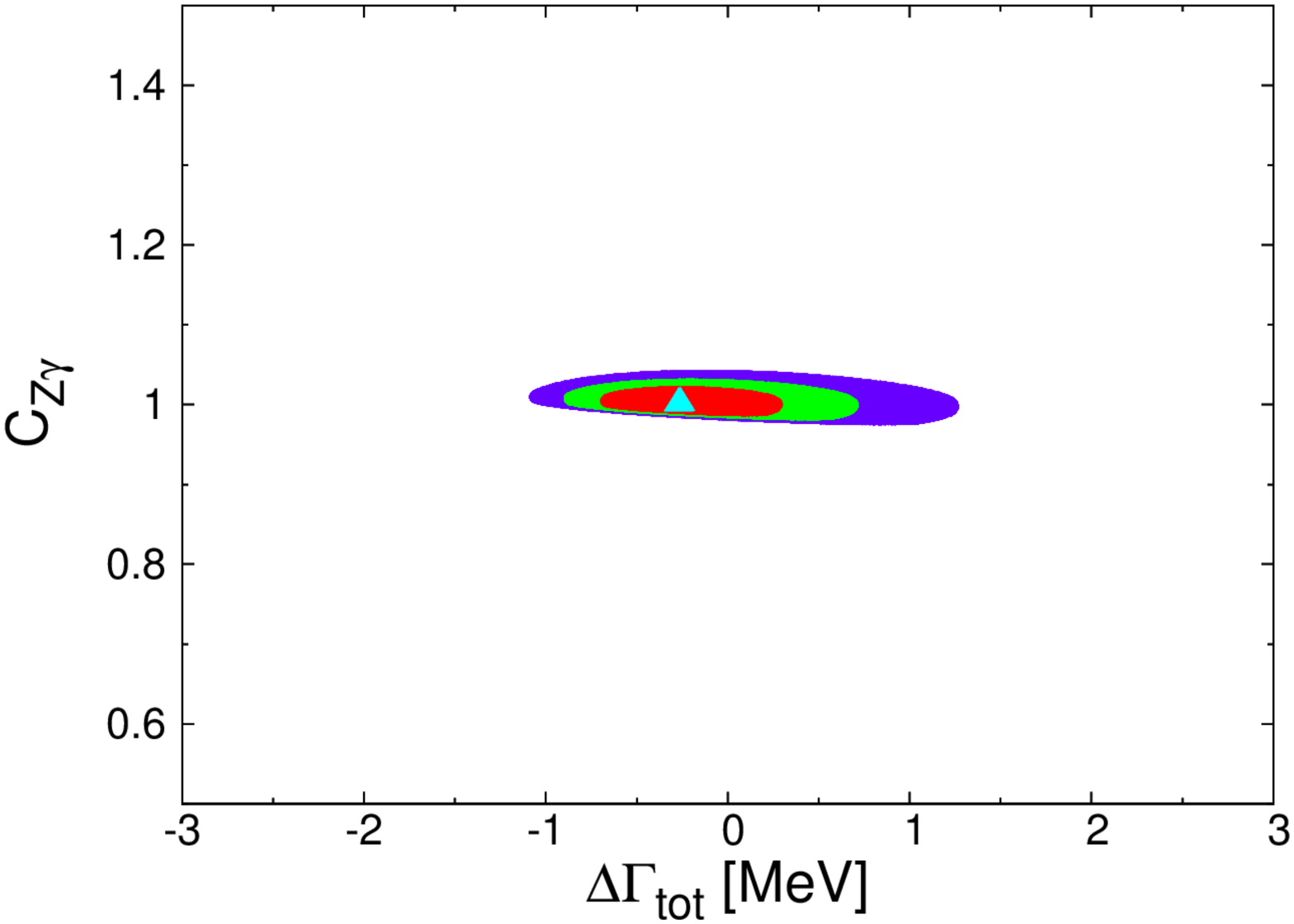}
\caption{\small \label{czgamma}
Predictions for $C_{Z\gamma}$ for the scenarios in which $\Delta S^\gamma=0$.
[Upper]: {\bf CPC1} (left), {\bf CPCN2} (middle), {\bf CPCX2} (right) ;
[Middle]: {\bf CPCX3} (left), {\bf CPC4} (middle), {\bf CPCX4} (right) ;
[Low]: {\bf CPV2} (left), {\bf CPV3} (middle), {\bf CPVN3} (right).
The color code is the same as in Fig.~\ref{CPC2} except {\bf CPC1} for which
$\Delta \chi^2 \le 1$ (red), $4$ (green), and $9$ (blue) above the minimum.
}
\end{figure}

\subsection{Predictions for $H \to Z\gamma$}
Before we close this section, we examine how large $C_{Z\gamma}$
can be in the scenarios with $\Delta S^{\gamma} =0$,
assuming the absence of additional particles running in the 
$H$-$\gamma$-$Z$ loop.
The results are shown in Fig.~\ref{czgamma}.
We observe that $C_{Z\gamma}$ can be as large as $1.2$ which may imply
$B(H\to Z\gamma) \lsim 1.4\, B(H_{\rm SM}\to Z\gamma)$.

%
%
\section{Conclusions}

We have performed global fits to the Higgs couplings to gauge bosons
and fermions, using all the data from the Tevatron, 7+8 TeV and 13 TeV
data from ATLAS and CMS. Overall, the allowed parameter space regions
shrink substantially from those in 2014. 
Notably, the data precision is now sensitive to the bottom-Yukawa coupling 
and the overall average signal strength shows a 2-$\sigma$ deviation 
from the SM value.

Let us summarize the major findings or improvements from previous results.
\begin{enumerate}

\item
 The combined average signal strength for the Higgs boson now stands
at a $2$-$\sigma$ deviation from the SM value, namely $\mu_{\rm exp} 
= 1.10 \pm 0.05$.

\item 
For the first time the bottom-Yukawa coupling shows statistical difference
between the positive and negative signs. Thanks to the discriminating power
of the Higgs-gluon vertex $S^g$ the positive sign of the bottom-Yukawa
is more preferred than the negative one.

\item 
Previously in 2014 the fits still allowed the negative sign of the
top-Yukawa coupling at 95\% CL.  Now with more precisely measured signal 
strengths together with the establishment of the associated production with
the top-quark pair, the negative island of the top-Yukawa is now entirely
ruled out, except in the scenarios with non-zero
$\Delta S^\gamma$.

\item
The nonstandard (or invisible decay) branching ratio of the Higgs boson
is now reduced to less than 8.4\%, 
which improves substantially from
the previous value of 19\%. This is obtained by varying only 
$\Delta \Gamma_{\rm tot}$. It would be relaxed if more parameters
are allowed to vary in the fit. 

\item
When we relax the custodial symmetry requirement ($C_w$ not necessarily
equal to $C_z$), we find that the coupling $C_w$ is larger than 
$C_z$ though still within $1\sigma$, and more constrained than $C_z$.

\item
We have also made the predictions for $H \to Z\gamma$ by showing the
effective coupling $C_{Z\gamma}$. In most scenarios, it is predicted to be
SM-like. The most extreme allowed value would $C_{Z\gamma} \simeq 1.2$,
which gives a branching ratio 40\% larger than the SM value.

\end{enumerate}

\section*{Acknowledgment}  
The work of K.C. was supported by the National Science
Council of Taiwan under Grants Nos. MOST-105-2112-M-007-028-MY3 and
MOST-107-2112-M-007-029-MY3.
The work of J.S.L. was supported by
the National Research Foundation of Korea (NRF) grant
No. NRF-2016R1E1A1A01943297.
The work of P.-Y.T. was supported by World Premier International Research 
Center Initiative (WPI), MEXT, Japan. 
%
%
We thank Jubin Park for 
useful discussions. 

\newpage

\section*{Appendix}
\def\theequation{\Alph{section}.\arabic{equation}}

\begin{appendix}

\section{Relation between our formalism and the kappa formalism}
  
Here in this appendix we compare the definitions of coupling modifiers 
taken in our formalism
to those defined by LHCHXSWG, and make the correspondence
taking the specific examples of the loop-induced 
$Hgg$ and $H\gamma\gamma$ couplings. 

In the LHCHXSWG YR3 \cite{YR3} and YR4 \cite{YR4}, as well as in a
very recent paper by ATLAS \cite{atlas-k},
the definition of $\kappa_g$ is given by 
\begin{equation}
    \kappa_g^2 (\kappa_b, \kappa_t, M_H) \equiv
    \frac{ \kappa_t^2 \cdot \sigma^{tt}_{ggH} (M_H) +
      \kappa_b^2 \cdot \sigma^{bb}_{ggH} (M_H) +
      \kappa_t \kappa_b \cdot \sigma^{tb}_{ggH} (M_H) }
     { \sigma^{tt}_{ggH} (M_H) + \sigma^{bb}_{ggH} (M_H) +\sigma^{tb}_{ggH} (M_H) }
\end{equation}
where $\sigma^{tt}_{ggH} (M_H)$ and $\sigma^{bb}_{ggH} (M_H)$
denote the squares of the top and bottom contributions
to the $gg\to H$ production, respectively,
and $\sigma^{tb}_{ggH} (M_H)$ the top-bottom interference.
On the other hand, $\kappa_g$ can be also defined through the 
$H\to gg$ decay process:
\begin{equation}
\kappa_g^2 \equiv \frac{\Gamma(H \to gg)} {\Gamma^{\rm SM}( H\to gg)}\,.
\end{equation}
%

The LHCHXSWG is performing analyses 
through $gg\to H$ production at 8 TeV beyond the leading order
and, taking $M_H = 125$ GeV, they find~\cite{YR3,YR4}
\begin{eqnarray}
  \kappa^2_{g,{\rm YR3}} &=& 1.058 \kappa_t^2 + 0.007 \kappa_b^2 - 0.065\kappa_t \kappa_b
  \nonumber \\
  \kappa^2_{g,{\rm YR4}} &=& 1.042 \kappa_t^2 + 0.002 \kappa_b^2 - 0.040\kappa_t \kappa_b
  - 0.005\kappa_t \kappa_c + 0.0005\kappa_b \kappa_c + 0.00002 \kappa_c^2
  \;.
\end{eqnarray}
The difference between $\kappa^2_{g,{\rm YR3}}$ and $\kappa^2_{g,{\rm YR4}}$
can be attributed to the choices of the QCD and factorization scales and the PDF set,
the different remormalization
scheme for the masses of the fermions entering into the loops, etc.
On the other hand, in the recent paper by ATLAS, 
the simpler $\kappa_g^2$ based on  
the Higgs decay $H\to gg$ is taken and they find~ \cite{atlas-k}:
\begin{equation}
\kappa^2_{g,{\rm ATLAS}} =
1.11 \kappa_t^2 + 0.01\kappa_b^2  - 0.12 \kappa_t \kappa_b
\end{equation}
which is closer to $\kappa^2_{g,{\rm YR3}}$.

In our work, we only perform LO analysis and $\kappa_g^2$ is given by
\begin{equation}
\kappa_g^2=C^2_g={\frac{\left|S^g(M_H)\right|^2+\left|P^g(M_H)\right|^2}
{\left|S^g_{\rm SM}(M_H)\right|^2}}
\end{equation}
independently of whether one considers $gg\to H$ production or $H\to gg$ decay.
Using the numerical expression Eq.~(\ref{eq:hgg})
which is obtained by taking $M_H = 125.09$ GeV and $\Delta S^g=P^g=0$, we have
\begin{equation}
\kappa^2_{g,{\rm OURS}} \simeq 
1.11 C_t^2 + 0.01 C_b^2 - 0.12 C_t C_b \;,
\end{equation}
which is very consistent with $\kappa^2_{g,{\rm ATLAS}}$ based on $H\to gg$ decay
with the identification of $C_f=\kappa_f$.

One of the important findings of our work is the preferred sign of the
bottom-Yukawa coupling $C_b$. The observable effect on signal strength
due to flipping of the sign of $C_b$ comes from the {\it interference} term
($\propto C_t C_b$ or $\kappa_t \kappa_b$), but not the square of the
bottom-Yukawa term.
Thus, all three expressions of $\kappa^2_{g,{\rm YR3}}$,
$\kappa^2_{g,{\rm YR4}}$, and $\kappa^2_{g,{\rm ATLAS}}$
and our expression of $\kappa^2_{g,{\rm OURS}}$ yield a similar change 
of order $O(10)\%$
due to the flipping of the sign of the bottom-Yukawa coupling.

Similarly, LHCHXSWG gives the definition for $\kappa_\gamma$:
\begin{eqnarray}
\kappa_\gamma^2 \equiv  \frac{ \sum_{i,j} \kappa_i \kappa_j
  \Gamma^{ij}_{\gamma\gamma} (M_H)}{ \sum_{i,j} \Gamma^{ij}_{\gamma\gamma} (M_H)} 
 \simeq 
 1.59 \kappa^2_W+0.07 \kappa^2_t-0.67 \kappa_W \kappa_t \nonumber
\end{eqnarray}
where the pairs
$(i, j)$ are $bb$, $tt$, $\tau\tau$, $WW$, $bt$, $b\tau$, $bW$, $t\tau$,
$tW$, and $\tau W$. 
The above numerical expression is compared with
that obtained by the use of Eq.~(\ref{eq:haa}) after
taking $M_H = 125.09$ GeV and $\Delta S^\gamma=P^\gamma=0$:
\begin{eqnarray}
\kappa^2_\gamma &\simeq & 
 1.583 C^2_W+0.070 C^2_t-0.667 C_W C_t + 0.006 C_W C_b \nonumber \\
&& + 0.009 C_W C_\tau
 +0.003 C_W C_c-0.001 C_t C_b -0.002 C_t C_\tau  \,, \nonumber
\end{eqnarray}
and we find excellent consistency.
We note that our formalism takes on the advantage that it can admit pseudoscalar form
factor $P^g$ and $P^\gamma$ into the effective $Hgg$ and $H\gamma\gamma$
vertices.

\setcounter{equation}{0}
\section{13 TeV Data: Tables~\ref{aa}, \ref{zz}, \ref{ww},
\ref{bb}, \ref{tautau}, and \ref{tth}}

In this appendix, we list all the details of 13 TeV Higgs 
signal strengths used in our global fitting.

\begin{table}[h!]
\caption{\small \label{aa}
{\bf (LHC: 13 TeV)} Data on signal strengths of $H\rightarrow \gamma \gamma$
by the ATLAS and CMS after ICHEP 2018.
The $\chi^2$ of each data with respect to the SM is shown in the last column.
The sub-total $\chi^2$ of this decay mode is shown at the end. \smallskip
}
\begin{ruledtabular}
\begin{tabular}{cccr}
Channel & Signal strength $\mu$ & $M_H$(GeV)   & $\chi^2_{\rm SM}$(each)\\
        & c.v $\pm$ error       &             & \\
\hline
\multicolumn{4}{c}
{ATLAS (79.8${\rm fb}^{-1}$(13TeV)): 
Fig.8 of \cite{ICHEP_haa_ATLAS}(Jul. 2018)}\\
\hline
ggF & $0.97^{+0.15}_{-0.14}$\cite{ICHEP_haa_ATLAS}  & 125.09 & 0.04 \\
VBF & $1.40^{+0.43}_{-0.37}$\cite{ICHEP_haa_ATLAS}  & 125.09 & 1.17 \\
VH  & $1.08^{+0.59}_{-0.54}$\cite{ICHEP_haa_ATLAS}  & 125.09 & 0.02 \\
ttH & $1.12^{+0.43}_{-0.37}$\cite{ICHEP_haa_ATLAS}  & 125.09 & 0.11 \\
\hline
\multicolumn{4}{c}
{CMS (35.9${\rm fb}^{-1}$(13TeV)): FIG.17 of \cite{Sirunyan:2018ouh} (Apr. 2018) }\\
\hline
ggF & $1.10^{+0.20}_{-0.18}$\cite{Sirunyan:2018ouh} & 125.4 & 0.31 \\
VBF & $0.8^{+0.6}_{-0.5}$\cite{Sirunyan:2018ouh}    & 125.4 & 0.11 \\
VH  & $2.4^{+1.1}_{-1.0}$\cite{Sirunyan:2018ouh}    & 125.4 & 1.96 \\
ttH & $2.2^{+0.9}_{-0.8}$\cite{Sirunyan:2018ouh}    & 125.4 & 2.25 \\
\hline
&&& subtot: 5.97 

\end{tabular}
\end{ruledtabular}
\end{table}

\begin{table}[h!]
\caption{\small \label{zz}
{\bf (LHC: 13 TeV)} The same as Table~\ref{aa} but for $H\rightarrow Z Z^{(\ast)}$.
$^\dagger$This data point is not included in our $\chi^2$ analysis. \smallskip }
\begin{ruledtabular}
\begin{tabular}{cccr}
Channel & Signal strength $\mu$ & $M_H$(GeV)  & $\chi^2_{\rm SM}$(each)\\
        & c.v $\pm$ error       &             & \\
\hline
\multicolumn{4}{c}
{ATLAS (79.8${\rm fb}^{-1}$ at 13TeV): Tab.9 of \cite{ICHEP_hzz_ATLAS}(Jun. 2018}\\
\hline
ggF & $1.04^{+0.16}_{-0.16}$\cite{ICHEP_hzz_ATLAS}   & 125 & 0.06 \\
VBF & $2.8^{+0.94}_{-0.94}$\cite{ICHEP_hzz_ATLAS}   & 125 & 3.67 \\
VH  & $0.9^{+1.01}_{-1.01}$\cite{ICHEP_hzz_ATLAS}   & 125 & 0.01 \\
ttH\,$^\dagger$ & $<4.04(95\%)$\cite{ICHEP_hzz_ATLAS}           & 125  & - \\
\hline
\multicolumn{4}{c}
{CMS (77.4${\rm fb}^{-1}$ at 13TeV): FIG.10 of \cite{ICHEP_hzz_CMS} (Jul. 2018) }\\
\hline
ggF & $1.15^{+0.18}_{-0.16}$\cite{ICHEP_hzz_CMS}   & 125.09    & 0.88\\
VBF & $0.69^{+0.75}_{-0.57}$\cite{ICHEP_hzz_CMS}   & 125.09    & 0.17 \\
VH$_{\rm had}$  & $0.00^{+1.16}_{-0.00}$\cite{ICHEP_hzz_CMS} & 125.09   & 0.74 \\
VH$_{\rm lep}$  & $1.25^{+2.46}_{-1.25}$\cite{ICHEP_hzz_CMS} & 125.09   & 0.04 \\
ttH & $0.00^{+0.53}_{-0.00}$\cite{ICHEP_hzz_CMS}   & 125.09    & 3.56 \\
\hline
&&& subtot: 9.13 

\end{tabular}
\end{ruledtabular}
\end{table}

\begin{table}[h!]
\caption{\small \label{ww}
{\bf (LHC: 13 TeV)} The same as Table~\ref{aa} but for $H\rightarrow W^+W^-$. \smallskip}
\begin{ruledtabular}
\begin{tabular}{cccr}
Channel & Signal strength $\mu$ & $M_H$(GeV)   & $\chi^2_{\rm SM}$(each)\\
        & c.v $\pm$ error       &            &  \\
\hline
\multicolumn{4}{c}
{ATLAS (36.1${\rm fb}^{-1}$(13TeV)): page 8 of \cite{ATLAS:2018gcr}(Mar. 2018)}\\
\hline
ggF & $1.21^{+0.22}_{-0.21}$\cite{ATLAS:2018gcr}  & 125 & 1.00 \\
VBF & $0.62^{+0.37}_{-0.36}$\cite{ATLAS:2018gcr}  & 125 & 1.05 \\
\hline
\multicolumn{4}{c}
{CMS (35.9${\rm fb}^{-1}$ at 13TeV): Fig.9 of \cite{CMS:2018xuk} (Mar. 2018) }\\
\hline
ggF & $1.38^{+0.21}_{-0.24}$\cite{CMS:2018xuk}   & 125.09 & 2.51 \\
VBF & $0.29^{+0.66}_{-0.29}$\cite{CMS:2018xuk}   & 125.09 & 1.16 \\
WH  & $3.27^{+1.88}_{-1.70}$\cite{CMS:2018xuk}   & 125.09 & 1.78 \\
ZH  & $1.00^{+1.57}_{-1.00}$\cite{CMS:2018xuk}   & 125.09 & 0.00 \\
\hline
&&& subtot: 7.50 

\end{tabular}
\end{ruledtabular}
\end{table}

\begin{table}[h!]
\caption{\small \label{bb}
{\bf (LHC: 13 TeV)} The same as Table~\ref{aa} but for $H\rightarrow b\bar{b}$.
\smallskip}
\begin{ruledtabular}
\begin{tabular}{cccr}
Channel & Signal strength $\mu$ & $M_H$(GeV)   & $\chi^2_{\rm SM}$(each)\\
        & c.v $\pm$ error       &            &  \\
\hline
\multicolumn{4}{c}
{ ATLAS (79.8${\rm fb}^{-1}$(13TeV)) Fig.3 of
 \cite{Aaboud:2018zhk} (Aug. 2018)}\\
\hline
WH & $1.08^{+0.47}_{-0.43}$\cite{Aaboud:2018zhk}  & 125.0 & 0.03 \\
ZH & $1.20^{+0.33}_{-0.31}$\cite{Aaboud:2018zhk}  & 125.0 & 0.42 \\
\hline
\multicolumn{4}{c}
{CMS (77.2${\rm fb}^{-1}$(13TeV)) \cite{ICHEP_hbb_CMS,Sirunyan:2018kst} (Aug. 2018)}\\
\hline
ggH       & $2.51^{+2.43}_{-2.01}$\cite{ICHEP_hbb_CMS}    & 125.09 & 0.56 \\
VH        & $1.06^{+0.26}_{-0.26}$\cite{Sirunyan:2018kst} & 125.09 & 0.05 \\
ttH       & $0.91^{+0.45}_{-0.43}$\cite{ICHEP_hbb_CMS}    & 125.09 & 0.04\\
\hline
&&& subtot: 1.11
\end{tabular}
\end{ruledtabular}
\end{table}
\begin{table}[h!]
\caption{\small \label{tautau}
{\bf (LHC: 13 TeV)} The same as Table~\ref{aa} but for $H\rightarrow \tau^+\tau^-$.
\smallskip}
\begin{ruledtabular}
\begin{tabular}{cccr}
Channel & Signal strength $\mu$ & $M_H$(GeV)   & $\chi^2_{\rm SM}$(each)\\
        & c.v $\pm$ error       &            &  \\
\hline
\multicolumn{4}{c}
{ATLAS (36.1${\rm fb}^{-1}$ at 13TeV)\cite{ICHEP_htautau_ATLAS} (Jun. 2018)}\\
\hline
ggH & $0.98^{+0.62}_{-0.51}$\cite{ICHEP_htautau_ATLAS}  & 125.09 & 0.00 \\
VBF & $1.18^{+0.59}_{-0.55}$\cite{ICHEP_htautau_ATLAS}  & 125.09 & 0.11 \\
\hline
\multicolumn{4}{c}
{CMS (35.9${\rm fb}^{-1}$ at 13TeV)Fig.6 of \cite{ICHEP_htautau_CMS} (Jun. 2018)}\\
\hline
ggH      & $1.12^{+0.53}_{-0.50}$\cite{ICHEP_htautau_CMS} & 125.09 & 0.06 \\
VBF      & $1.13^{+0.45}_{-0.42}$\cite{ICHEP_htautau_CMS} & 125.09 & 0.10 \\
WH      & $3.39^{+1.68}_{-1.54}$\cite{ICHEP_htautau_CMS} & 125.09 & 2.41 \\
ZH      & $1.23^{+1.62}_{-1.35}$\cite{ICHEP_htautau_CMS} & 125.09 & 0.03 \\
\hline
&&& subtot: 2.70
\end{tabular}
\end{ruledtabular}
\end{table}

\begin{table}[h!]
\caption{\small \label{tth}
{\bf (LHC: 13 TeV)} The same as Table~\ref{aa} but for exclusive $ttH$ production mode. 
$^\dagger$This data point is not included in our $\chi^2$ analysis. \smallskip}
\begin{ruledtabular}
\begin{tabular}{cccr}
Channel & Signal strength $\mu$ & $M_H$(GeV)   & $\chi^2_{\rm SM}$(each)\\
        & c.v $\pm$ error       &            & \\
\hline
\multicolumn{4}{c}
{ATLAS (79.8${\rm fb}^{-1}$(13TeV)): Fig.5 of \cite{Aaboud:2018urx}(Jun. 2018),
(36.1${\rm fb}^{-1}$(13TeV)): Fig.16 of \cite{Aaboud:2017jvq}(Apr. 2018)}\\
\hline
$\gamma\gamma$ & $1.39^{+0.48}_{-0.42}$(79.8${\rm fb}^{-1}$)\cite{Aaboud:2018urx}  & 125.09 & 0.86 \\
$ZZ^{(\ast)}$\,$^\dagger$ & $<1.77(68\%)$(79.8${\rm fb}^{-1}$)\cite{Aaboud:2018urx}  & 125.09 & - \\
$WW^{(\ast)}$ & $1.5^{+0.6}_{-0.6}$(36.1fb$^{-1}$)\cite{Aaboud:2017jvq}  & 125.09 & 0.69 \\
bb & $0.84^{+0.64}_{-0.61}$(36.1fb$^{-1}$)\cite{Aaboud:2017rss}  & 125.09 & 0.06 \\
$\tau\tau$ & $1.5^{+1.2}_{-1.0}$(36.1fb$^{-1}$)\cite{Aaboud:2017jvq}  & 125.09 & 0.25 \\
\hline
\multicolumn{4}{c}
{CMS (35.9${\rm fb}^{-1}$(13TeV)) }\\
\hline
$WW^{(\ast)}$ & $1.69^{+0.68}_{-0.61}$\cite{ICHEP_tth_CMS_VVtautau}    & 125.09 & 1.28 \\
bb({\rm hadronic}) & $0.9^{+1.5}_{-1.5}$\cite{ICHEP_tth_CMS_bb1}       & 125.09 & 0.00 \\
bb({\rm leptonic}) & $0.72^{+0.45}_{-0.45}$\cite{ICHEP_tth_CMS_bb2}    & 125.09 & 0.39 \\
$\tau\tau$ & $0.15^{+1.07}_{-0.91}$\cite{ICHEP_tth_CMS_VVtautau}       & 125.09 & 0.63 \\
\hline
&&& subtot: 4.17 
\end{tabular}
\end{ruledtabular}
\end{table}

\end{appendix}

\clearpage


\end{document}